 \documentclass[final,5p,times,twocolumn]{elsarticle}

\usepackage{amssymb}
\usepackage{amsthm}
\usepackage{amsmath}

\journal{Computer Physics Communications}

\newcommand{\refeq}[1]{(\ref{#1})}
\begin{document}

\begin{frontmatter}


\title{The field line map approach for simulations of magnetically confined plasmas}
\author[label1]{Andreas Stegmeir}
\ead{Andreas.Stegmeir@ipp.mpg.de}
\author[label1]{David Coster}
\author[label1]{Omar Maj}
\author[label1]{Klaus Hallatschek}
\author[label1]{Karl Lackner}

\address[label1]{Max-Planck-Institut f\"{u}r Plasmaphysik, EURATOM Association, 85748 Garching, Germany}


\begin{abstract}
In the presented field line map approach the simulation domain of a tokamak is covered with a cylindrical grid, which is Cartesian within poloidal planes. Standard finite-difference methods can be used for the discretisation of perpendicular (w.r.t.~magnetic field lines) operators. The characteristic flute mode property $\left(k_{\parallel}\ll k_{\perp}\right)$ of structures is exploited computationally by a grid sparsification in the toroidal direction. A field line following discretisation of parallel operators is then required, which is achieved via a finite difference along magnetic field lines. This includes field line tracing and interpolation or integration. The main emphasis of this paper is on the discretisation of the parallel diffusion operator. Based on the support operator method a scheme is constructed which exhibits only very low numerical perpendicular diffusion. The schemes are implemented in the new code GRILLIX, and extensive benchmarks are presented which show the validity of the approach in general and GRILLIX in particular. The main advantage of the approach is that it does not rely on field/flux-aligned, which become singular on the separatrix/X-point. Most tokamaks are based on the divertor concept, and the numerical treatment of the separatrix is therefore of importance for simulations of the edge and scrape-off layer.  
\end{abstract}


\end{frontmatter}


\section{Introduction}\label{sec_introduction}
The modelling of the edge and scrape-off layer of tokamaks is in many ways more difficult than the core \cite{dendy_obrien:tokamak93}. However, this region is of high importance since it may have a significant influence also on the core region, e.g.~plasma and impurity densities are often largely set by edge conditions and in important operating conditions the edge plays a key role in the improvement of confinement \cite{stangeby:boundaryplasma90}. Moreover, a prediction of heat fluxes on the divertor plates for future tokamaks is of high importance from the engineering point of view \cite{dietz:diviter93,dietz:divheatflux95}. 

\begin{figure}
a)\hspace{0.24\textwidth}b)\newline
\includegraphics[height=0.33\textwidth]{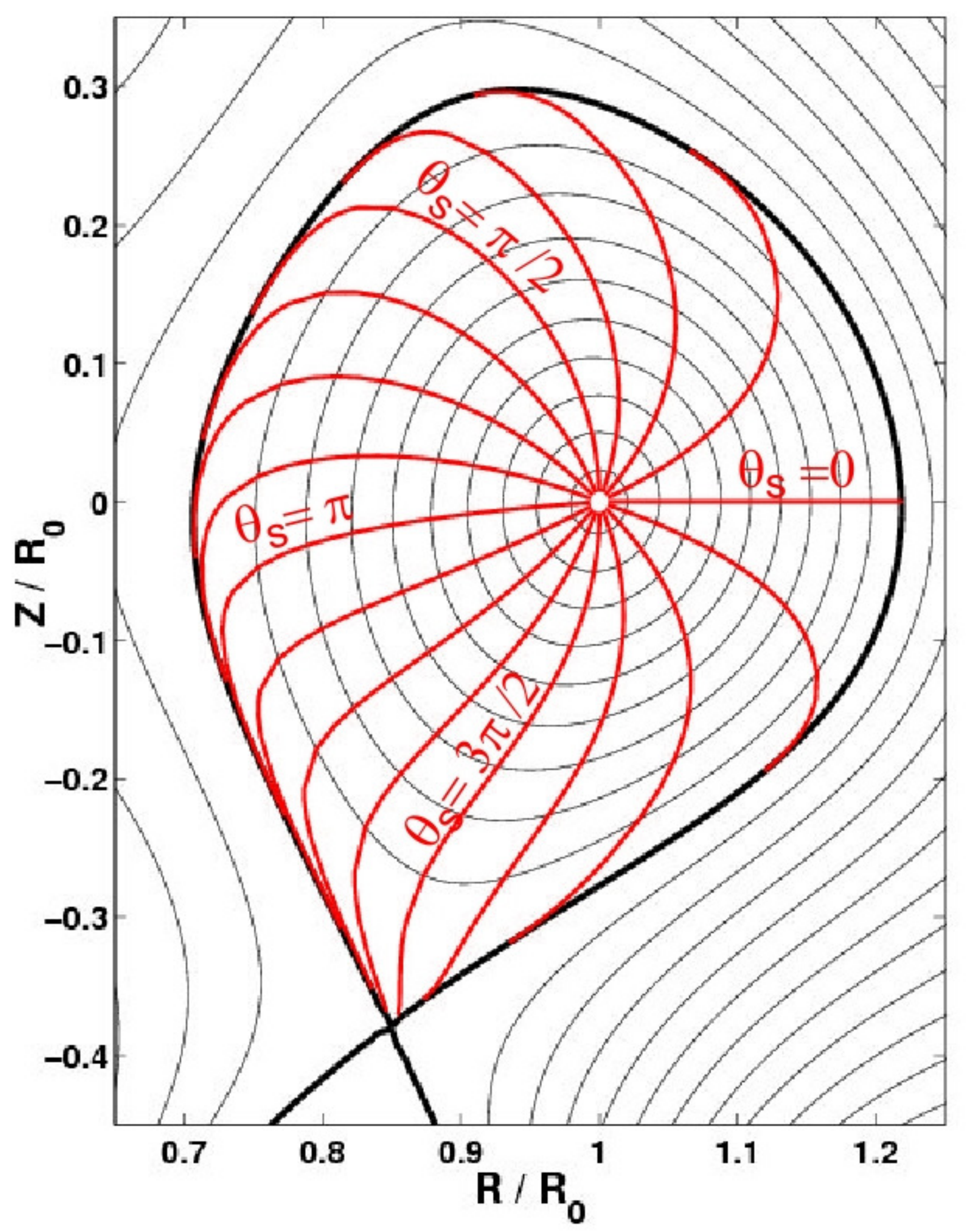}
\includegraphics[height=0.33\textwidth]{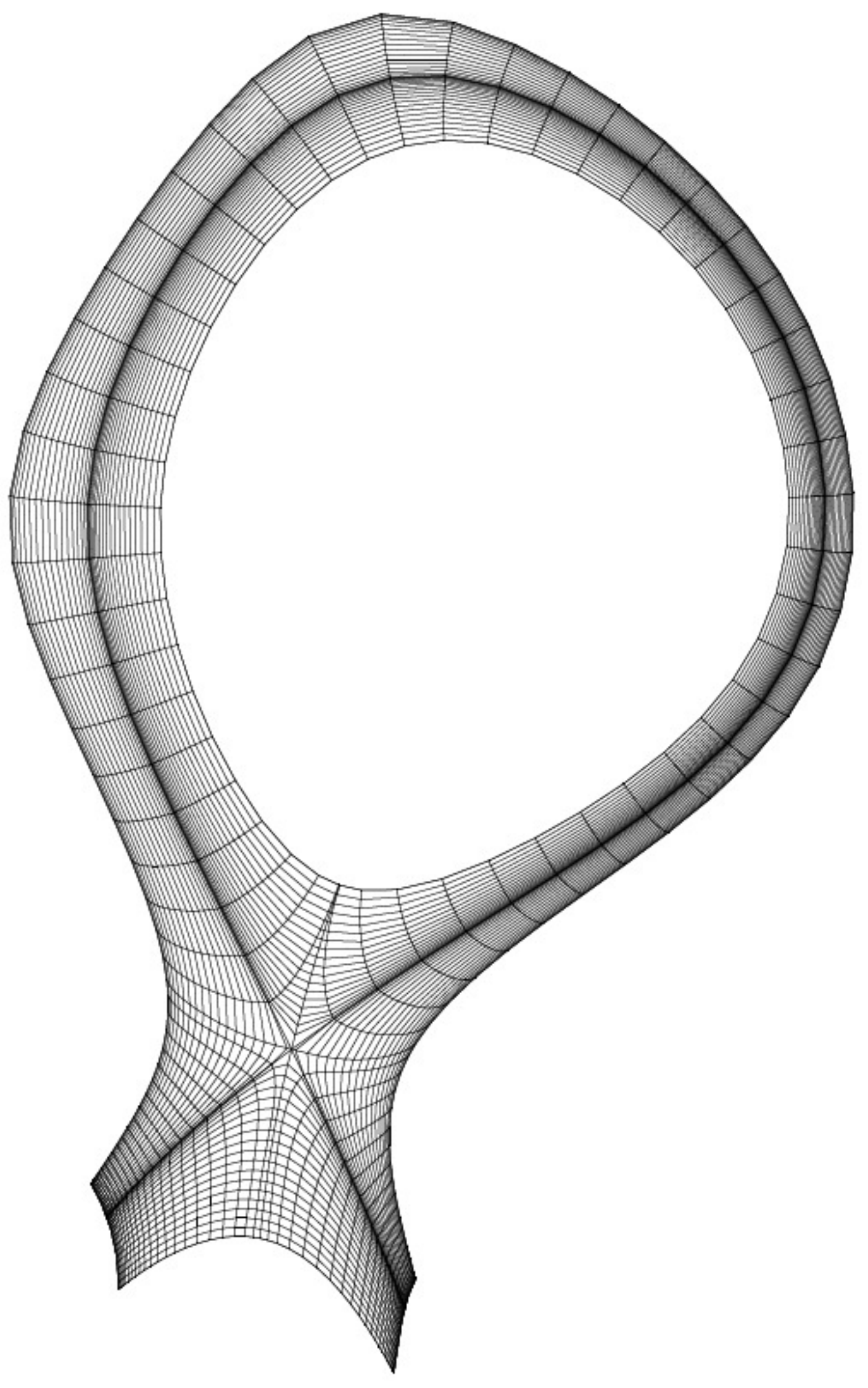}
\caption{a) Contours of flux label $\rho$ (black) and poloidal straight field line angle $\theta_s$ (red, symmetry coordinates) in diverted geometry for equilibrium from \cite{mccarthy:equi99}. b) Example for flux aligned mesh. The contours of the poloidal coordinate are orthogonal to flux surfaces. The X-point is connected to eight cells instead of the usual four cells.}
\label{fig_contours_straight}
\end{figure}

A major complexity at the modelling is introduced by the complex geometry of diverted machines. Field-aligned coordinates are often employed in simulations, since they allow for a convenient way to computationally exploit the characteristic flute mode property $\left(k_{\parallel}\ll k_{\perp}\right)$ of the structures. However, field-aligned coordinates become singular on the separatrix and simulations cannot span a domain across it. Any set of poloidal $\left(\theta_s\right)$ and toroidal $\left(\varphi_s\right)$ straight field line angles has to satisfy along magnetic field lines the condition \cite{haseleer:ccord90}:
\begin{align}
\frac{d\theta_s}{d\varphi_s}=\frac{1}{q(\psi)}.
\label{straight_angles}
\end{align}
At the X-point the poloidal magnetic field vanishes and therefore on the separatrix the safety factor $q$ diverges. The straight field line angles, which have to satisfy condition \refeq{straight_angles}, cannot span the whole separatrix. As exemplified in fig.~\ref{fig_contours_straight}a the contours of $\theta_s$ are sucked into the X-point (see also \cite{strumberger:coordexample04}).

Also often employed are coordinates, where the field-alignment property is given up, but which is still aligned with flux surfaces, i.e.~$\rho(\psi)$ is retained as a radial coordinate. However, flux-aligned coordinate systems are still singular on points, where $\nabla\psi=0$, i.e.~at O-points and X-points \cite{mattor:xcoord95}. Although these singularities can be cured numerically, O- and X-points remain somewhat exceptional points of the numerical grid (see fig.~\ref{fig_contours_straight}b). This could in the worst case even lead to numerical artefacts. Moreover, structured flux-aligned meshes have a huge resolution imbalance within the poloidal plane due to the flux expansion near the X-point \cite{strauss:xsim96,dudson:bout++15}. Simulations might suffer from this as perpendicular operators arising in practically any plasma model (e.g.~$\nabla_{\perp}^2,\mathbf{v}_E\cdot\nabla)$ act approximately isotropically within poloidal planes of tokamaks.   

The field line map approach is presented in section \ref{sec_fieldlinemap}. Although field/flux-aligned coordinates may become singular, the operators appearing in plasma models are still well defined, of course. The idea behind the approach is that the flute mode property can also be exploited at the discretisation step without any need for construction of a field/flux-aligned coordinate system. The approach consists of a cylindrical or Cartesian grid with a field line following discretisation for parallel operators. A separatrix can be treated as well as a magnetic axis, where X/O-points are treated like any other grid point and no resolution imbalance arises. Ultimately, the result is similar to the flux-coordinate independent (FCI) \cite{ottaviani:coord11,hariri:fenicia13,hariri:xcoord14} approach. However, as the motivation for the FCI approach was initially still based on field-aligned coordinate systems, the derivation is here performed completely without any reference to field- or flux-aligned coordinate systems.

As the discretisation of perpendicular operators in the field line map approach is straight forward, the main emphasis in this paper is on the discretisation of parallel operators, especially the parallel diffusion operator in section \ref{sec_pardiff}. Since an interpolation or integration is involved at the discretisation, parallel operators exhibit also numerical perpendicular 'diffusion'. Motivated by previous work from \cite{guenter:numpoll05,guenter:numpoll07}, a numerical scheme is developed which exhibits very low numerical diffusion. The discussion extends previous work from \cite{stegmeir:cpp14}. Several model problems are also discussed in \ref{appendix}  

The developed numerical methods are implemented in the new code GRILLIX. In section \ref{sec_benchmarks} extensive benchmarks performed with GRILLIX are presented, which show the validity of the field line map approach in general and GRILLIX in particular. 

The paper is concluded with a summary and final remarks in section \ref{sec_summary}.

\section{Field line map approach}\label{sec_fieldlinemap}

\subsection{Overview}
The field line map approach is described in the following for the case of a toroidal configuration $\left(R,Z,\varphi\right)$, but it can be applied also to axial periodic configurations $\left(x,y,z\right)$, where $z$ is the axial coordinate. The transition should be trivial.

For a tokamak a cylindrical coordinate system is well defined everywhere, except for the toroidal symmetry axis which is outside the domain of interest. We span the simulation domain with a cylindrical grid $R_i,Z_j,\varphi_k$ (For axial configurations a Cartesian grid $x_i,y_j,z_k$ is used). Within each poloidal plane $k$ the grid $\left(R_i,Z_j\right)$ is Cartesian and bounded by extreme flux surfaces, which is the only dependence on flux surfaces of the approach. Based on the assumption of a strong toroidal field $\left(B^{tor}\gg B^{pol}\right)$, any perpendicular operator is approximated by derivatives with respect to only $R$ and $Z$, but not $\varphi$. In order to exploit the flute mode property $k_{\parallel}\ll k_{\perp}$, a dense resolution is chosen within poloidal planes, whereas the grid is sparsified along the $\varphi$ direction. The low resolution in $\varphi$ requires a field line following discretisation for parallel operators to achieve a sufficient directional accuracy for parallel operators. 

It is noted that the assumption of a strong toroidal field is not a strictly necessary condition. If this assumption breaks down, a dense resolution also in $\varphi$ would have to be retained, and the perpendicular operators would have to be adjusted to take into account also derivatives with respect to $\varphi$. Ultimately, the field line map approach would go over into a discretisation on a dense cylindrical grid, where the flute mode property would not be exploited any more. However, the method would still retain its validity.

\subsection{Perpendicular operators}
Under the assumption of a strong toroidal field, any perpendicular operator can be approximated with derivatives with respect to only $R$ and $Z$, e.g.~the perpendicular Laplace operator becomes:
\begin{align}
\nabla_\perp^2 =&\sum\limits_{x^n,x^m=R,Z,\varphi}\frac{1}{R}\frac{\partial}{\partial x^n}\left[R\left(g^{nm}-b^nb^m\right)\frac{\partial}{\partial x^m}\right] \notag \\
\approx&\frac{\partial^2}{\partial R^2}+\frac{\partial^2}{\partial Z^2}+\frac{1}{R}\frac{\partial}{\partial R},
\end{align}
where $g^{nm}$ the inverse metric of the cylindrical coordinate system and $b^n$ the contravariant components of the unit vector of the magnetic field. Hence, the Stencil of any perpendicular operator remains within the Cartesian poloidal planes. Standard finite difference methods can be used for their discretisation (e.g.~\cite{ames:nummethpde92}), e.g.~a second order finite difference method yields for the discrete perpendicular Laplace operator $\mathbf{D}_\perp$:
\begin{align}
\left(\mathbf{D}_\perp\mathbf{u}\right)_{i,j,k}:=&\frac{u_{i+1,j,k}+u_{i-1,j,k}+u_{i,j+1,k}+u_{i,j-1,k}-4u_{i,j,k}}{h^2} \notag \\
&+\frac{u_{i+1,j,k}-u_{i-1,j,k}}{2R_{i}h},
\end{align}
where $h$ denotes the Cartesian grid spacing. Bold face is used on the discrete level for vectors and matrices, i.e.~$\mathbf{u}:=\left(u_{1,1,1},u_{2,1,1},\dots\right)^T$, and the subscript $i,j,k$ denotes the corresponding grid point.

\subsection{Parallel gradient}
The parallel gradient is:
\begin{align}
\nabla_{\parallel}u=\lim\limits_{\epsilon\rightarrow 0}\frac{u\left(\mathbf{x}^\epsilon\right)-u\left(\mathbf{x}\right)}{\epsilon},
\end{align}
where $\epsilon$ is the arc length along a magnetic field line and:
\begin{align}
\frac{d\mathbf{x}^\epsilon}{d\epsilon}=&\mathbf{b}, &  \mathbf{x}^\epsilon(0)=\mathbf{x}.
\end{align}
This motivates the discretisation of the parallel gradient via a finite difference along magnetic field lines. The Stencil will cover neighbouring poloidal planes, and since the grid is sparsified along the toroidal direction, a field line following discretisation has to be performed. Field lines are traced from each grid point towards the neighbouring poloidal planes according to:
\begin{align}
R_{i,j}^{\pm}=R_{i}+\int\limits_{0}^{\pm\Delta\varphi}\frac{B^R}{B^\varphi}\,d\varphi, \quad Z_{i,j}^{\pm}=Z_{j}+\int\limits_{0}^{\pm\Delta\varphi}\frac{B^Z}{B^\varphi}\,d\varphi,
\end{align}
with $B^R,B^Z,B^\varphi$ the contravariant components of the magnetic field. The integrations start from grid points $\left(R_{i},Z_j\right)$, and $'\pm'$ denotes co/counter-direction with respect to magnetic field. For axisymmetric equilibria the penetration points $\left(R_{i,j}^\pm,Z_{i,j}^\pm\right)$ are independent of the toroidal grid index $k$. Also the lengths along field lines are computed according to:
\begin{align}
\Delta s^{\pm}_{i,j}=\int\limits_{0}^{\pm\Delta\varphi}\sqrt{1+\frac{\left(B^R\right)^2+\left(B^Z\right)^2}{\left(B^{tor}\right)^2}}R\,d\varphi,
\end{align}
where $B^{tor}=\sqrt{B^\varphi B_\varphi}=R\,B^\varphi$ is the toroidal field strength.

\begin{figure}
a)\newline
\includegraphics[width=0.48\textwidth]{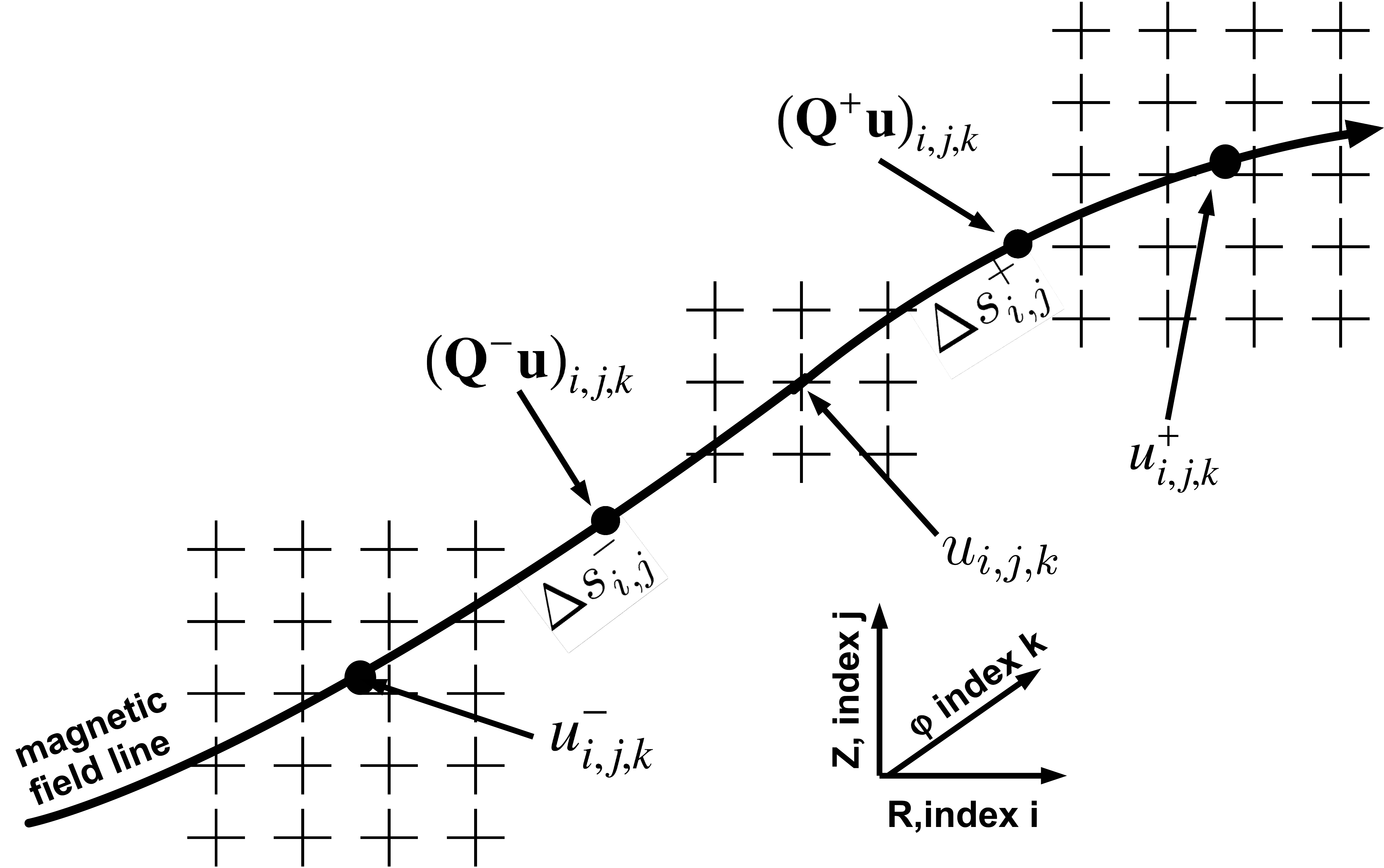}\newline
b)\newline
\includegraphics[width=0.48\textwidth]{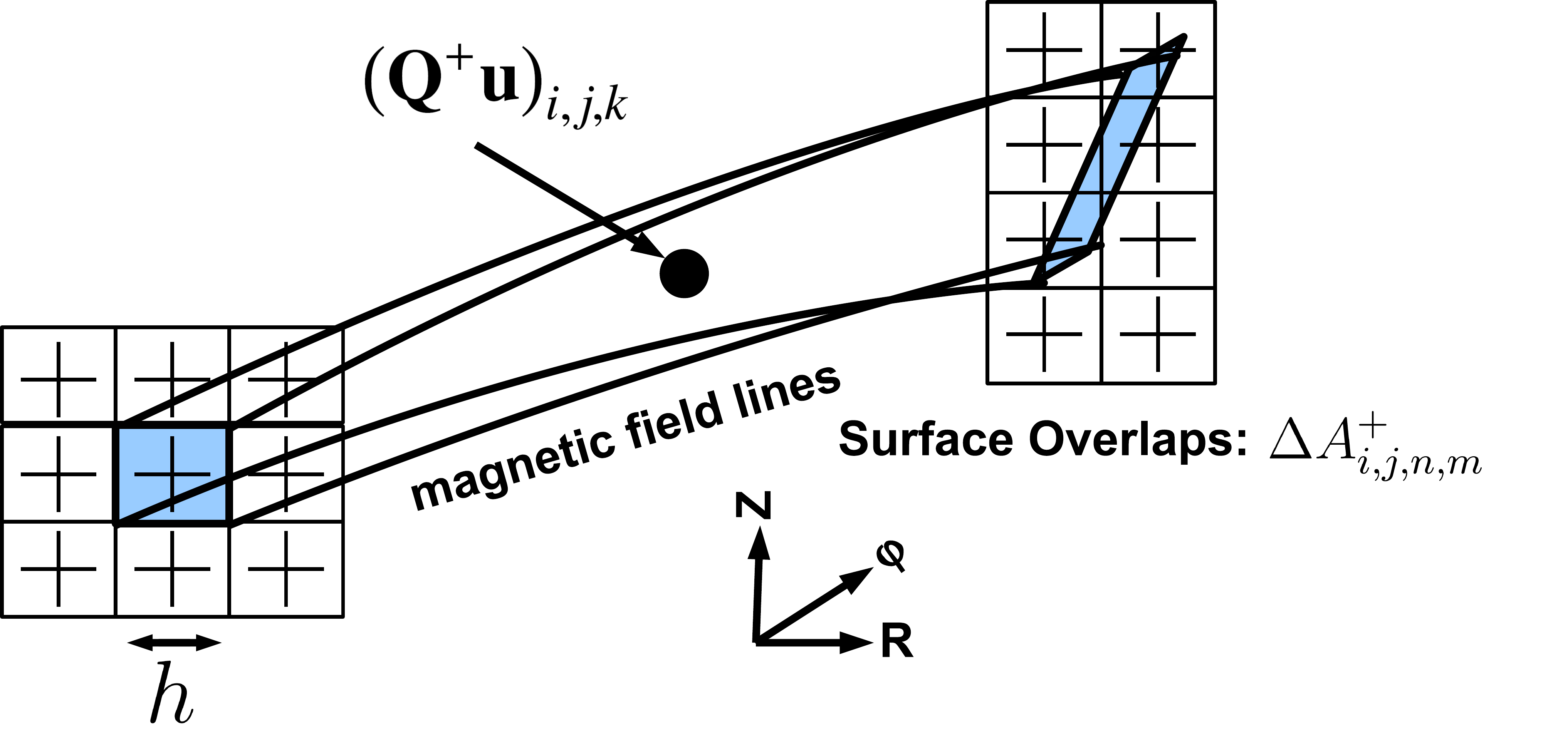}
\caption{Sketch of discrete parallel gradient: a) Interpolation method, b) integration method.}
\label{fig_par_grad}
\end{figure}

A sketch of the discretisation of the parallel gradient is shown in fig.~\ref{fig_par_grad}a. The discrete parallel gradient is collocated at positions half way along the magnetic field line towards the neighbouring poloidal planes. This gives rise to two possible discretisations ($'+'$ and $'-'$). Since the penetration points do not in general coincide with grid points,  a 2D-interpolation has to be performed within the Cartesian grid, to obtain the values for some quantity $u$ at the penetration points. The interpolated values are denoted with $u_{i,j,k}^{\pm}$. The discrete parallel gradient operators $\mathbf{Q}^\pm$ are defined as:
\begin{align}
\left(\mathbf{Q}^{\pm}\mathbf{u}\right)_{i,j,k}:=\pm\frac{u_{i,j,k}^{\pm}-u_{i,j,k}}{\Delta s_{i,j}^\pm}
\label{par_grad_interpolation}
\end{align}
The parallel gradient at the considered grid point itself could be obtained via a further linear interpolation of the discrete parallel gradient along the magnetic field line. However, since this work will concentrate on the discretisation of the parallel diffusion operator, this issue is left for future work. Results on the parallel gradient can also be found in \cite{hariri:fenicia13}.

Later on, especially for field lines which are strongly distorted, another discretisation of the parallel gradient is useful, which is based on integration:
\begin{align}
\nabla_\parallel u=\frac{1}{B}\nabla\cdot\left(u\mathbf{B}\right)=\frac{1}{B}\lim\limits_{V\rightarrow 0}\frac{1}{V}\int\limits_{\partial V}u\mathbf{B}\cdot d\mathbf{S}.
\label{coordfree_continuouslevel}
\end{align}
The surface integration can be mimicked on the discrete level. Flux boxes around magnetic field lines are taken as discrete finite volumes (see fig.~\ref{fig_par_grad}b), such that the only contributions to the surface integral come from the toroidal ends of the flux box:
\begin{align}
&\left(\mathbf{Q}^\pm\mathbf{u}\right)_{i,j,k}:=\pm\frac{1}{\Delta\mathcal{V}_{i,j}^\pm \bar{B}^\mathcal{V}_{i,j}}\left[\sum\limits_{n,m}\left(u_{n.m,k\pm1}B^{tor}_{n,m}\Delta A_{i,j,n,m}^\pm\right)-u_{i,j,k}B^{tor}_{i,j}h^2\right],
\label{par_grad_coordfree}
\end{align}
where $\Delta\mathcal{V}_{i,j}^\pm$ are the discrete volumes, $\bar{B}^\mathcal{V}_{i,j}$ is the magnetic field strength in the center of the flux box and $\Delta A_{i,j,n,m}^\pm$ is the surface overlap of grid point $\left(n,m,k\pm 1\right)$ with  the toroidal end of the flux box surface of grid point $\left(i,j,k\right)$ (Toroidal index $k$ can be dropped in $\Delta A_{i,j,n,m}^\pm$ due to axisymmetry). Based on the fact $\nabla\cdot\mathbf{B}=0$, the flux box volumes $\Delta\mathcal{V}_{i,j}^\pm$ can be computed with high accuracy during the field line tracing process.

The introduction of discrete fluxes $\mathbf{q}^\pm=\left(q^\pm_{1,1,1},q^\pm_{2,1,1}\dots\right)^T$ is useful for later purposes: 
\begin{align}
\mathbf{q}^\pm:=\mathbf{Q}^\pm\mathbf{u},
\end{align}
where $\mathbf{Q}^\pm$ are matrices according to equation \refeq{par_grad_interpolation} or \refeq{par_grad_coordfree}, which contain also the interpolating coefficients or the surface overlaps.

\subsection{Parallel diffusion}\label{sec_pardiff}
In this section the discretisation of the parallel diffusion operator, defined as
\begin{align}
\mathcal{D}_{\parallel}u:=\nabla\cdot\left[\mathbf{b}\nabla_{\parallel}u\right],
\end{align}
is discussed. 

The interpolation or integration process at the discretisation of the parallel gradient introduces an erroneous numerical perpendicular coupling among distinct field lines. Therefore, the discrete parallel diffusion operator will exhibit a spurious numerical perpendicular 'diffusion', which depends in general on resolution and the applied interpolation/integration method. In a magnetised plasma the dynamics can be strongly anisotropic, i.e.~very fast along magnetic field lines and very slow perpendicular. The numerical perpendicular diffusion arising from the discretisation error of the parallel diffusion operator might then compete or even overwhelm the real perpendicular diffusion or dynamics. Therefore, it is important to construct numerical schemes which exhibit only very low numerical perpendicular diffusion. In the following, two schemes for the discrete parallel diffusion operator are presented, a naive discretisation and a discretisation according to the support operator method \cite{shashkov:support95,shashkov:support96}. The support scheme was motivated from \cite{guenter:numpoll05,guenter:numpoll07}, and it will turn out that it exhibits a much lower numerical diffusion. In many cases, where an increase of resolution might not be possible due to hardware constraints, the naive discretisation might fail, whereas the support schemes still work well.

\subsubsection{Naive discretisation}
Under the assumption $\nabla\cdot\mathbf{b}\ll k_{\parallel}$, the parallel diffusion operator can be approximated as:  
\begin{align}
\mathcal{D}_{\parallel}\approx\nabla_{\parallel}^2.
\end{align} 
This motivates a discretisation via a further finite difference of the parallel gradient along magnetic field lines. The discrete parallel diffusion operator $\mathbf{D}_\parallel^{naive}$ is proposed as:
\begin{align}
\left(\mathbf{D}_{\parallel}^{naive}\mathbf{u}\right)_{i,j,k}:=\frac{2}{\Delta s_{i,j}^+ + \Delta s_{i,j}^-}\left[\left(\mathbf{Q}^+\mathbf{u}\right)_{i,j,k}-\left(\mathbf{Q}^-\mathbf{u}\right)_{i,j,k}\right]
\label{naive_diffusion}
\end{align}

\subsubsection{Support operator method}
The naive scheme yields a consistent discretisation, but it does not conserve any 'good' property of the parallel diffusion operator. Based on the support operator method \cite{shashkov:support95,shashkov:support96} a scheme is derived which conserves the self-adjointness of the parallel diffusion operator on the discrete level. 

Let $u,v$ be two real valued arbitrary scalar fields, we define a scalar product via an integration over the whole domain:
\begin{align}
\left\langle u,v\right\rangle:=\int\limits_V u\, v\,dV.
\end{align}
It holds that:
\begin{align}
\left\langle u,\mathcal{D}_{\parallel}v\right\rangle=\int\limits_V u\nabla\cdot\left[\nabla_{\parallel}v\mathbf{b}\right]\,dV=\int\limits_{\partial V}u\nabla_{\parallel}v\,\mathbf{b}\cdot d\mathbf{S}-\int\limits_{V}\nabla_{\parallel}u\nabla_{\parallel}v\,dV
\end{align}
If we further assume that the domain is periodic ($\varphi$-direction) and/or the quantities vanish at the boundaries, the following integral equality holds:
\begin{align}
\left\langle u,\mathcal{D}_{\parallel}v\right\rangle=-\left\langle\nabla_{\parallel}u,\nabla_{\parallel}v\right\rangle,
\label{integral_equality}
\end{align}
It is immediately obvious that:
\begin{align}
\nabla_{\parallel}^\dag=-\nabla\cdot\left[\mathbf{b}\circ\right], \quad \mathcal{D}_{\parallel}^\dag=\mathcal{D}_{\parallel}.
\end{align}
The method of support operators gives a procedure for the construction of discrete second order operators which conserve certain integral equalities on the discrete level. Within the support operator method a first order operator, the prime operator, is discretised, and derived operators are constructed via a discrete analogue of certain integral equalities. In our case we will derive the discrete parallel diffusion operator with the parallel gradient as the prime operator and via conservation of integral equality \refeq{integral_equality} on the discrete level. We discuss the construction of the discrete parallel diffusion operator for the case of of the $'+'$ discretisation in the following. The $'-'$ discretisation follows analogously and a combination of both is given at the end of this section.

\begin{figure}
\includegraphics[height=0.27\textwidth]{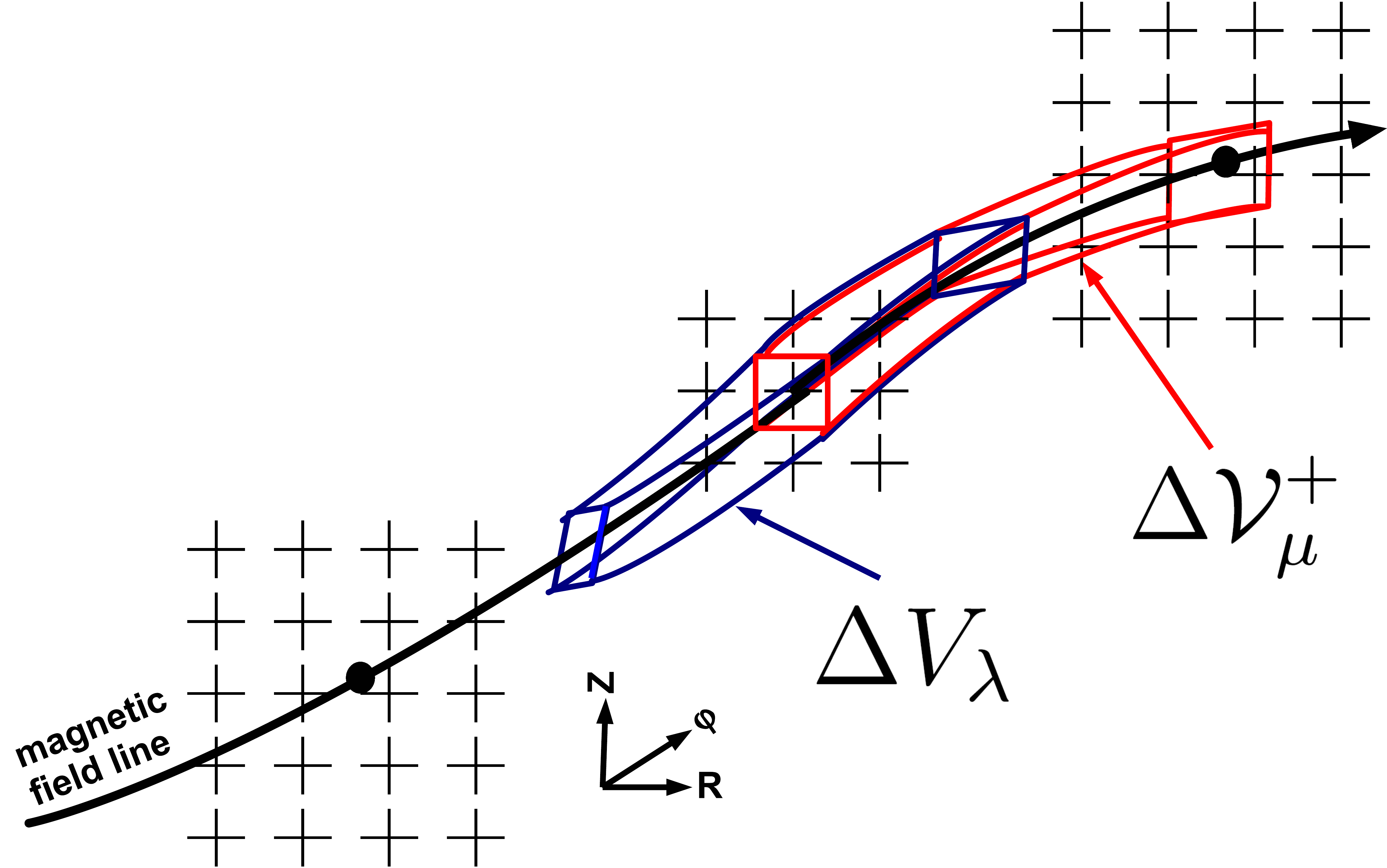}
\caption{Illustration of discrete volumes. $\Delta V_\lambda$ are flux boxes toroidally limited by $\left[\varphi_{k(\lambda)}-\Delta\varphi/2,\varphi_{k(\lambda)}+\Delta\varphi/2\right]$ and $\Delta\mathcal{V}_\mu^{\pm}$ toroidally limited by$ \left[\varphi_{k(\mu)},\varphi_{k(\mu)}\pm\Delta\varphi\right]$}
\label{fig_par_volumes}
\end{figure}

Let us start from the discrete parallel gradient according to equation \refeq{par_grad_interpolation} or \refeq{par_grad_coordfree}, which can both be represented with a matrix $\mathbf{Q}^+$. Scalar functions $\mathbf{u},\mathbf{v}$ are collocated on the basic cylindrical grid. Gradients $\mathbf{q}^+,\mathbf{p}^+$ are collocated on points half way along magnetic field lines towards neighbouring poloidal planes (see again fig.~\ref{fig_par_grad}). The discrete parallel gradient maps from the cylindrical grid $\left(SG\right)$ to that gradient's grid $\left(FG^+\right)$:
\begin{align}
\mathbf{Q}^+:SG\rightarrow FG^+
\end{align}
In the next step we want to mimic the integral equality \refeq{integral_equality} on the discrete level. However, on the discrete level the left hand side of equation \refeq{integral_equality} denotes a scalar product on the space $SG$, whereas the right hand side on the space $FG^+$. Therefore, we have to define two discrete scalar products, which both mimic an integration over the same whole domain:
\begin{align}
\left\langle \mathbf{u},\mathbf{v}\right\rangle_{SG}:=\sum_{\lambda}u_\lambda\, v_\lambda\, \Delta V_\lambda,\quad \left\langle \mathbf{p}^+,\mathbf{q}^+\right\rangle_{FG^+}:=\sum_{\mu}p^+_\mu\, q^+_\mu\, \Delta\mathcal{V}_\mu^+,
\label{scalarprods}
\end{align}
where Greek letters indicate summations over all grid points. The finite volumes are chosen as finite flux boxes around magnetic field lines as illustrated in fig.~\ref{fig_par_volumes}. 
The discrete parallel diffusion operator,
\begin{align}
\mathbf{D}_\parallel^{supp,+}:SG\rightarrow SG,
\end{align}
can now be derived from the discrete parallel gradient, and by imposing the integral equality \refeq{integral_equality} on the discrete level. On the one hand:
\begin{align}
\left\langle \mathbf{u}, \mathbf{D}^{supp,+}_{\parallel}\mathbf{v}\right\rangle_{SG}=\sum\limits_{\lambda,\sigma}u_\lambda\, \mathbf{D}_{\parallel\,\lambda,\sigma}^{supp,+}\,v_{\sigma}\,\Delta V_{\lambda},
\label{pardiffsupp_lhs}
\end{align}
On the other hand:
\begin{align}
- \left\langle \mathbf{Q}^+\mathbf{u}, \mathbf{Q}^+\mathbf{v}\right\rangle_{FG^+}=-\sum\limits_{\mu,\nu,\tau}\mathbf{Q}^+_{\mu,\nu}u_\nu\mathbf{Q}^+_{\mu,\tau}v_{\tau}\Delta\mathcal{V}_\mu^+.
\label{pardiffsupp_rhs}
\end{align}
The conservation of integral equality \refeq{integral_equality} on the discrete level demands the equality of expressions \refeq{pardiffsupp_lhs} and \refeq{pardiffsupp_rhs} for arbitrary $\mathbf{u},\mathbf{v}$. Via relabelling of the indices $\nu\rightarrow\lambda,\, \tau\rightarrow\sigma$ the discrete parallel diffusion operator can finally be obtained:
\begin{align}
\mathbf{D}_{\parallel,\lambda,\sigma}^{supp,+}=-\sum_{\mu}\mathbf{Q}^+_{\mu,\lambda}\mathbf{Q}^+_{\mu,\sigma}\frac{\Delta\mathcal{V}^+_{\mu}}{\Delta V_{\lambda}}.
\end{align}
In axial geometry the volume factors vanish $\left(\Delta V_{\lambda}=\Delta \mathcal{V}_{\mu}=h^2\Delta z\right)$ and the discrete parallel diffusion operator becomes:
\begin{align}
\mathbf{D}_{\parallel}^{supp,+}=-\left(\mathbf{Q}^+\right)^T\mathbf{Q}^+,
\label{supp_transpose}
\end{align}
which expresses the self-adjointness property on the discrete level. For toroidal geometries, due to the general geometry and the $1/R$ dependence of the toroidal field strength, the volume factors account for effects arising from $\nabla\cdot\mathbf{b}\neq 0$. In addition the volume factors matter if the parallel distances $\Delta s_{i,j}^+$ vary, i.e.~if the grid is not equidistant in the parallel sense.

Finally, a small modification is applied to the scheme. A discrete parallel diffusion operator can also be derived analogously with the $'-'$ discretisation of the parallel gradient. This will end up in a generally different but consistent scheme by itself. However, it is desirable that the final scheme is independent of this initial arbitrary choice. To achieve this, the average between both schemes is taken:
\begin{align}
\mathbf{D}^{supp}_{\parallel}:=\frac{1}{2}\left(\mathbf{D}^{supp,+}_{\parallel}+\mathbf{D}^{supp,-}_{\parallel}\right).
\label{supp_diffusion}
\end{align}
This modification does not alter the self-adjointness property on the discrete level, since:
\begin{align}
&\left\langle\mathbf{u},\mathbf{D}^{supp}_{\parallel}\mathbf{v}\right\rangle_{SG}
=\frac{1}{2}\left\langle\mathbf{u},\mathbf{D}^{supp,+}_{\parallel}\mathbf{v}\right\rangle_{SG}+\frac{1}{2}\left\langle\mathbf{u},\mathbf{D}^{supp,-}_{\parallel}\mathbf{v}\right\rangle_{SG} \notag \\
&=\frac{1}{2}\left\langle\mathbf{D}^{supp,+}_{\parallel}\mathbf{u},\mathbf{v}\right\rangle_{SG}+\frac{1}{2}\left\langle\mathbf{D}^{supp,-}_{\parallel}\mathbf{u},\mathbf{v}\right\rangle_{SG} 
=\left\langle\mathbf{D}^{supp}_{\parallel}\mathbf{u},\mathbf{v}\right\rangle_{SG}.
\end{align}

A further remark concerns numerical stability. The support operator discretisation excludes possible numerical instabilities arising from the interpolation, since it guarantees a strict decrease of the $L^2$ norm:
\begin{align}
\left\langle \mathbf{u},\mathbf{D}^{supp}_{\parallel}\mathbf{u}\right\rangle_{SG}=-\frac{1}{2}\left\langle \mathbf{Q}^+\mathbf{u},\mathbf{Q}^+\mathbf{u}\right\rangle_{FG^+}-\frac{1}{2}\left\langle \mathbf{Q}^-\mathbf{u},\mathbf{Q}^-\mathbf{u}\right\rangle_{FG^-}\leq 0,
\end{align}
which is in generally not fulfilled with the naive discretisation method.

\subsection{A two-dimensional model problem} \label{sec_model2d}
\begin{figure}[!htb]
a)\newline
\includegraphics[width=1.0\linewidth]{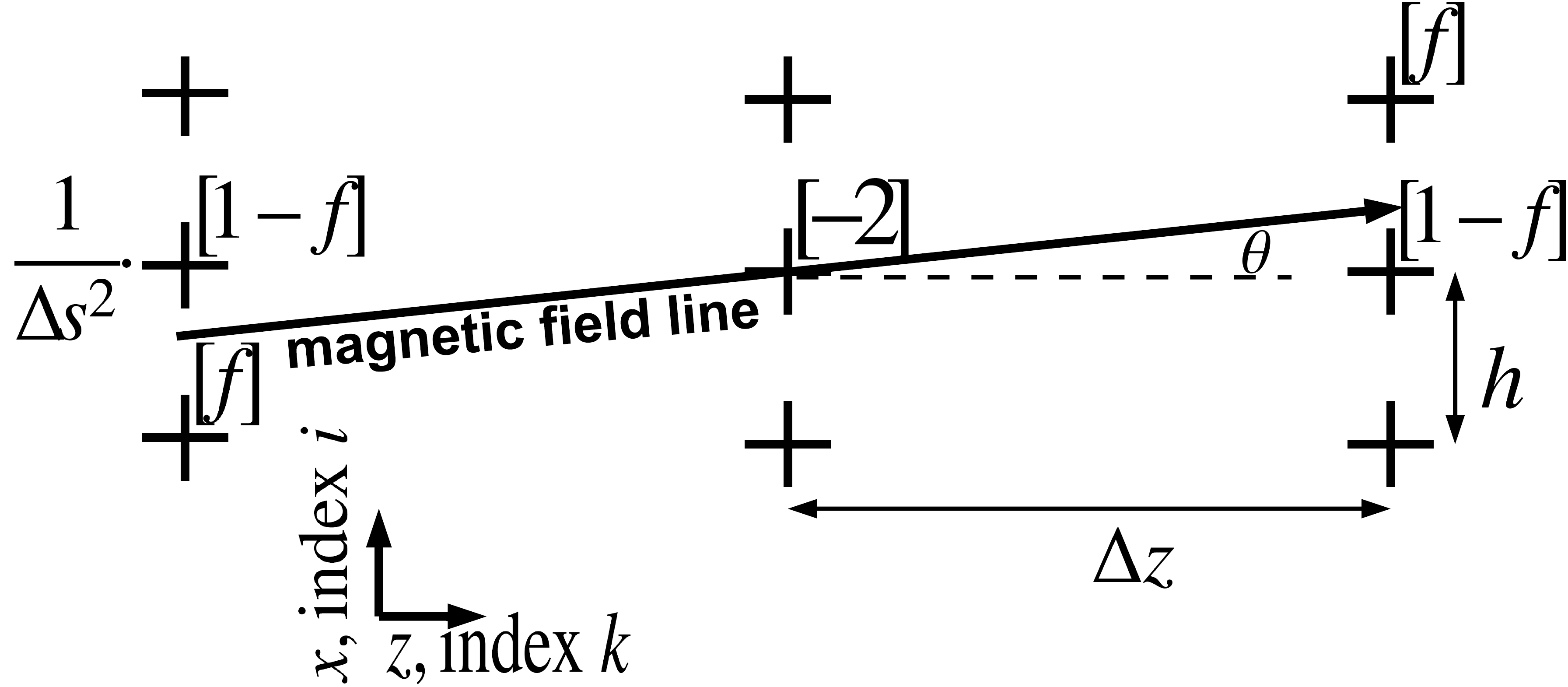}\newline
b)\newline
\includegraphics[width=1.0\linewidth]{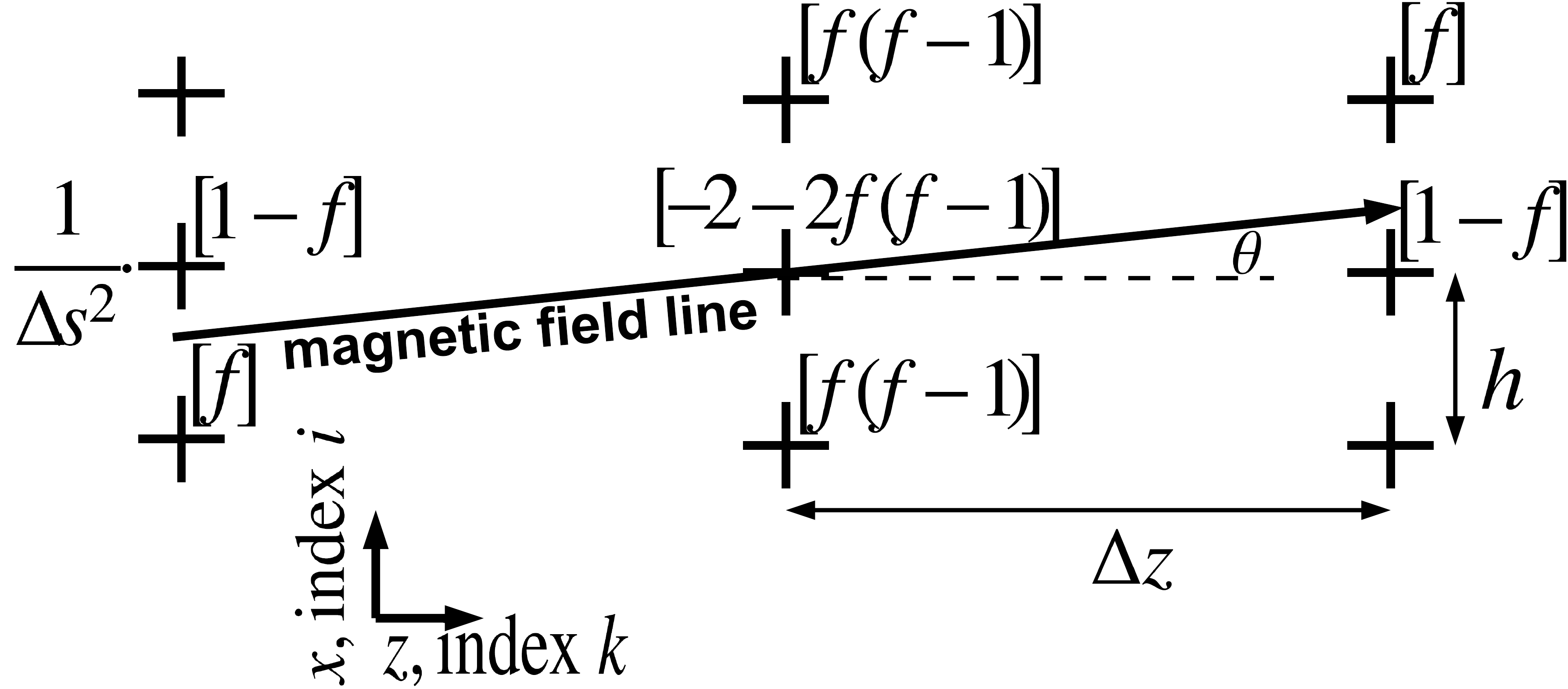}
\caption{Model problem for homogeneous magnetic field. Discrete parallel diffusion operator for a) naive scheme, b) for support scheme.} 
\label{fig_model_hom}
\end{figure}

To illustrate the difference between the naive and the support scheme, and to enlighten the advantage of the support scheme a two-dimensional model problem is considered. The coordinate $x$ plays the role of a coordinate within poloidal planes and $z$ the role of a periodic toroidal/axial coordinate, and the domain is spanned by a regular grid $x_i,z_k$, with grid spacing $h$ in the $x$-direction and grid spacing $\Delta z$ in the $z$-direction. The magnetic field is uniform with an inclination with respect to the grid:
\begin{align}
\mathbf{b}=\sin\theta\mathbf{e}_x+\cos\theta\mathbf{e}_z
\end{align}
The displacement of the penetration points with respect to the grid point can be expressed by a factor $f$:
\begin{align}
x_{i}^\pm=x_i\pm fh, \quad f=\frac{\Delta z}{h}\tan\theta,
\end{align}
Without loss of generality only cases of slight inclination with respect to the grid are considered in the following, i.e.~$\left|f\right|\leq 1$. Using a linear interpolation the values at the penetration points are determined as:
\begin{align}
u_{i,k}^\pm= (1-f)\,u_{i,k\pm 1} + f\,u_{i\pm1,k\pm1}.
\end{align}
The parallel distances are all equal $\Delta s_\lambda^\pm=\Delta s=\sqrt{\left(fh\right)^2+\Delta z^2}$, as also the finite volumes $\Delta V_{\lambda}=\Delta\mathcal{V}_\mu=h\Delta z$. To illustrate also the construction of the scheme we give for this case also explicit expressions for the matrices $Q^\pm$ on a $3\times3$ grid: 
\begin{align}
\mathbf{Q}^+=-\left(\mathbf{Q}^-\right)^T=\frac{1}{\Delta s}
\left(\begin{smallmatrix}
  -1  &     &     & 1-f &  f  &     &     &     &     \\
      &-1   &     &     & 1-f &  f  &     &     &     \\
      &     &  -1 &   &     & 1-f &     &     &     \\
      &     &     &  -1 &     &     & 1-f &  f  &     \\
      &     &     &     &  -1 &     &     & 1-f &  f  \\
      &     &     &     &     & -1  &  &     & 1-f \\
 1-f  & f   &     &     &     &     & -1  &     &     \\
      & 1-f &   f &     &     &     &     &  -1 &     \\
      &     & 1-f &     &     &     &     &     & -1   
\end{smallmatrix}
\right),
\label{matrixq_2dprob}
\end{align}
The only inner grid point is $i=2,\,k=2$, which corresponds to the fifth row of the matrices $\mathbf{Q}^\pm$. The discrete parallel diffusion operator at the inner grid point is illustrated in fig.~\ref{fig_model_hom}a for the naive scheme according to equation \refeq{naive_diffusion} and in fig.~\ref{fig_model_hom}b for the support scheme according to equation \refeq{supp_diffusion}. It is apparent that the Stencil of the support scheme is larger. In the limit that the penetration points coincide with grid points, i.e.~$f=0,1$, both schemes yield the standard second order finite difference expression. 

Now we investigate the action of the discrete parallel diffusion operators on a Fourier mode $u=\exp\left(ik_xx+ik_zz\right)$. The analytic result is: 
\begin{align}
\mathcal{D}_\parallel u=-k_{\parallel}^2u, \quad\text{with:}\quad k_\parallel:=k_x\sin\theta +k_z\cos\theta. 
\end{align}
Performing a Taylor expansion in $\left(k_xh,k_z\Delta z\right)$ yields:
\begin{align}
\mathbf{D}_\parallel^{naive}\mathbf{u}\approx &\left[-k_{\parallel}^2-\frac{f(1-f)(k_xh)^2}{\Delta s^2} +\mathcal{O}\left(\frac{(k_xh,k_z\Delta z)^4}{\Delta s^2}\right)\right]\mathbf{u}, 
\label{taylor_naive_lin}\\
\mathbf{D}_\parallel^{supp}\mathbf{u}\approx & \left[-k_{\parallel}^2+\mathcal{O}\left(\frac{(k_xh,k_z\Delta z)^4}{\Delta s^2}\right)\right]\mathbf{u},
\label{taylor_supp_lin}
\end{align}
It is apparent that for finite displacement $\left(f\neq 0,1\right)$ the leading error for the naive scheme scales like $\left(k_xh\right)^2/\Delta s^2$. This reflects the error of the linear interpolation, which scales also like $\left(k_xh\right)^2$. The error term represents a numerical perpendicular diffusion with diffusivity coefficient:
\begin{align}
\chi_{\perp,num}^{naive}=\frac{f\left(1-f\right)h^2}{\Delta s^2}. 
\end{align}
For the support scheme an improved scaling of $\left(k_xh,k_z\Delta z\right)^4/\Delta s^2$ for the leading error is obtained. Therefore, the numerical diffusion becomes actually a hyperdiffusion. 

The same analysis performed here for linear interpolation is repeated in \ref{app_polint} with third order polynomial interpolation. Moreover, in \ref{app_inhom} a model problem for the case of an inhomogeneous magnetic field is discussed, and in \ref{app_integration} an example for the application of the integration method is given. 

\subsection{Map distortion} \label{sec_mapdistort}
Especially close to the X-point field lines become strongly distorted. To illustrate this, we consider the lowest order expansion for the magnetic field around the X-point \cite{farina:xgeom93}:
\begin{align}
\mathbf{B}=B_0\left[\mathbf{e}_z+\alpha\left(x\mathbf{e}_x-y\mathbf{e}_y\right)\right],
\end{align}
Two generic field lines, which have initially a poloidal distance $\delta_0$, will exponentially diverge:
\begin{align}
\delta\left(z\right)\sim \delta_0\exp\left(\alpha z\right) 
\label{ergodic_fieldlines}
\end{align}
We clarify now the effect of such a strong distortion on the numerical scheme. 

\begin{figure}[!htb]
\begin{tabular}{c c}
\includegraphics[height=0.33\linewidth]{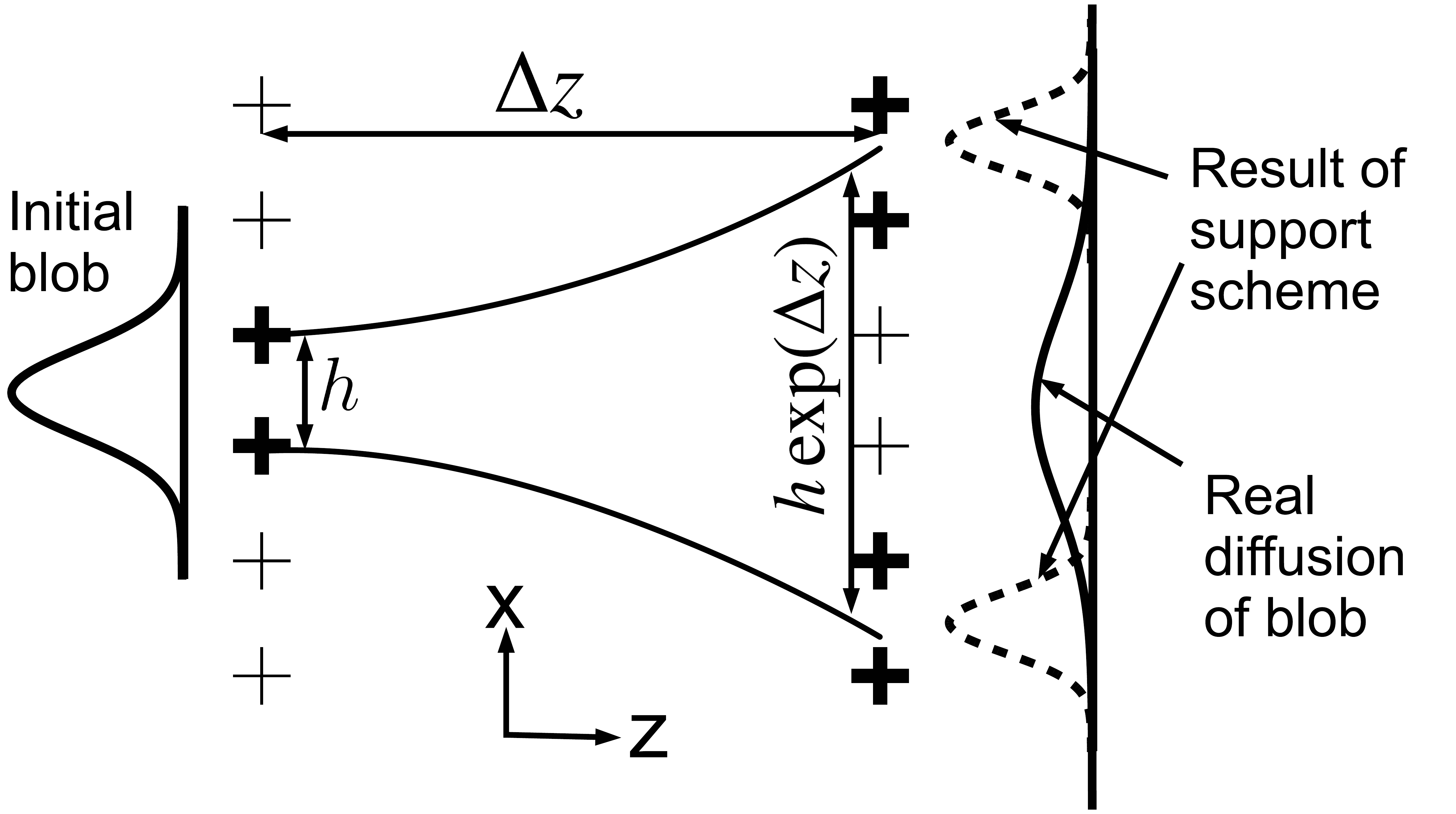} &
\includegraphics[height=0.33\linewidth]{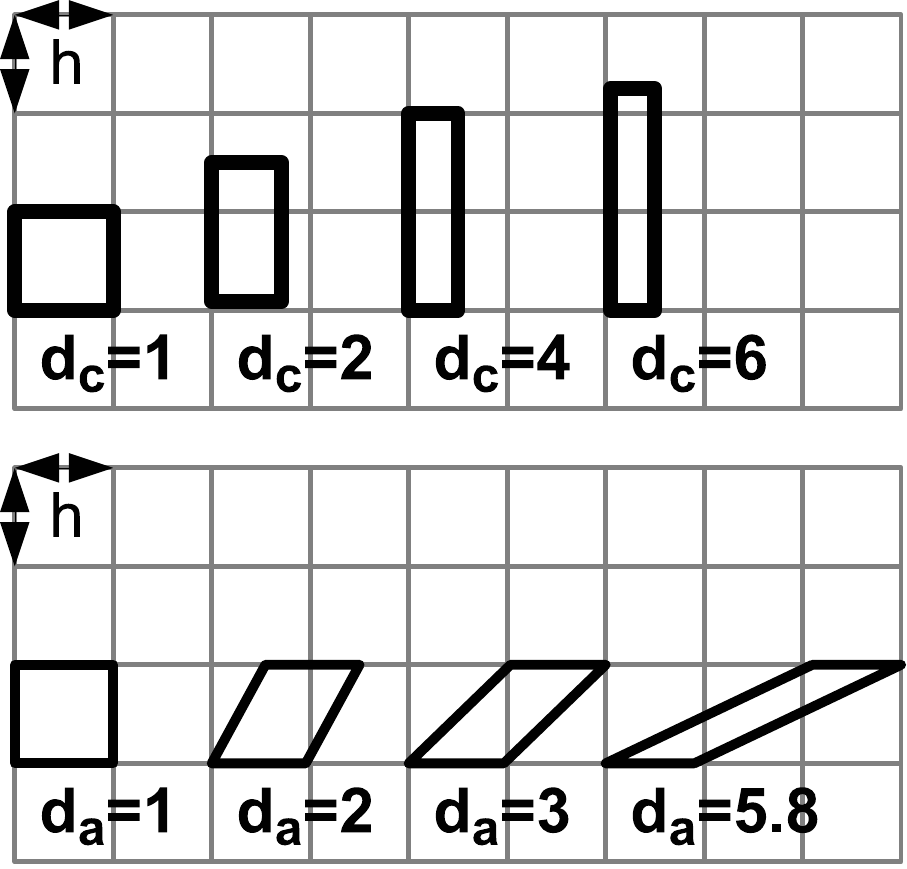}
\end{tabular}
\caption{a) Effects of map distortion on numerical scheme. In the illustration only thick points are connected by the support scheme (linear interpolation). A blob should in reality spread smoothly across many grid points. However the support scheme produces unphysical wiggles (dashed) in the neighbouring poloidal plane. b) Examples for conformal distortion (top) and angular distortion (bottom). Grey squares indicate base cells within the neighbouring poloidal plane.} 
\label{fig_wiggle2d}
\end{figure}

We consider again a two-dimensional setup as illustrated in fig.~\ref{fig_wiggle2d}a. The penetration points of two neighbouring field lines separate exponentially according to expression \refeq{ergodic_fieldlines}. Further on, we consider some blob with finite poloidal extent of order $h$ at some plane. In reality the blob should diffuse along magnetic field lines and thus spread across many grid points of the neighbouring poloidal plane. For the naive scheme this does not cause any problems, since the discrete parallel diffusion operator is computed by just 'taking' values from neighbouring poloidal planes. However, in the support scheme a value is not only 'taken' but also 'sent' towards neighbouring poloidal planes. In other words, if the parallel gradient is computed via interpolation at the penetration points, grid points which lie in between might not be connected to the original points by the scheme. The blob does not spread properly across the grid points, but only diffuses to points which are connected by the scheme. Finally, erroneous wiggles arise in the neighbouring poloidal plane.  

We define a criterion in order to exclude numerical artefacts arising from the map distortion: Starting with an initial square of lateral length $h$ at some plane, the square encounters a distortion, as its edges are traced. Therefore, also the map of the square at the neighbouring poloidal plane is distorted (see fig.~\ref{fig_par_grad}b). Neglecting the weak variation of $B^{tor}$, the area of the square is thereby conserved due to flux conservation. We can then distinguish two types of distortion \cite{ribeiro:confmap10}, a conformal where the square is stretched in one direction and squeezed in the other resulting in a quad with disparate lateral lengths, and an angular distortion where e.g.~two angles become acute and the other two obtuse resulting in a parallelogram. We quantify the distortion by: 
\begin{align}
d_c=&\max\limits_{i,j}\frac{\text{longest side of mapped quad }i,j}{\text{shortest side of mapped quad } i,j}, \\
d_a=&\max\limits_{i,j}\frac{\text{largest angle of mapped quad }i,j}{\text{smallest angle of mapped quad } i,j}.
\end{align}
The distortion decreases as the toroidal/axial resolution is increased, and a reasonable criterion is to require enough toroidal/axial resolution such that the mapped quad does not cover more than two squares in each direction within the neighbouring poloidal plane. Otherwise the map might jump across grid points. Applying this constraint, results in a threshold for the distortion (see fig.~\ref{fig_wiggle2d}b) of $d_c\leq 4$ and $d_a\leq3$, but we require for convenience the same threshold for both, i.e.:
\begin{align}
d_c,\,d_a\leq 4.
\label{map_distort_threshold}
\end{align} 

Another numerical artefact may arise, if for two neighbouring (perpendicular) points there is a jump in the points which are involved in the interpolation  (see example in \ref{app_inhom} around grid point $i=I$ and equation \refeq{model_dist_s1}). This may cause a small oscillation on the grid scale, which might be cured by adding a small amount of high order perpendicular diffusion.

Any problems related with a distortion of the map can be cured by using the integration method for the discrete parallel gradient (equation \refeq{par_grad_coordfree}). With this scheme the information is spread properly across grid points in the neighbouring planes. In the illustration of fig.~\ref{fig_par_grad}b the mapped area is obtained by tracing only the four corners of the base square resulting in a quad as mapped area. However, if the map distortion becomes really strong even this might not suffice and a more detailed contour of the base square might have to be traced which would yield a polygon as mapped area. Problems related with a jump of the points which are involved in the interpolation are also resolved with the integration method (see \ref{app_integration}). From a practical point of view the integration method might seem to be second choice compared to interpolation methods, since its implementation appears cumbersome though possible as will be shown with an example in section \ref{sec_verifymapdistort} (figure \ref{fig_wiggles}d). Nevertheless, it is presented here to show that even complicated cases with strong map distortion do not pose problems of principle to the support scheme. Moreover, there might be cases where its application might be necessary, if no sufficient toroidal resolution can be supplied to bring the distortion below the requested threshold given in \refeq{map_distort_threshold}.

\section{Benchmarks and examples}\label{sec_benchmarks}
The numerical methods presented in section \ref{sec_fieldlinemap} are implemented in the new code GRILLIX. In this section we present benchmarks, which shall show the validity of the field line map approach in general and GRILLIX in particular. As a model problem the parallel diffusion equation
\begin{align}
\frac{\partial}{\partial t}u=\chi_{\parallel}\mathcal{D}_{\parallel}u
\label{pardiffeq}
\end{align}
is considered. Space scales are normalised to $R_0$ in toroidal geometries respectively $L_{ax}/2\pi$ in axial geometries, where $L_{ax}$ is the axial periodicity length. Time is measured in $R_0^2/\chi_{\parallel}$ respectively $L_{ax}^2/\left(4\pi^2\chi_{\parallel}\right)$. 

Five possible discretisation schemes for the parallel diffusion operator are investigated:
\begin{itemize}
\item \textbf{N-1}: Naive scheme with bilinear interpolation
\item \textbf{N-3}: Naive scheme with third order bipolynomial interpolation
\item \textbf{S-3}: Support scheme with bilinear interpolation
\item \textbf{S-3}: Support scheme with third order bipolynomial interpolation
\item \textbf{S-C}: Support scheme with integration method for parallel gradient
\end{itemize}
Each bilinear interpolation involves four grid points and each third order bipolynomial interpolation 16 grid points symmetrically arranged around the considered penetration point (see e.g.~\cite{press:numrec07}). For the \textbf{S-C} method the overlaps $\Delta A_{i,j,n,m}^\pm$ arising in the definition of the discrete parallel gradient (expression \refeq{par_grad_coordfree}) are computed via the routine given in \cite{zerzan:overlaproutine89}. 

\subsection{Axial circular equilibria}
For axial circular equilibria the problem \refeq{pardiffeq} can be solved analytically. We will consider in the following flux shells, i.e.~radially bounded domains $\rho\in\left[\rho_{min},\rho_{\max}\right]$, with $\rho:=\sqrt{x^2+y^2}$ a flux label. Without loss of generality, we consider only a single radial, poloidal and axial mode $\left(r,m,n\right)$ as initial state:
\begin{align}
&u(t=0,\mathbf{x})=\sin\left(\pi r\frac{\rho-\rho_{min}}{\rho_{max}-\rho_{min}}\right)\sin\left(m\theta+n z\right),
\end{align}
with $\tan\theta=y/x$ being the poloidal angle. The analytic solution is:
\begin{align}
&u(t,\mathbf{x})=u(t=0,\mathbf{x})\exp\left(-\frac{t}{t_{m,n}(\rho)}\right), \notag \\ &t_{m,n}^{-1}(\rho)=\chi_{\parallel}\frac{m+n\,q(\rho)}{q(\rho)^2+\rho^2},
\label{axicirc_analyticsolution}
\end{align}
with $q(\rho)=B^z/B^\theta$ the safety factor.  A characteristic time for the decay of the mode is given by $t_{m,n}(\rho_0)$ with $\rho_0:=\left(\rho_{max}+\rho_{min}\right)/2$. 

The benchmarks, presented in the following, were performed with $\rho_{min}=0.1$ and $\rho_{max}=0.2$. Since the goal is the investigation of the spatial discretisation error, time steps were chosen sufficiently small, such that the temporal discretisation error played always only a subdominant role. The numerical error will be quantified in the $L^2$ norm:
\begin{align}
e_2(t):=\frac{\|u_{analytic}(t)-u_{numeric}(t)\|_2}{\|u_{analytic}(t)\|_2},
\end{align}
where the norm is computed on the discrete level as $\| u \|_2:=\sqrt{\left\langle u,u \right\rangle_{SG}}$ according to equation \refeq{scalarprods}.

\subsubsection{$k_{\parallel}\neq 0$ modes}\label{sec_axialcircconverg}
Firstly, we consider modes which have a non-vanishing parallel wavevector $k_{\parallel}\propto m+n\,q(\rho)$. 

\begin{figure}[!htb]
\includegraphics[width=1.0\linewidth]{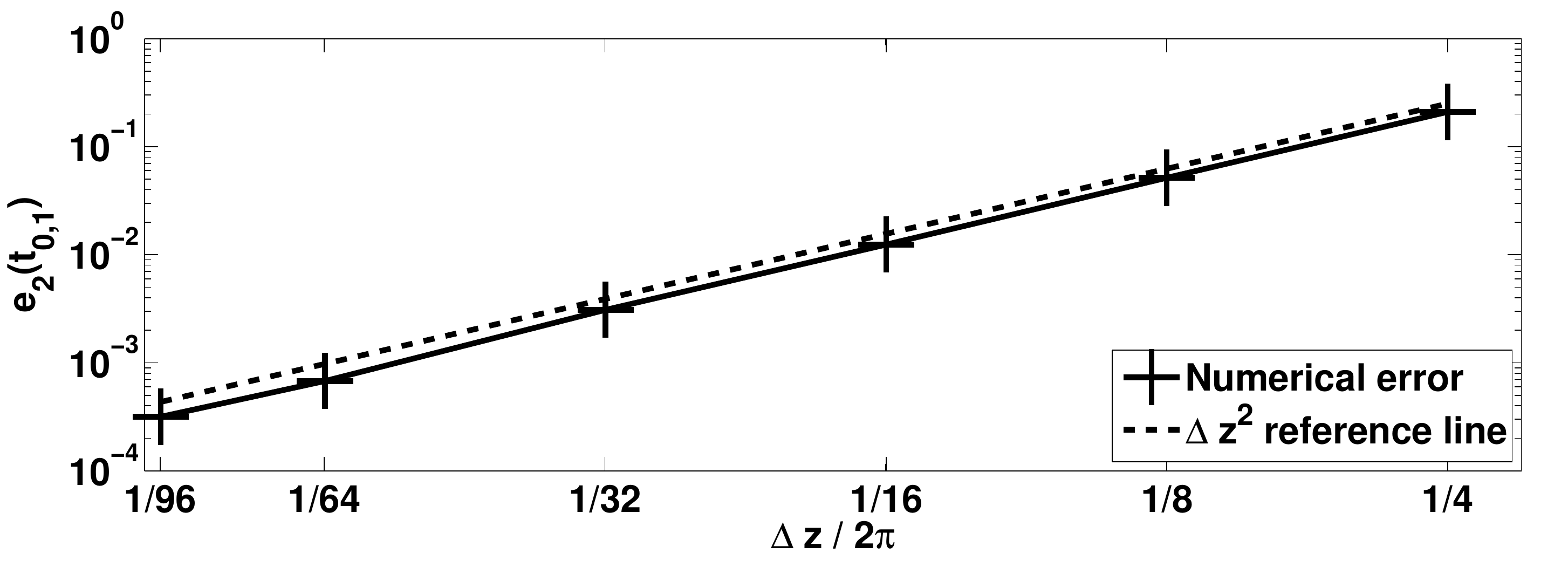}
\caption{Numerical error (for all schemes exactly overlapped) in dependence on axial resolution for a $\left(r=1,\,m=0,\,n=1\right)$ mode with $q=\infty$.} 
\label{fig_qinfty}
\end{figure}

In the case of a pure axial magnetic field $\left(q\rightarrow\infty\right)$ the penetration points coincide with grid points and the interpolation becomes exact. The discrete parallel diffusion operator reduces to the standard second order finite-difference expression for all schemes. Hence, this benchmark serves merely as a first basic test on the correctness of the implementation of the schemes in GRILLIX. Fig.~\ref{fig_qinfty} shows that the expected second order accuracy with axial (=parallel) resolution was obtained with GRILLIX. 

\begin{figure}[!htb]
\centering
\includegraphics[width=0.48\linewidth]{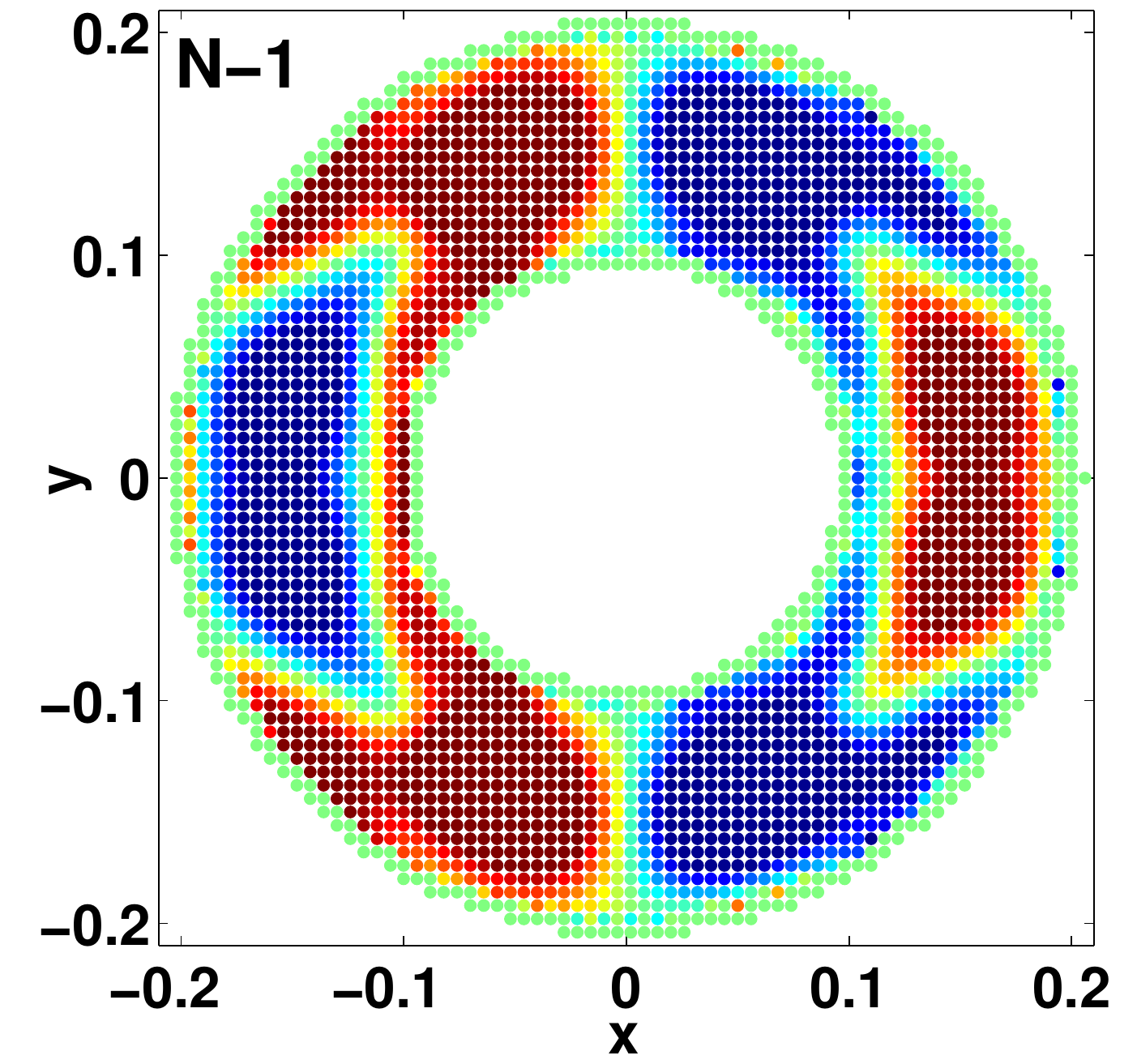}
\includegraphics[width=0.48\linewidth]{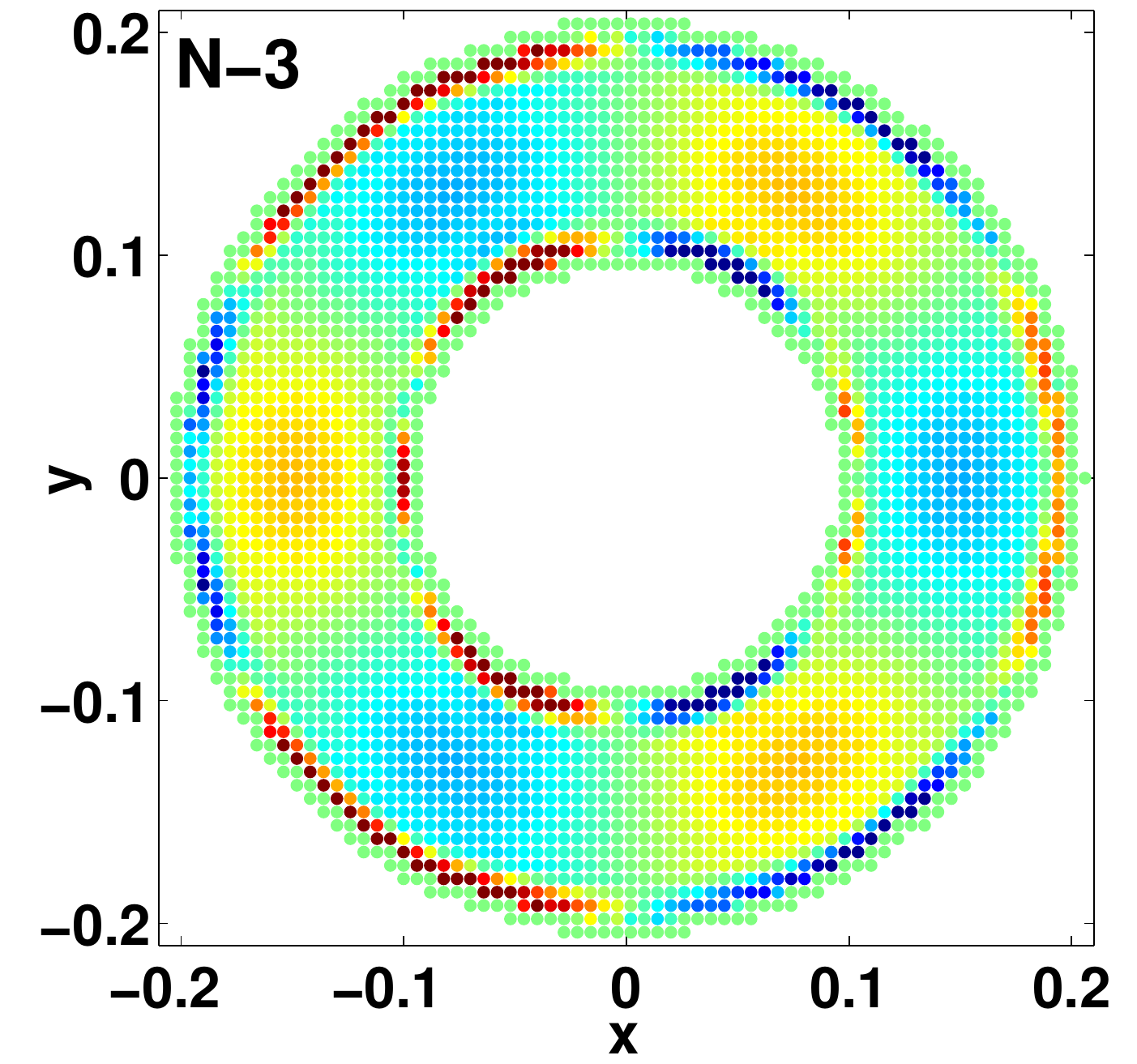}
\includegraphics[width=0.48\linewidth]{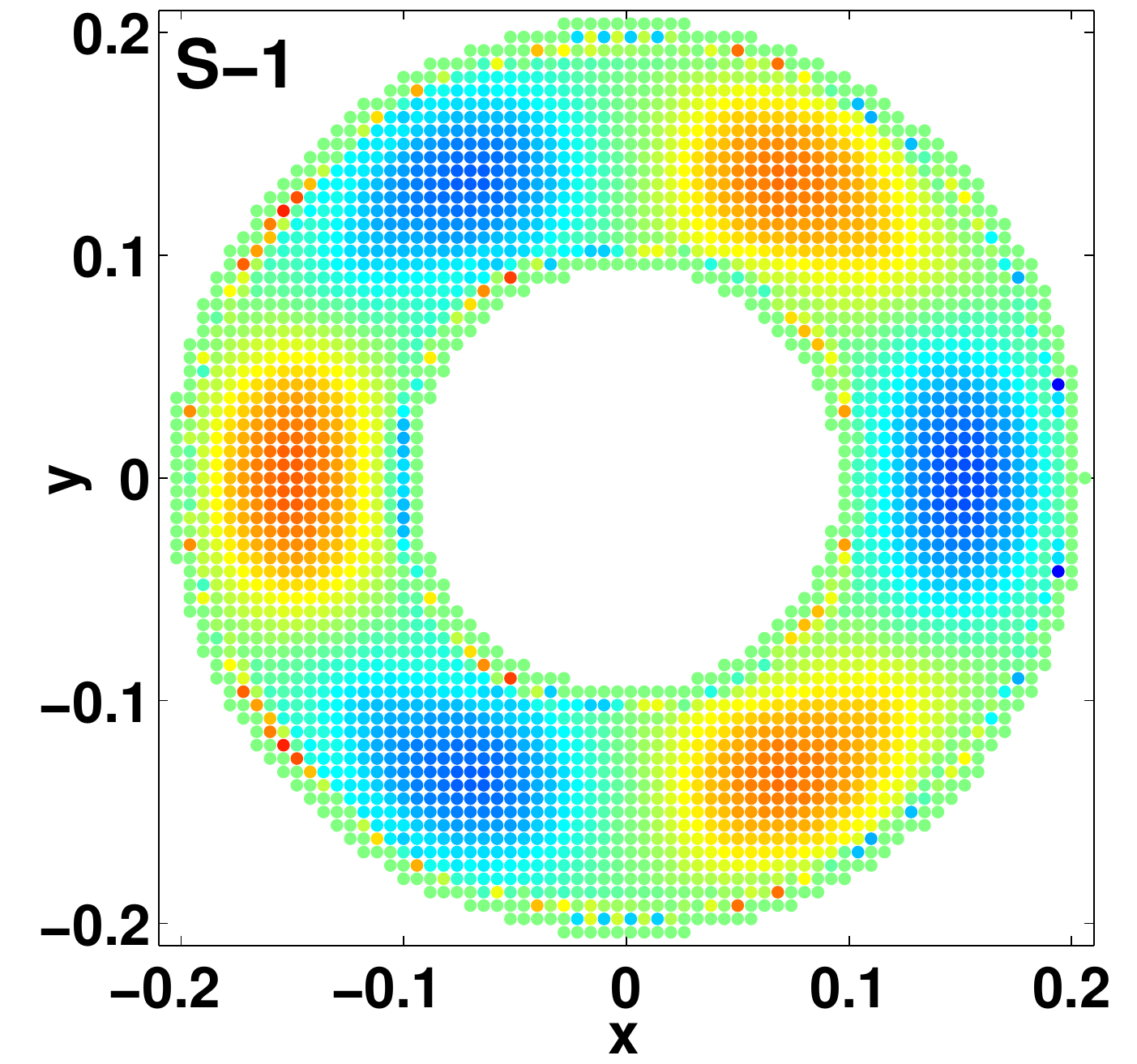}
\includegraphics[width=0.48\linewidth]{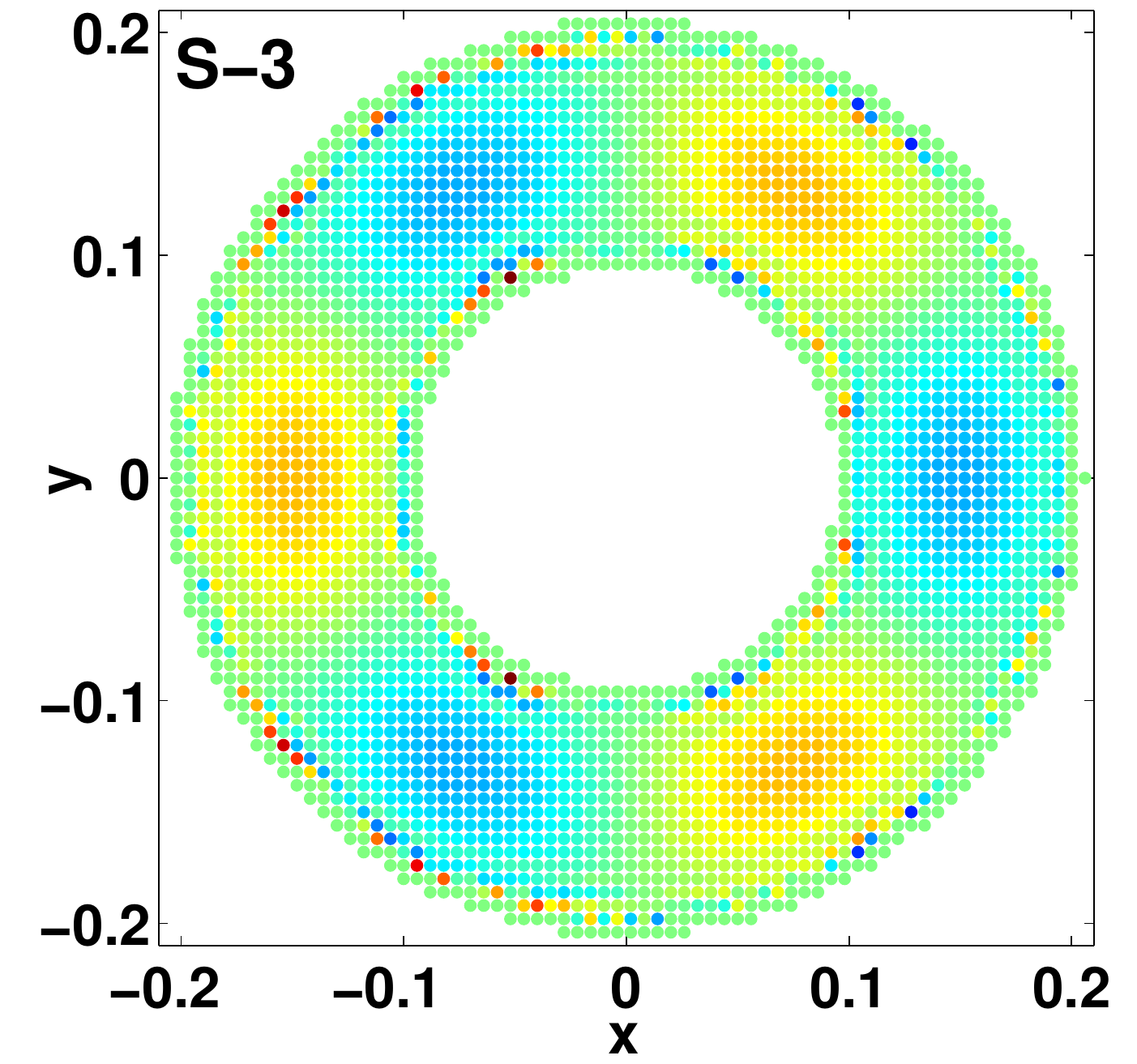}
\includegraphics[width=0.7\linewidth]{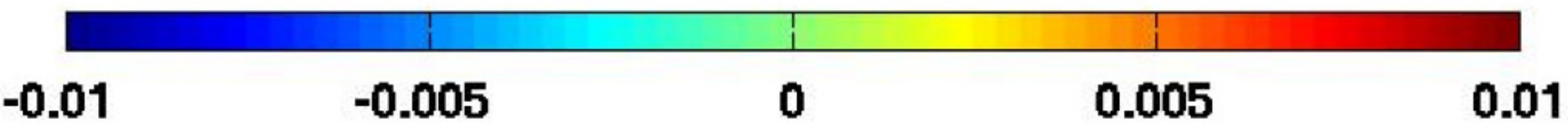}
\caption{Difference between numeric and analytic solution for a $\left(r=1,\,m=3,\,n=1\right)$ mode with $q=3.4$ at $t=t_{3,1}$ (for definition of $t_{3,1}$ see equation \refeq{axicirc_analyticsolution}) on plane $z=0$ for various schemes. Resolution was $h=6\cdot10^{-3},\,\Delta z/2\pi=1/32$.} 
\label{fig_errsnaps}
\end{figure}

If a finite safety factor $q\neq \infty$ is considered, the penetration points do in general not coincide with grid points any more, and the numerical error is dependent on the poloidal resolution $h$ and the scheme. In fig.~\ref{fig_errsnaps} the difference between the numeric solution and the corresponding analytic solution is shown for various different schemes. The basic behaviour is consistent with the predictions from the two-dimensional model of section  \ref{sec_model2d} and \ref{app_polint}. The overall error level is largest for the \textbf{N-1} scheme of order $\mathcal{O}\left(h^2\right)$ (see equation \refeq{taylor_naive_lin}), whereas the error level is lower of order $\mathcal{O}\left(h^4\right)$ for \textbf{N-3} (see equation \refeq{taylor_naive_pol}), \textbf{S-1} (see equation \refeq{taylor_supp_lin}) and \textbf{S-3} (see equation \refeq{taylor_supp_pol}). Significant errors arise at the radial boundaries of the domain. Though the parallel diffusion equation \refeq{pardiffeq} does not require radial boundary conditions, some assumption about the quantity must be made also outside the domain for the interpolation. Within GRILLIX it is implicitly assumed that $u=0$ outside the domain. This results in radially discontinuous derivatives at the boundary, and therefore especially for higher order interpolations large errors arise close to it. 

\begin{figure}[!htb]
a)\newline
\includegraphics[width=1.0\linewidth]{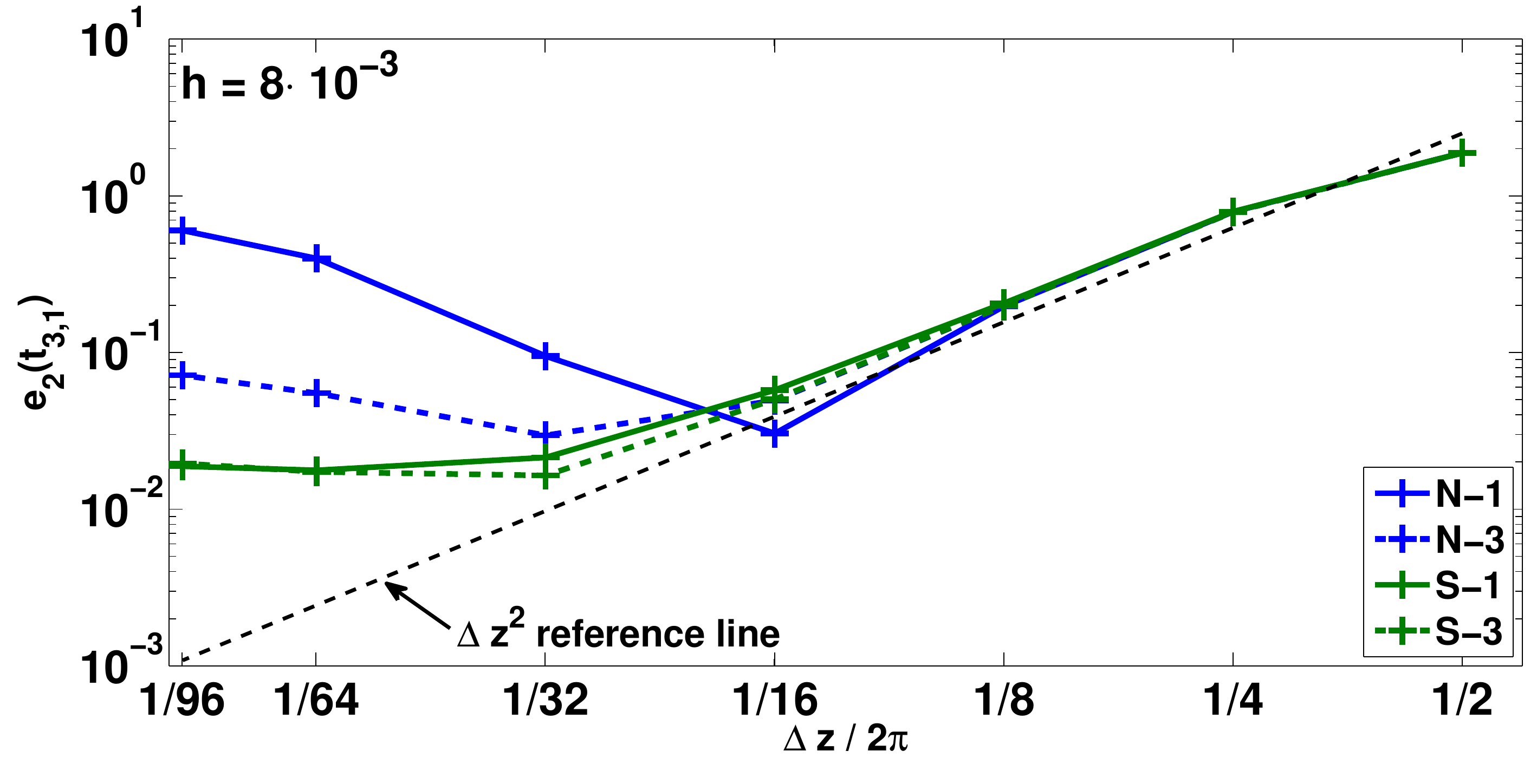}
b)\newline
\includegraphics[width=1.0\linewidth]{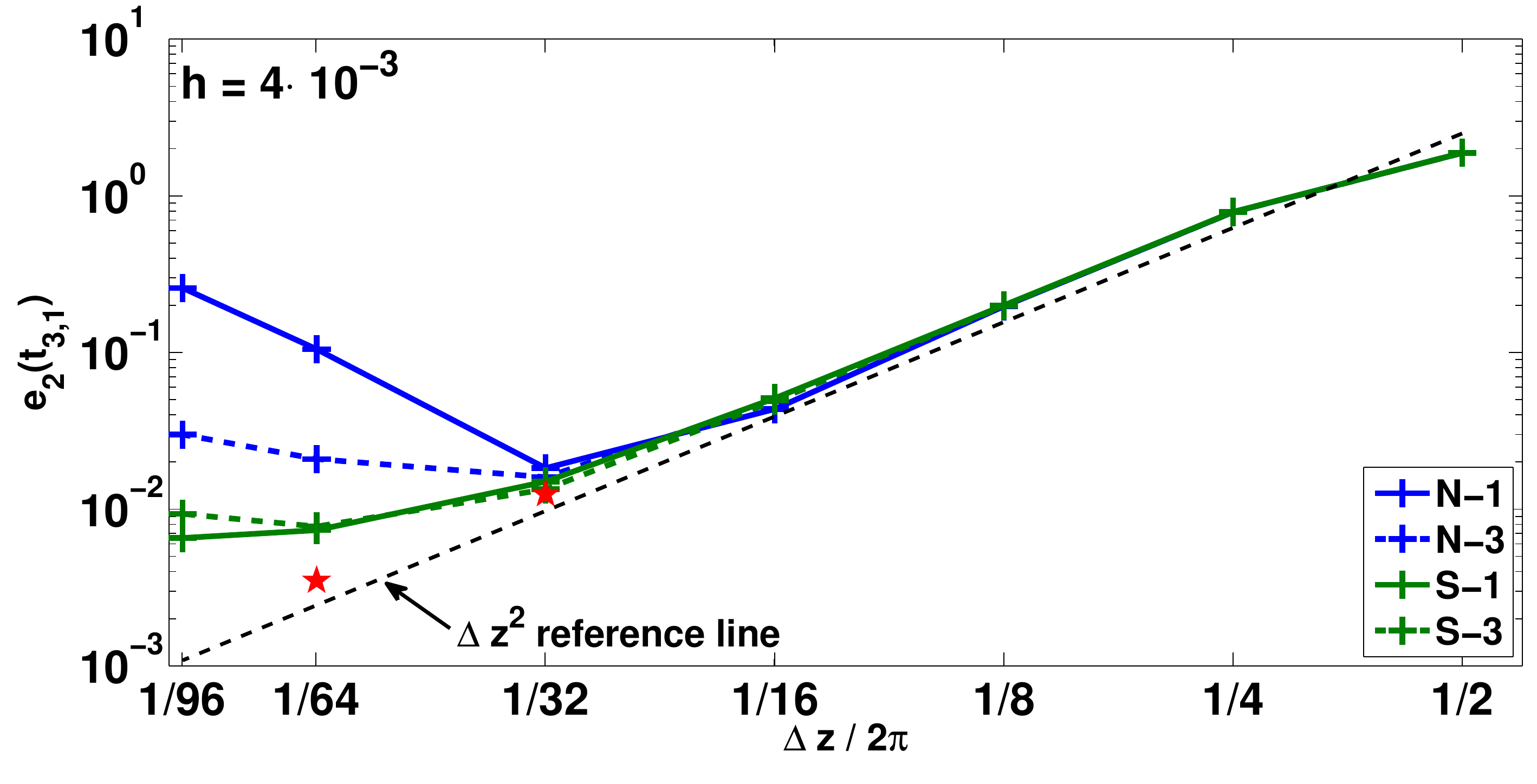}
\caption{Numerical error in dependence on axial resolution for a $\left(r=1,\,m=3,\,n=1\right)$ mode with $q=3.4$ for fixed poloidal resolution of a) $h=8\cdot 10^{-3}$ and b) $h=4\cdot 10^{-3}$. Red stars indicate values obtained in the limit $h\rightarrow0$ (see fig.~\ref{fig_numerr_h}).} 
\label{fig_numerr_npol}
\end{figure}

\begin{figure}[!htb]
a)\newline
\includegraphics[width=1.0\linewidth]{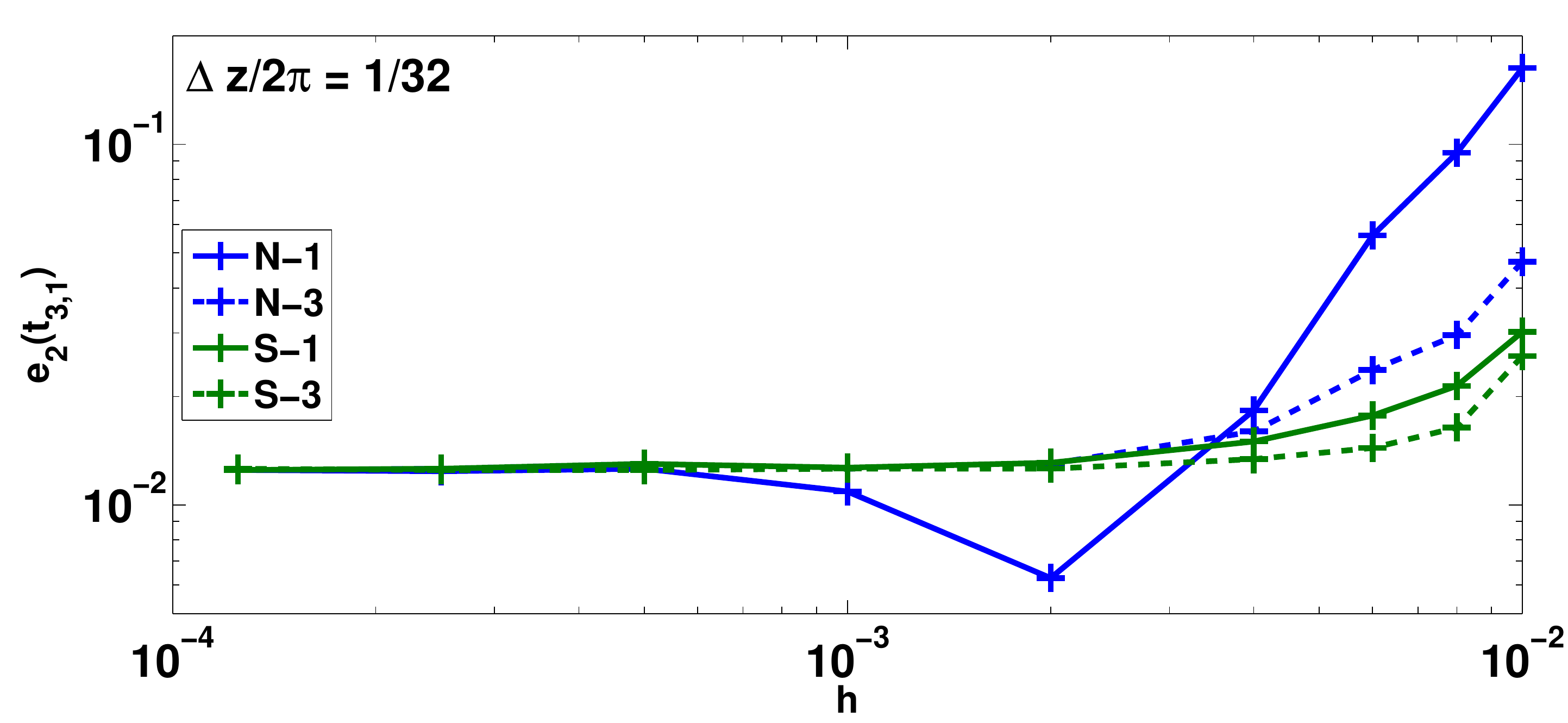}
b)\newline
\includegraphics[width=1.0\linewidth]{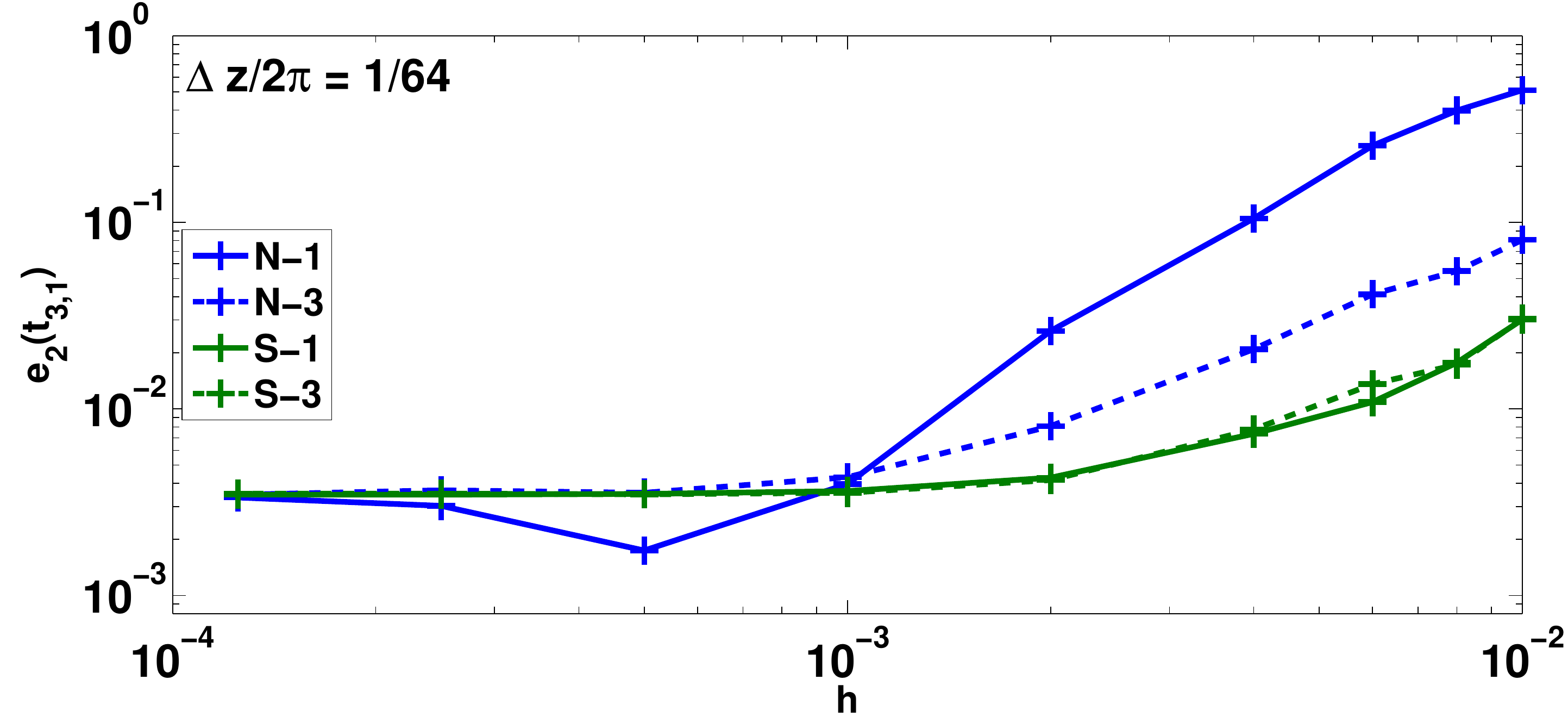}
\caption{Numerical error in dependence on poloidal resolution for a $\left(r=1,\,m=3,\,n=1\right)$ mode with $q=3.4$ for fixed axial resolution of a) $\Delta z/2\pi=1/32$ and b) $\Delta z/2\pi=1/64$.} 
\label{fig_numerr_h}
\end{figure}

The numerical error in dependence on axial resolution for two fixed poloidal resolutions is shown in fig.~\ref{fig_numerr_npol}. For low axial resolutions the error follows a $\Delta z^{2}$ law showing the second order accuracy of the schemes along magnetic field lines $\left(\Delta z\approx\Delta s\right)$. For higher axial resolutions the error is dominated by the poloidal resolution $h$ and deviates from the $\Delta z^2$ line but increases. Again in agreement with the scaling derived from the two-dimensional model from section \ref{sec_model2d} and \ref{app_polint}, the deviation occurs first for the \textbf{N-1} scheme and occurs later for the other schemes. In fig.~\ref{fig_numerr_h} the error in dependence on the poloidal resolution for two fixed axial resolutions is shown. Again the convergence is slowest for the \textbf{N-1} scheme, whereas the others exhibit similar convergence rates. Note that though the \textbf{S-1} scheme is based only on bilinear interpolation, it seems to perform even slightly better than the \textbf{N-3} scheme. All schemes converge to the same value of $e_2\approx1.1\cdot 10^{-2}$ at $\Delta z/2\pi=1/32$ (fig.~\ref{fig_numerr_h}a) and $e_2\approx3.5\cdot 10^{-3}$ at $\Delta z/2\pi=1/64$ (fig.~\ref{fig_numerr_h}b). These values are settled by axial resolution (see red stars in fig.~\ref{fig_numerr_npol}b).

\subsubsection{$k_{\parallel}=0$ structures}\label{sec_axialcircnumdiff}
\begin{figure}[!htb]
\centering
\begin{flushleft}a)\end{flushleft}
\includegraphics[width=0.48\linewidth]{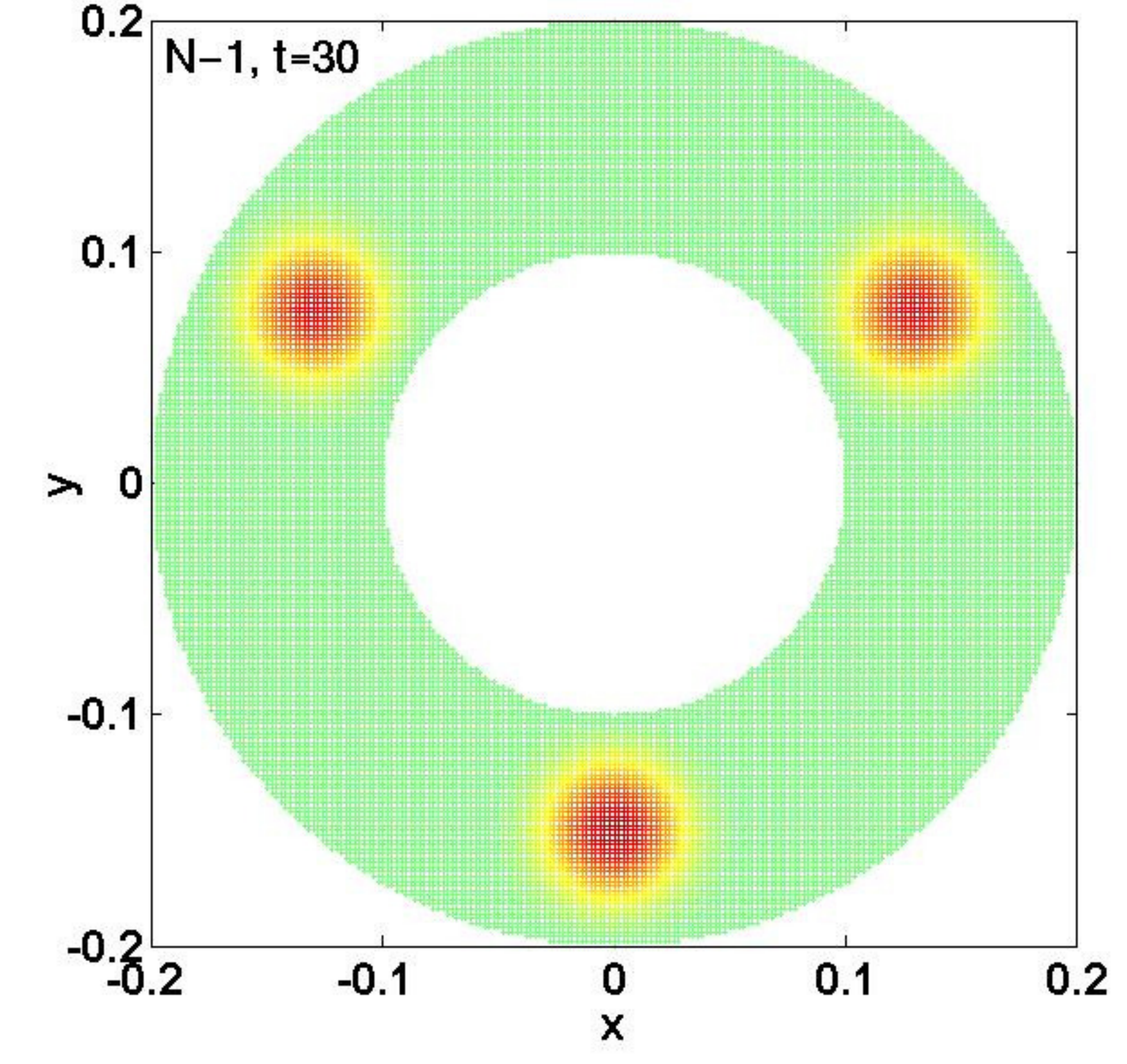}
\includegraphics[width=0.48\linewidth]{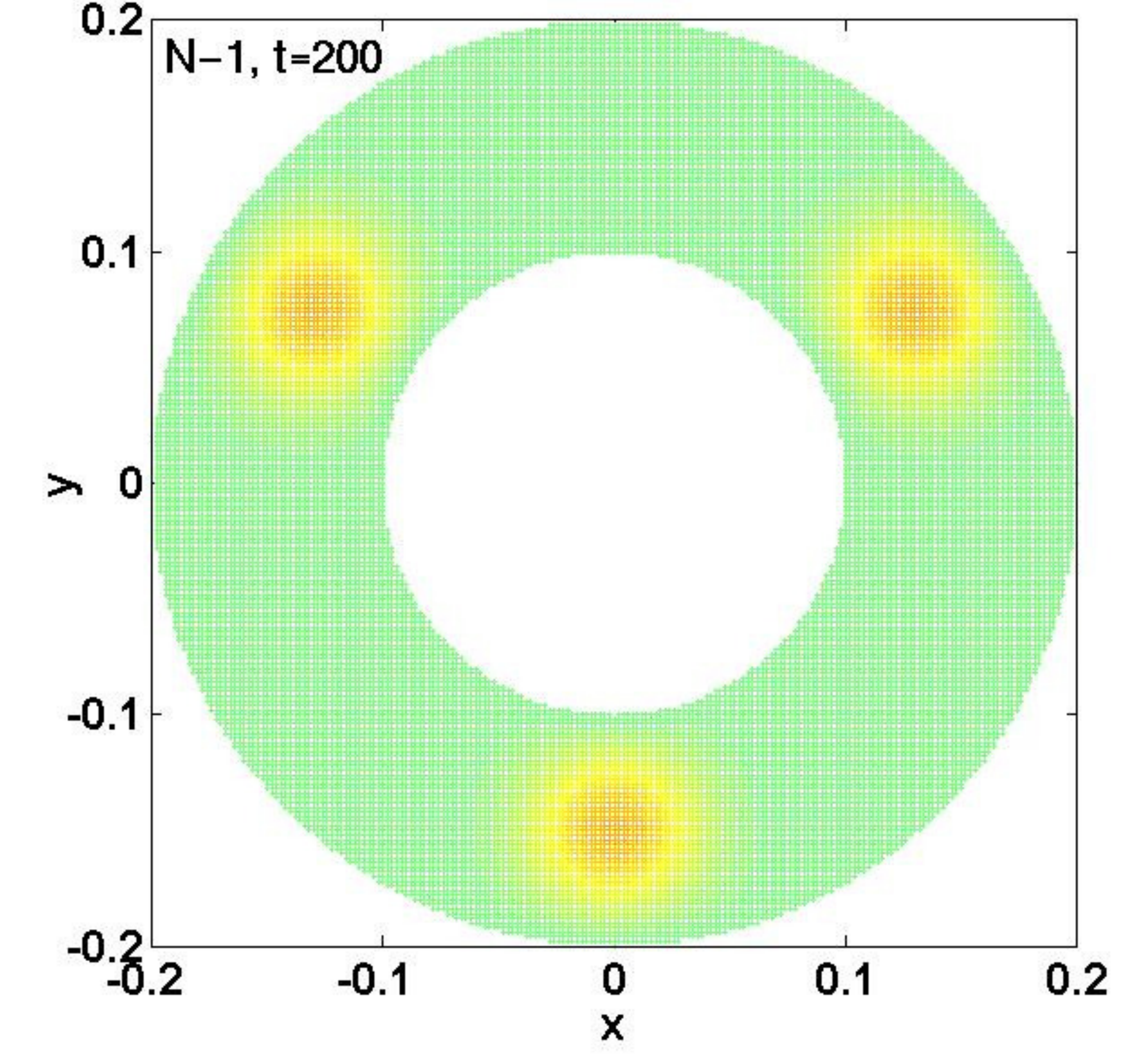}
\includegraphics[width=0.7\linewidth]{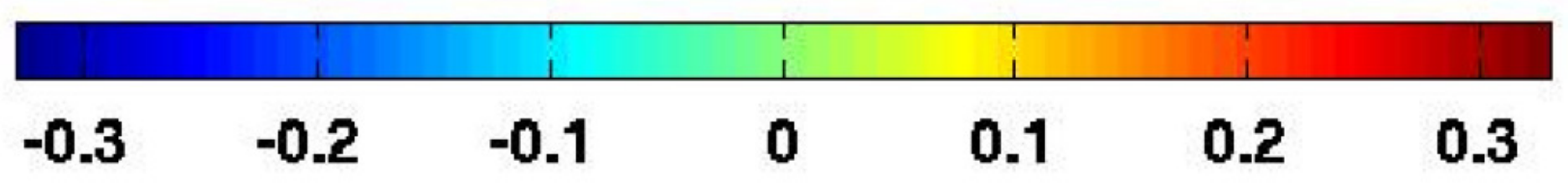}
\begin{flushleft}b)\end{flushleft}
\includegraphics[width=0.48\linewidth]{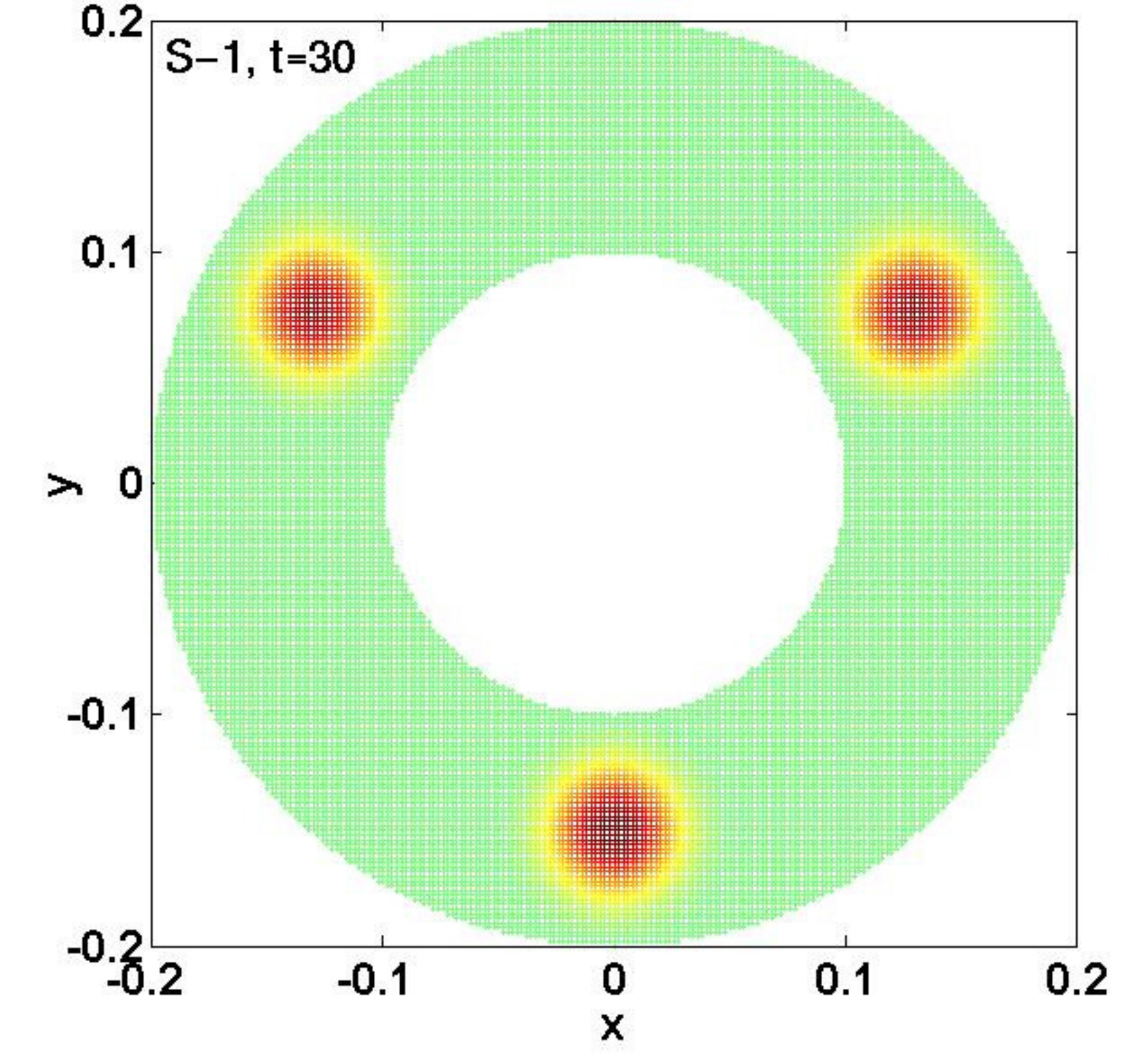}
\includegraphics[width=0.48\linewidth]{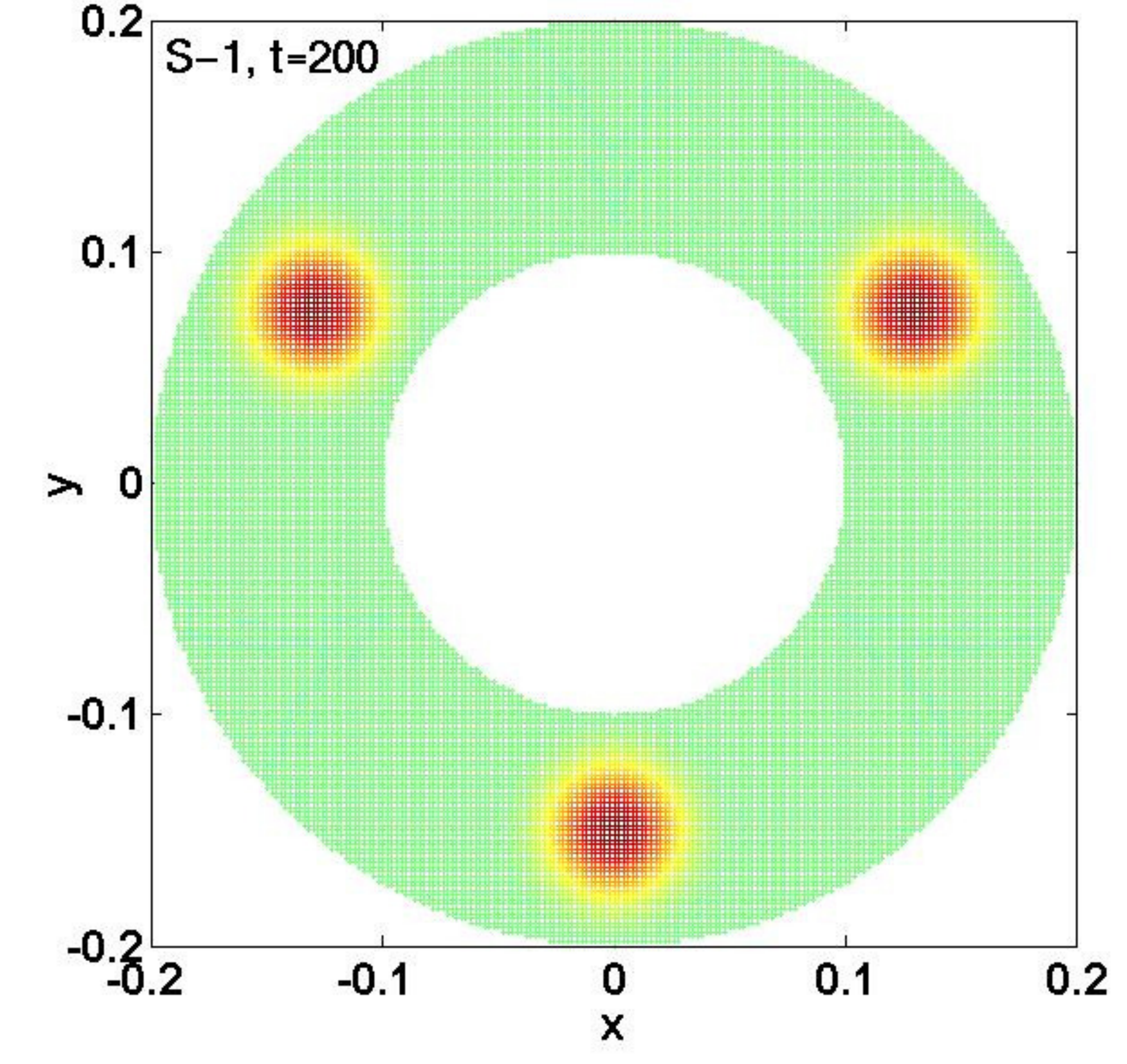}
\includegraphics[width=0.7\linewidth]{cbar2-eps-converted-to.pdf}
\caption{Snapshots of diffusion of blob at $t=30$ (left) and $t=200$ (right) on plane $z=0$ simulated with a) \textbf{N-1} scheme and b) \textbf{S-1} scheme. A radially constant safety factor of $q=3$ was chosen. Resolution was $h=2\cdot10^{-3}$, $\Delta z/2\pi=1/8$. Initial blob was located at the bottom ($x_g=0,\,y_g=-0.15,\,w_x=w_y=0.025$ according to  equation \refeq{blobdef}).} 
\label{fig_blobdiff}
\end{figure}

We consider an initial state which is Gaussian in the poloidal plane and a delta function in the axial direction, i.e.:
\begin{align}
u(t=0,\mathbf{x})=\exp\left[-\left(x-x_g\right)^2/w_x^2+\left(y-y_g\right)^2/w_y^2\right]\,\delta\left(z\right),
\label{blobdef}
\end{align} 
where the delta function is modelled on the discrete level with a Kronecker delta $\delta\left(z\right)\rightarrow \delta_{0\,k}\,2\pi/\Delta z$. This structure diffuses along magnetic field lines until it has established a state with $k_{\parallel}=0$, which should remain stable. In fig.~\ref{fig_blobdiff} the temporal evolution of such a Gaussian blob is shown computed with the \textbf{N-1} and \textbf{S-1} scheme. The example illustrates that after the structure has established a state in which $k_{\parallel}=0$ at $t\approx30$, it remains stable for a long time with the support operator scheme, whereas it decays fast with the naive scheme due to numerical perpendicular diffusion.

\begin{figure}[!htb]
a)\newline
\includegraphics[width=1.0\linewidth]{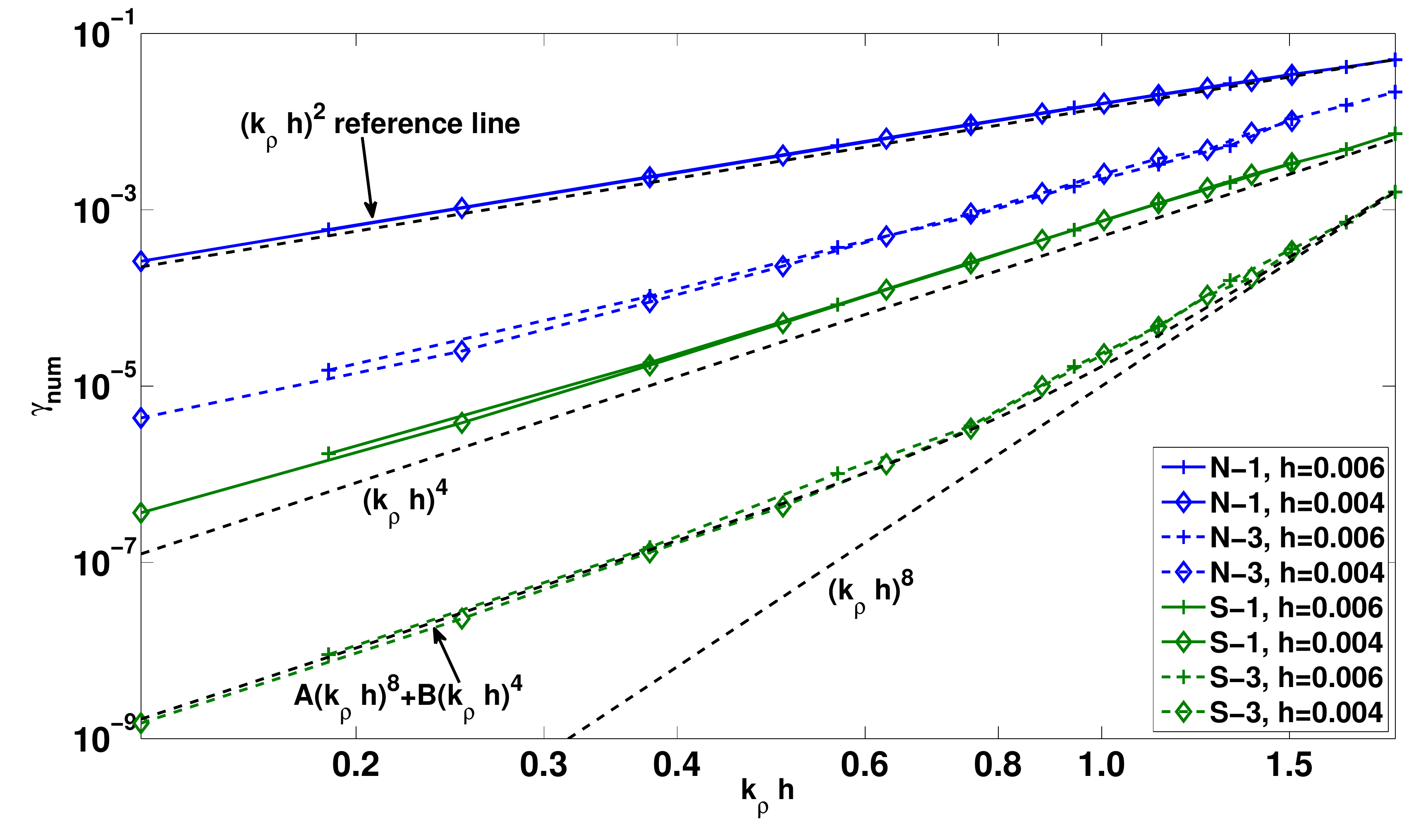}
b)\newline
\includegraphics[width=1.0\linewidth]{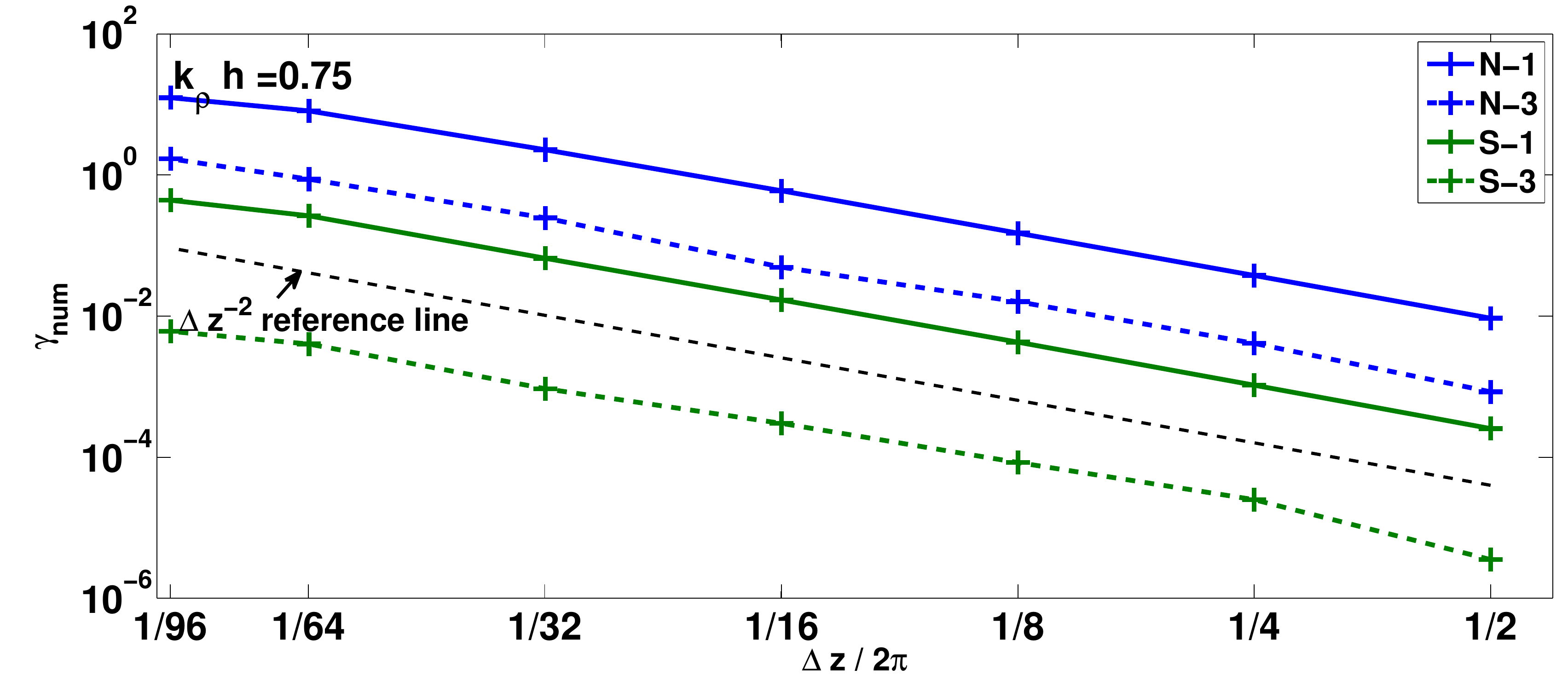}
\caption{Numerical decay rate in dependence on a) poloidal resolution of mode for fixed axial resolution $\Delta z=2\pi/2$. $k_{\rho}:=\pi r/(\rho_{max}-\rho_{min})$ is the radial wavevector. For the establishment of this plot the radial mode number $r$ has been varied at two distinct poloidal resolutions of $h=6\cdot10^{-3}$ and $h=4\cdot10^{-3}$ have been used. b) Numerical decay rate in dependence on axial resolution for fixed poloidal resolution of mode $\left(k_{\rho}h=0.75\right)$.} 
\label{fig_numdecayperp}
\end{figure}

Any decay of zonal modes $\left(r,\,m=0,\,n=0\right)$ arises from erroneous numerical perpendicular diffusion. The numerical diffusion can be quantified by estimating the decay rate of the $L^2$-norm of zonal modes. In fig.~\ref{fig_numdecayperp}a the measured decay rate in dependence on the poloidal resolution of the zonal modes is plotted for fixed axial resolution. In fig.~\ref{fig_numdecayperp}b the numerical decay rate in dependence on the axial resolution for a fixed poloidal resolution of a zonal mode is plotted. Summarizing the numerical decay exponent scales like:
\begin{align}
\gamma_{num}\propto\begin{cases}
\left(k_{\rho}h\right)^2/\Delta z^2, \quad\text{for \textbf{N-1}}, \\
\left(k_{\rho}h\right)^4/\Delta z^2, \quad\text{for \textbf{N-3}}, \\
\left(k_{\rho}h\right)^4/\Delta z^2, \quad\text{for \textbf{S-1}}, \\
\left[A\left(k_{\rho}h\right)^8+B\left(k_{\rho}h\right)^4\right]/\Delta z^2, \quad\text{for \textbf{S-3}}.  
\end{cases}
\label{scaling_numdecay}
\end{align}

These results confirm again the scalings of the two-dimensional model (see equations \refeq{taylor_naive_lin}, \refeq{taylor_supp_lin}, \refeq{taylor_naive_pol}, \refeq{taylor_supp_pol}). Only for the \textbf{S-3} scheme a break in the scaling is observed, which arises from the boundaries, where the conditions for the interpolation become worse (It has also been observed that the numerical decay sets first in near the boundaries). The performed numerical measurements and the discussion of the two-dimensional model problem (see section \ref{sec_model2d} and \ref{app_polint}) may suggest the general rule that, if $p$ is the order of the polynomial interpolation, the numerical decay exponent scales like:
\begin{align}
\gamma_{num}\propto\begin{cases}
\left(k_{\rho}h\right)^{p+1}/\Delta z^2, \quad\text{for naive schemes}, \\
\left(k_{\rho}h\right)^{2\left(p+1\right)}/\Delta z^2, \quad\text{for support schemes}, \\
\end{cases}
\label{numdecay_general}
\end{align}
apart from effects introduced by the boundary, which introduce a break into the scaling at high poloidal resolutions. The investigation of numerical perpendicular diffusion for other interpolation techniques is left for future work. 

\subsection{Realistic toroidal geometry}
We consider an equilibrium, which is an analytic solution to the Grad-Shafranov equation \cite{mccarthy:equi99}. In order to prove the applicability of the scheme also to complicated and realistic tokamak geometries, we consider a flux shell in the edge ($\rho_{min}=0.90$, $\rho_{max}=0.95$, the normalized radial flux label is defined as $\rho:=\sqrt{\left(\psi-\psi_s\right)/\left(\psi_0-\psi_s\right)}$, where $\psi_0,\,\psi_s$ the poloidal magnetic flux at the magnetic axis, respectively at the separatrix).

\subsubsection{Verification of map distortion}\label{sec_verifymapdistort}
\begin{figure}[!htb]
\centering
\begin{flushleft}a)\hspace{0.23\textwidth}b)\end{flushleft}
\includegraphics[width=0.48\linewidth]{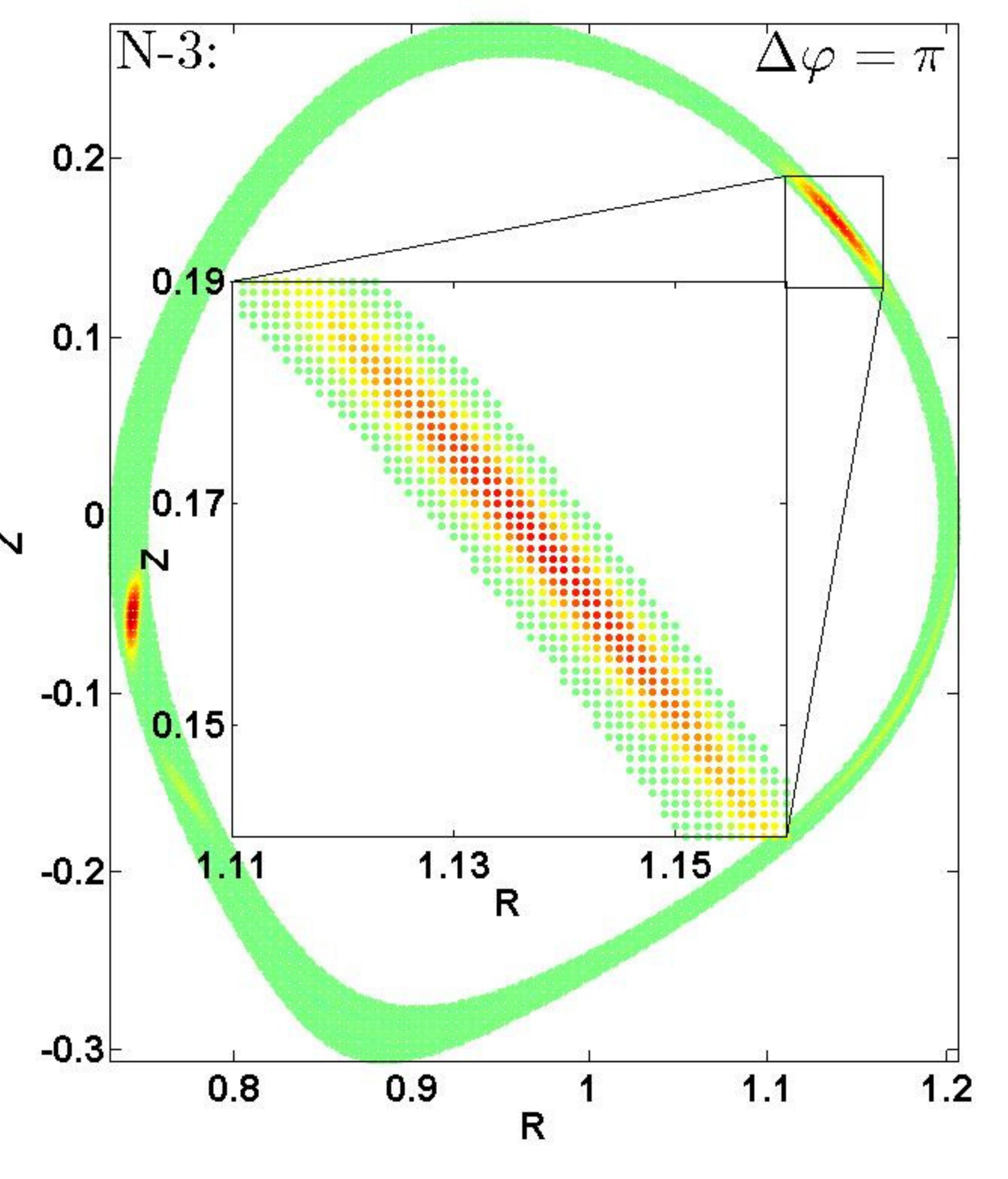}
\includegraphics[width=0.48\linewidth]{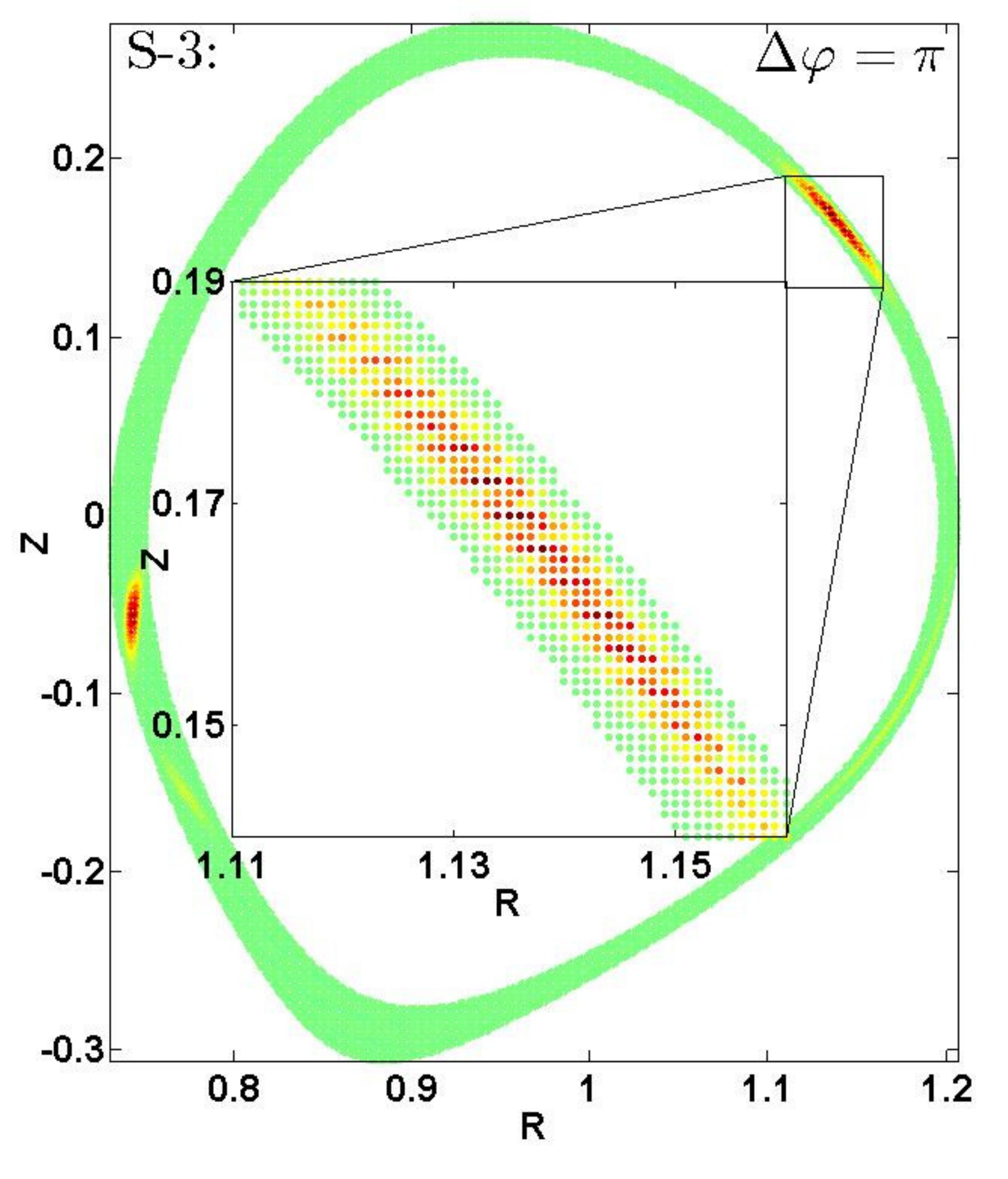}
\begin{flushleft}c)\hspace{0.23\textwidth}d)\end{flushleft}
\includegraphics[width=0.48\linewidth]{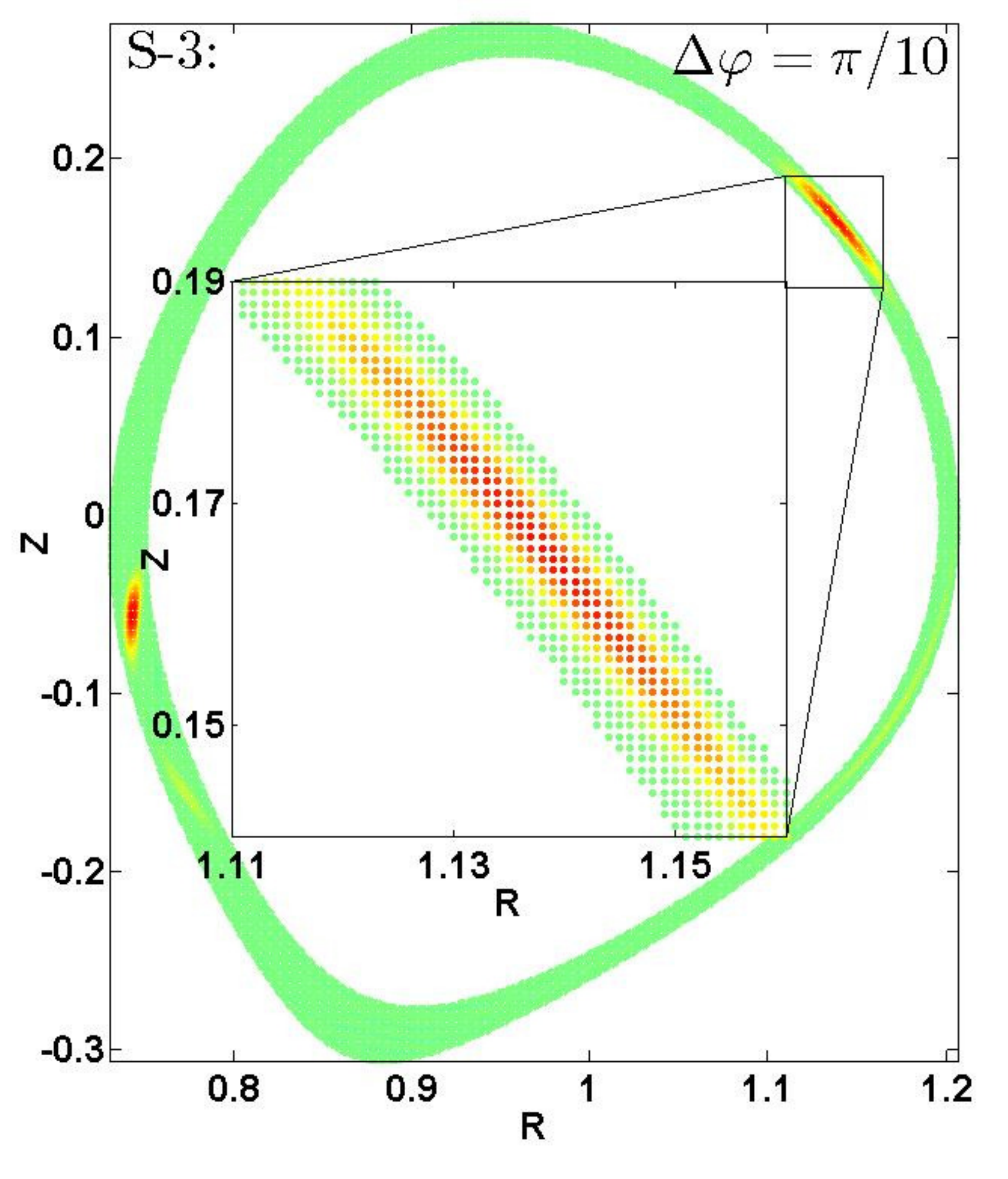}
\includegraphics[width=0.48\linewidth]{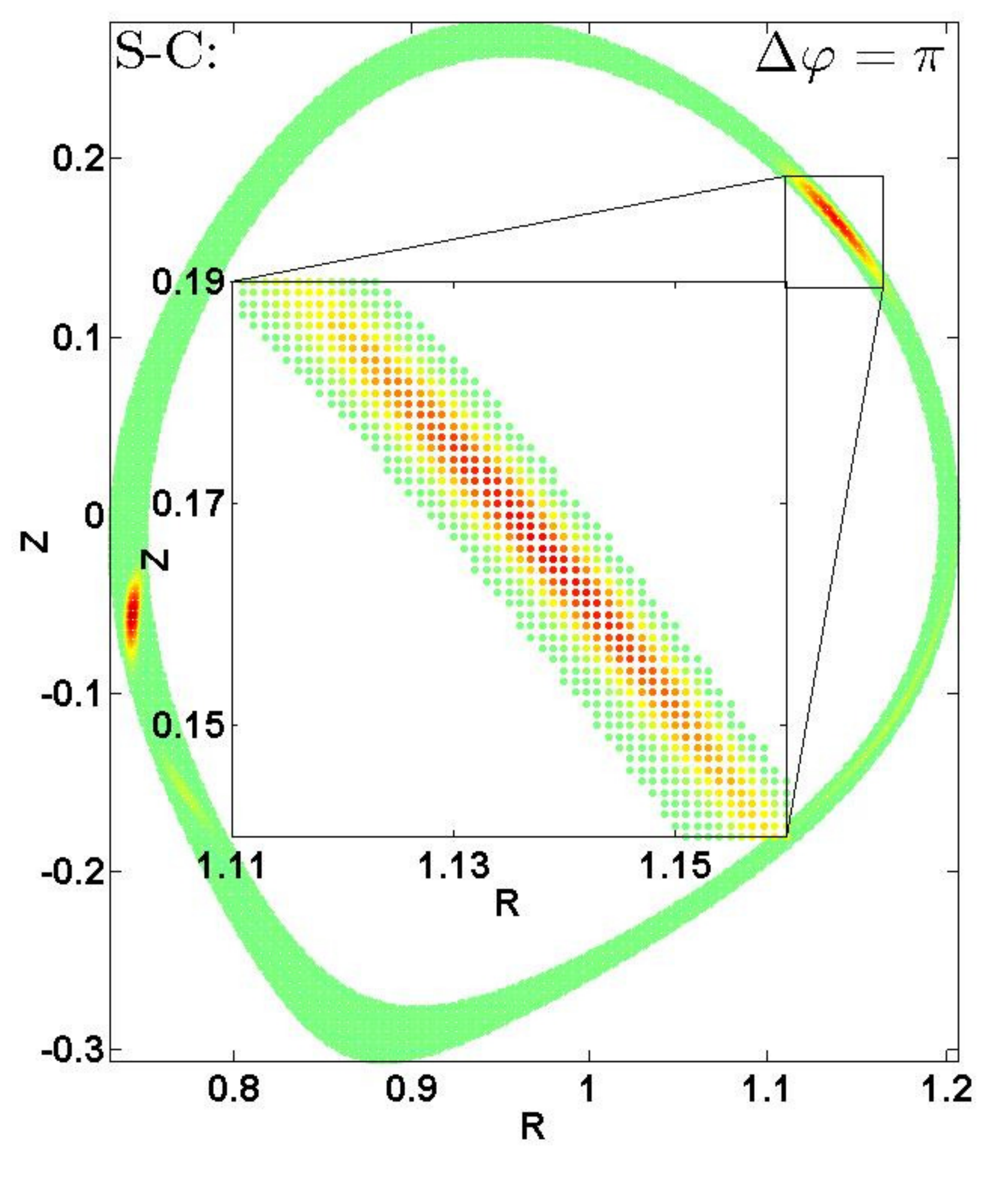}
\includegraphics[width=0.7\linewidth]{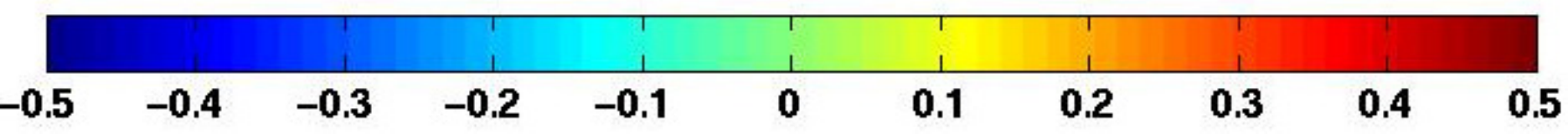}
\caption{Snapshots of diffusion of Gaussian blob at $t=10$, $\varphi=\pi$. Initial state was Gaussian blob $\left(R_g=0.88,\,Z_g=-0.29,\,w_R=w_Z=8\cdot10^{-3}\right)$ and poloidal resolution was $h=1\cdot10^{-3}$. a) Computed with \textbf{N-3} scheme and $\Delta\varphi/2\pi=1/2$, b) with \textbf{S-3} scheme and $\Delta\varphi/2\pi=1/2$. The strong map distortion causes unphysical wiggles. This can be cured by increasing toroidal resolution as was done in c) (\textbf{S-3} scheme with $\Delta\varphi/2\pi=1/20$) or d) by using the integration method for the parallel gradient (\textbf{S-C} scheme with $\Delta\varphi/2\pi=1/2$).} 
\label{fig_wiggles}
\end{figure}

To illustrate the effects of the map distortion on the schemes, we consider the diffusion of a Gaussian blob situated at the bottom of the flux shell in proximity to the X-point. In fig.~\ref{fig_wiggles}a,b snapshots are shown of solutions computed with the naive scheme \textbf{N-3} and the support scheme \textbf{S-3}. The simulations were performed only with two poloidal planes $\Delta\varphi/2\pi=1/2$. On a coarse scale the blob appears at similar positions, but on a finer scale strong unphysical wiggles arise in the simulation performed with the support scheme. As discussed in section \ref{sec_mapdistort}, the erroneous wiggles are a result of the strong map distortion, which was $d_c=44$ and $d_a=138$. These problems can be cured by requiring enough toroidal resolution, as shown in fig.~\ref{fig_wiggles}c where the simulation was repeated with a higher toroidal resolution of $\Delta\varphi/2\pi=20$. The distortion remained with values of $d_c=1.7$ and $d_a=1.9$ well below the proposed threshold of $4$ (see expressions \refeq{map_distort_threshold}). The problems with the map distortion can also be cured via the integration method for the parallel gradient (equation \refeq{par_grad_coordfree}). In fig.~\ref{fig_wiggles}d the same simulation was performed with again a low toroidal resolution of $\Delta\varphi/2\pi=1/2$, but with the \textbf{S-C} scheme. No erroneous wiggles but smooth structures arise.

\begin{figure}[!htb]
a)\newline
\includegraphics[width=1.0\linewidth]{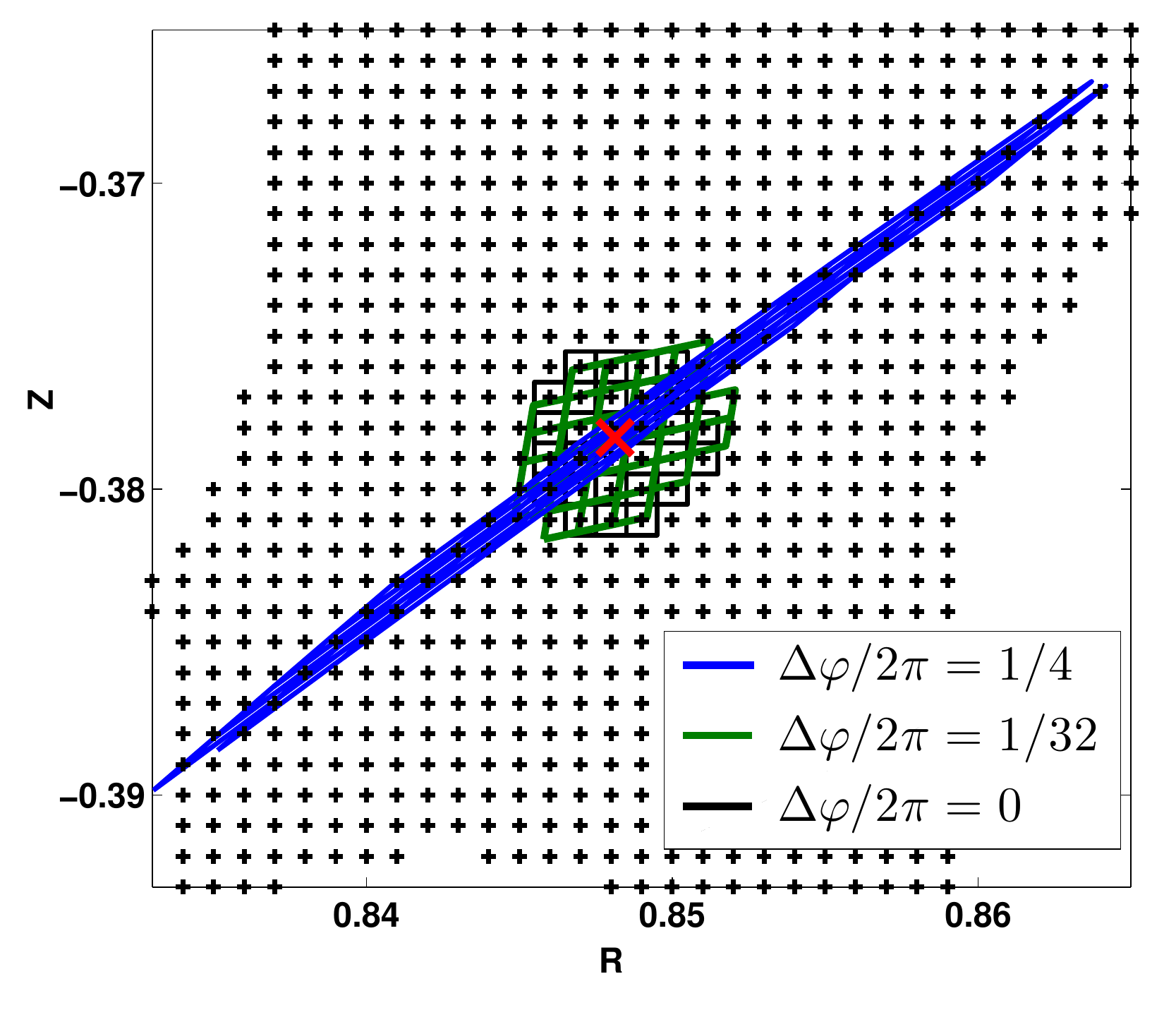}\newline
b)\newline
\includegraphics[width=1.0\linewidth]{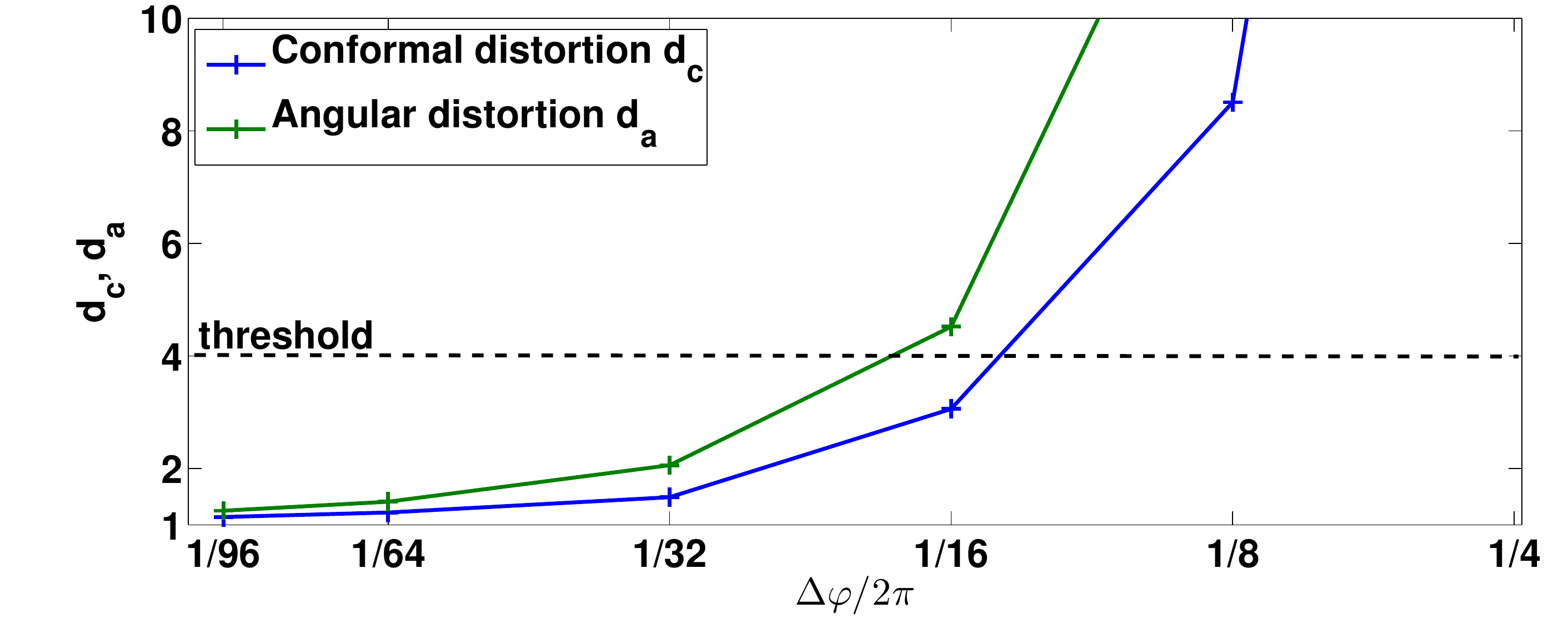}
\caption{a) Mapped quads ('+'-direction) for some sample grid points around X-point (red cross) for different toroidal resolutions. b) Map distortion around X-point in dependence on toroidal resolution.} 
\label{fig_mapdistortx}
\end{figure}

The strongest map distortion occurs around the X-point. For the given equilibrium the mapped quads around the X-point for different toroidal resolutions are illustrated in fig.~\ref{fig_mapdistortx}a. For too low toroidal resolutions (blue) the mapped quads are strongly distorted and extend over many grid points. The quality of the grid would be much too poor for support schemes (interpolation based, i.e.~\textbf{S-1}, \textbf{S-3}). At higher toroidal resolutions (green) the mapped quads are only weakly distorted and the support schemes would work well. In fig.~\ref{fig_mapdistortx}b the map distortion in dependence on toroidal resolution is plotted. A slightly higher toroidal resolution than $\Delta\varphi/2\pi=1/16$ would have to be used, to fulfil the proposed requirement \refeq{map_distort_threshold}.

\subsubsection{Convergence behaviour}
For realistic geometry rigorous convergence checks -as it was done for axial circular geometry in section \ref{sec_axialcircconverg}- are elaborate and computationally quite costly. Effects of a complex magnetic field structure on the schemes (map distortion) have been discussed in detail in section \ref{sec_mapdistort} and have been verified in section \ref{sec_verifymapdistort}. Therefore, only a rough convergence test for the schemes is presented here for realistic geometry.

For the diffusion of the Gaussian blob, which was presented in section \ref{sec_verifymapdistort}, a spatial resolution scan was performed. However, we only consider the obtained maximum value on plane $\varphi=\pi$ at time $t=10.0$, i.e.:
$\|u\left(t=10,R,Z,\varphi=\pi\right)\|_{\infty}$. Effects of numerical diffusion will manifest themselves in a drop of this value. A reduction of the timestep below values where the temporal discretisation error would be negligible turned also out to be computationally to costly. Therefore, the timestep was held constant at $\Delta t=1\cdot10^{-2}$ regardless of spatial resolution, in order to obtain the behaviour of the spatial discretisation error. 

\begin{figure}[!htb]
a)\newline
\includegraphics[width=1.0\linewidth]{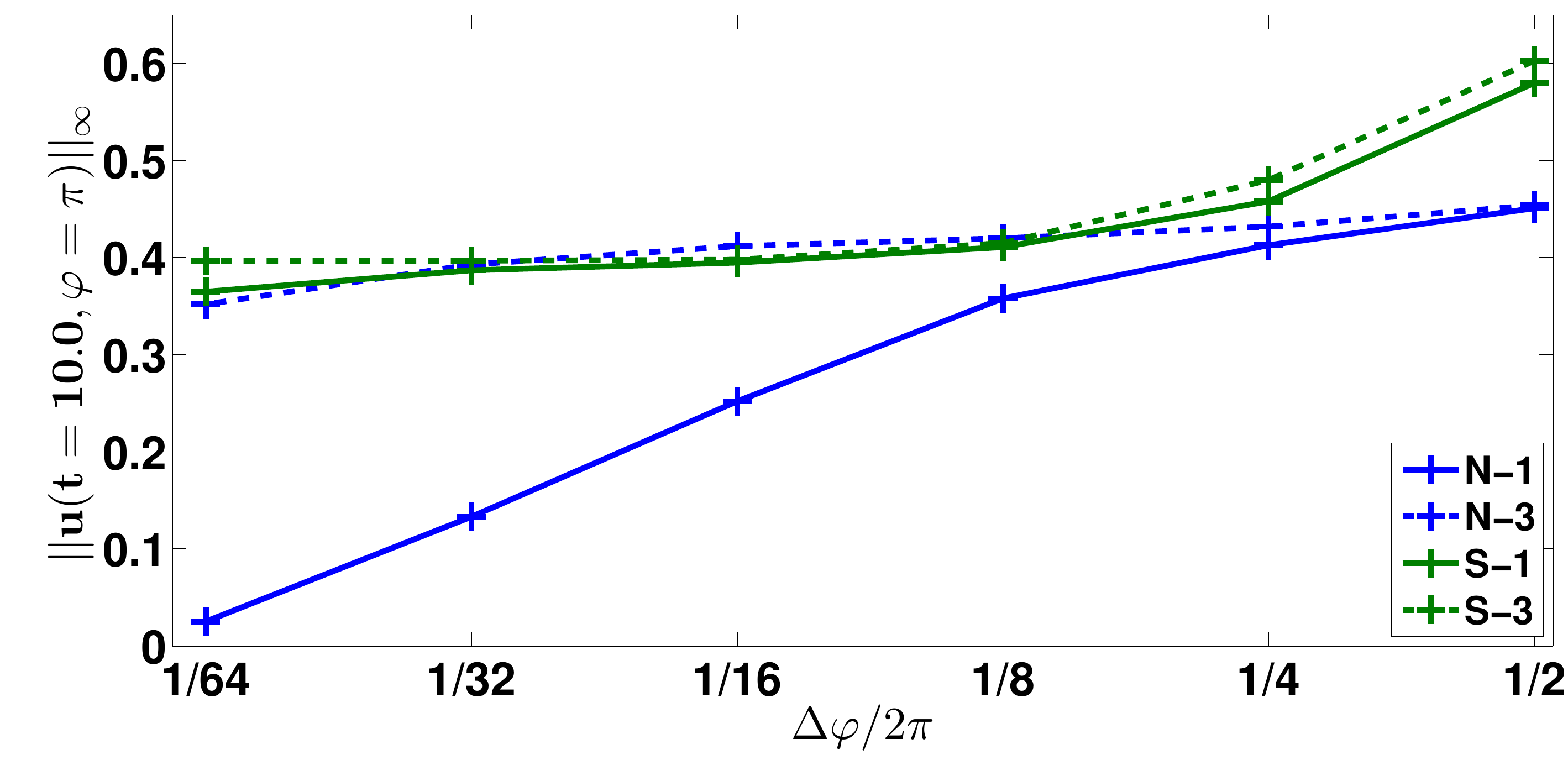}
b)\newline
\includegraphics[width=1.0\linewidth]{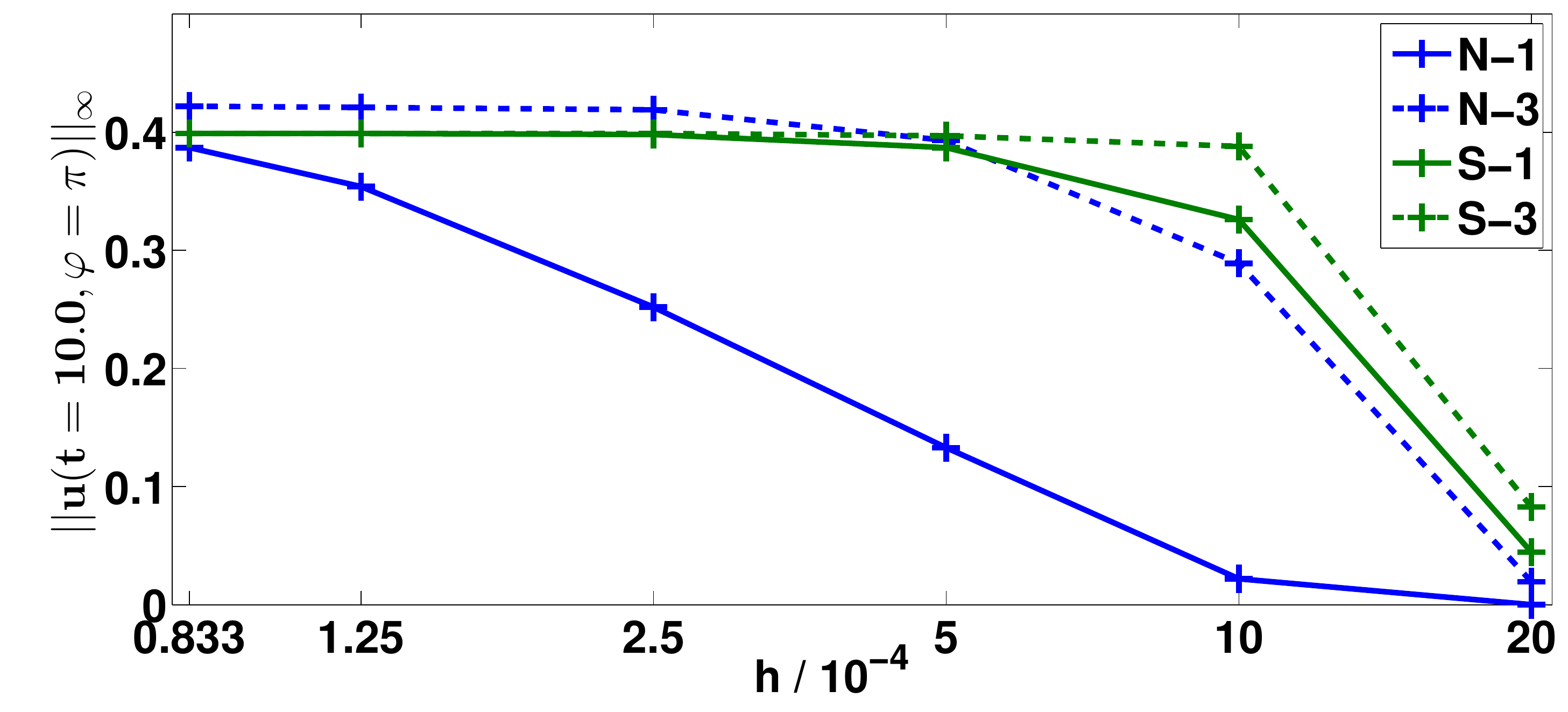}
\caption{Spatial resolution scan of diffusion of Gaussian blob $\left(R_g=0.88,\,Z_g=-0.29,\,w_R=w_Z=8\cdot10^{-3}\right)$. Obtained maximum value on plane $\varphi=\pi$ at time $t=10.0$ (For examples of snapshots see also fig.~\ref{fig_wiggles}) in dependence on a) toroidal resolution for fixed poloidal resolution of $h=5\cdot10^{-4}$ and b) in dependence on poloidal resolution for fixed toroidal resolution of $\Delta\varphi/2\pi=1/32$.} 
\label{fig_blobconverg}
\end{figure}

In fig.~\ref{fig_blobconverg}a the convergence behaviour in dependence on toroidal resolution is shown. At low toroidal resolutions the support and the naive schemes yield different values, since the map distortion is above the requested threshold given in \refeq{map_distort_threshold} and unphysical wiggles arise with the \textbf{S-1} and \textbf{S-3} schemes. At higher toroidal resolutions $\Delta\varphi/2\pi\leq1/8$ the \textbf{N-3}, \textbf{S-1} and \textbf{S-3} schemes agree quite well, whereas the numerical perpendicular diffusion for the \textbf{N-1} scheme is large which causes a decay of the blob. At the highest toroidal resolution $\Delta\varphi/2\pi=1/64$ the value drops slightly for the \textbf{N-3} and \textbf{S-1} schemes, which is an indication for numerical perpendicular diffusion. The convergence behaviour in dependence on poloidal resolution is shown in fig.~\ref{fig_blobconverg}b. It is apparent that the convergence for the \textbf{N-1} scheme is slowest and could even not be achieved satisfactory with the available resolution, whereas \textbf{N-3}, \textbf{S-1} and \textbf{S-3} exhibit similar convergence rates.
The naive scheme convergences against a slightly different value $\left(\approx0.42\right)$ than the support schemes $\left(\approx0.40\right)$. The reason for this is that also slightly different operators are discretised with the naive scheme $\left(\nabla_{\parallel}^2\right)$ and the support schemes $\left(\nabla\cdot\left[\mathbf{b}\nabla_\parallel\circ\right]\right)$. 

\subsubsection{Numerical diffusion}\label{sec_shapenumdiff}
Performing a resolution scan of numerical decay rates for zonal modes is again computationally elaborate. Moreover, the points per radial wavelength of a zonal mode varies within the poloidal planes due to flux expansion. Therefore, we consider only a single example and investigate if the numerical decay rates for realistic geometry correspond with the results from section \ref{sec_axialcircnumdiff} obtained in axial circular geometry.

We consider a zonal mode $r=2,\,m=0,\,n=0$ and choose a resolution of $h=5\cdot10^{-4}$ and $\Delta \varphi/2\pi=1/20$. The poloidal resolution of the mode varies between $k_\perp h \approx 3\cdot10^{-1}$ at outboard midplane and $k_\perp h \approx 8\cdot10^{-2}$ at the bottom. From the scaling \refeq{scaling_numdecay} and the numerical measurement given in fig.~\ref{fig_numdecayperp}a we may expect decay exponents of:
\begin{align}
\gamma_{num}\approx\begin{cases}
\left[10^{-2},10^{-1}\right], \quad\text{for \textbf{N-1}}, \\
\left[10^{-4},10^{-2}\right],  \quad\text{for \textbf{N-3}}, \\
\left[10^{-5},10^{-3}\right], \quad\text{for \textbf{S-1}}, \\
\left[10^{-8},10^{-5}\right], \quad\text{for \textbf{S-3}}.  
\end{cases}
\end{align}
We measured in the realistic geometry numerical decay rates of around: 
\begin{align}
\gamma_{num}\approx\begin{cases}
8\cdot10^{-2}, \quad\text{for \textbf{N-1}}, \\
2\cdot10^{-3},  \quad\text{for \textbf{N-3}}, \\
4\cdot10^{-4}, \quad\text{for \textbf{S-1}}, \\
1\cdot10^{-5}, \quad\text{for \textbf{S-3}}, 
\end{cases}
\end{align}
which agrees with the estimates.

\section{Conclusion and final remarks}\label{sec_summary}
In the field line map approach field/flux-aligned coordinates, which become singular on the separatrix/X-point, are avoided. The concept is based on a cylindrical grid, which is sparse in the toroidal direction, and a field line following discretisation for parallel operators to exploit the characteristic flute mode property of the solutions. The discretisation of perpendicular operators is straight forward and simple, whereas in the discretisation of parallel operators an interpolation (see equation \refeq{par_grad_interpolation}) or integration (see equation \refeq{par_grad_coordfree}) is involved. Although it is not discussed here, there is no obvious reason, why the developed methods could not easily be generalized also to three-dimensional equilibria, i.e.~stellerators.

Two discretisation schemes for the parallel diffusion operator have been presented: A naive scheme, and a discretisation according to the support operator method, which conserves the self-adjointness property of the parallel diffusion operator on the discrete level. It has been shown in section \ref{sec_model2d} with a two-dimensional model problem that the support schemes exhibit a lower (better scaling) numerical 'diffusion' (see equations \refeq{taylor_naive_lin} and \refeq{taylor_supp_lin}). The effects of a strongly distorted map on the support scheme have been identified and verified. For interpolation based schemes (\textbf{S-1} and \textbf{S-3}) a minimum toroidal resolution has to be supplied, to reduce the map distortion below the threshold given in expression \refeq{map_distort_threshold}. The problems with distorted maps can also be cured via the integration method for the parallel gradient (\textbf{S-C}). 

The numerical methods are implemented in the new code GRILLIX and extensive benchmarks, which show the validity of the field line map approach in general and GRILLIX in particular, were presented in section \ref{sec_benchmarks}. Investigated were mainly the support and naive scheme each with a bilinear and a third order bipolynomial interpolation (\textbf{N-1}, \textbf{N-3}, \textbf{S-1}, \textbf{S-3}). Due to the interpolation, the convergence behaviour depends in general on the toroidal resolution $\Delta\varphi$ and also the poloidal resolution $h$. In agreement with the two-dimensional model (see section \ref{sec_model2d} and \ref{app_polint}), the convergence rate with respect to the poloidal resolution is slowest for the \textbf{N-1} scheme, whereas it is similar for the \textbf{N-3}, \textbf{S-1} and \textbf{S-3} scheme. Considering structures with $k_{\parallel}=0$, a general scaling for their numerical decay rate is suggested in equation \refeq{numdecay_general}. However, the presence of radial boundaries may introduce at high resolutions a break of this scaling for higher order interpolation methods. 

The scalings were derived only from a two-dimensional model and rigorous benchmarks have only been performed for axial circular geometry. However, the suggested results and scalings seem to be applicable also to realistic geometry, as long as the map distortion is below the proposed threshold \refeq{map_distort_threshold}, i.e.~a sufficient toroidal resolution has to be supplied. As shown for example in section \ref{sec_shapenumdiff}, numerical decay rates can be estimated also for general geometry from the scalings \refeq{numdecay_general} and the numerical measurement shown in fig.~\ref{fig_numdecayperp}.

From a practical point of view the \textbf{S-3} scheme might be preferable, since it exhibits the lowest numerical diffusion among the presented methods. Even higher order interpolation methods (e.g.~fifth order bipolynomial interpolation \textbf{S-5}) may be problematic, since their stencil can become large and they could cause unphysical oscillatory structures (Runge's phenomenon) especially near the radial boundaries. Although the implementation of the \textbf{S-C} scheme might seem cumbersome, its advantage is that any spurious effects arising from map distortion are absent. Future work might therefore address a combination of interpolation and integration, where the surface integral in equation \refeq{coordfree_continuouslevel} is executed by assuming interpolated polynomials as basis functions. 

In future work an application of the developed methods to a simple plasma turbulence model will be presented.

\section{Acknowledgements}
The authors want to thank Markus Held and Matthias Wiesenberger from University of Innsbruck for contributing to this work with fruitful comments. 

The research leading to these results has received funding from the European Research Council under the European Union’s Seventh Framework Programme (FP7/2007-2013)/ERC Grant Agreement No. 277870.

A part of this work was carried out using the HELIOS supercomputer system at Computational Simulation Centre of International Fusion Energy Research Centre (IFERC-CSC), Aomori, Japan, under the Broader Approach collaboration between Euratom and Japan, implemented by Fusion for Energy and JAEA.

\appendix

\section{Additional model problems} \label{appendix}
In this appendix additional two-dimensional model problems are discussed, where explicit expressions for the parallel diffusion operator are given.

As in the model problem of section \ref{sec_model2d} $x$ plays the role of a coordinate within poloidal planes and $z$ the role of a periodic toroidal/axial coordinate. However, the magnetic field is assumed to be more general now:
\begin{align}
\mathbf{B}=B_0\mathbf{e}_z+B^x(x)\mathbf{e}_x,
\end{align}
with $B^x\ll B_0$. Again the domain is spanned by a regular grid $x_i,z_k$, with grid spacing $h$ in the $x$-direction and grid spacing $\Delta z$ in the $z$-direction. From each grid point $x_i,z_k$ the penetration points at the planes $z_{k\pm1}$ are expressed via:
\begin{align}
x_i^\pm=x_i\pm h\left(n^\pm_i+f_i^\pm\right),
\end{align}  
where $n^\pm_i \in \mathbb{Z}$, and $0\leq f_i^\pm <1$ a dimensionless quantity which expresses the displacement of the penetration points with respect to the grid. For simplicity, we assume in the following that the lengths along field lines and the flux box volumes are all equal, i.e.~$\Delta s_i=\Delta s$ and $\Delta V_i=\Delta\mathcal{V}_i=h\Delta z$.

\subsection{Inhomogeneous magnetic field (Interpolation)}\label{app_inhom}
\begin{figure}[!htb]
a)\newline
\includegraphics[width=1.0\linewidth]{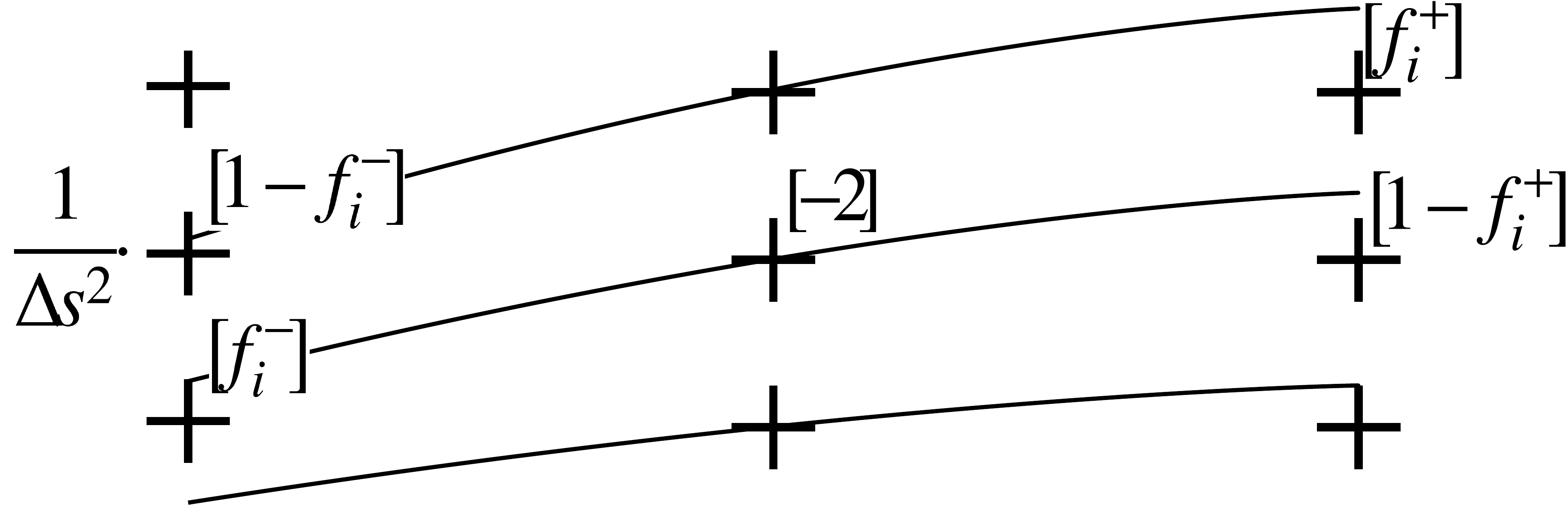}\newline
b)\newline
\includegraphics[width=1.0\linewidth]{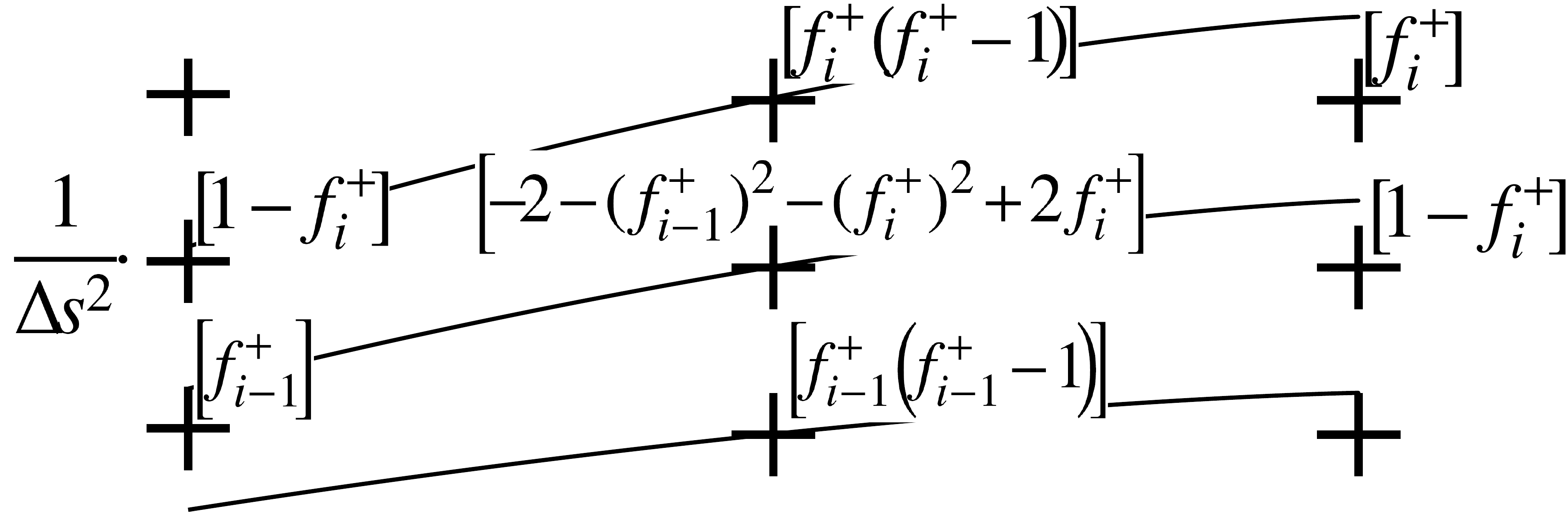}\newline
b)\newline
\includegraphics[width=1.0\linewidth]{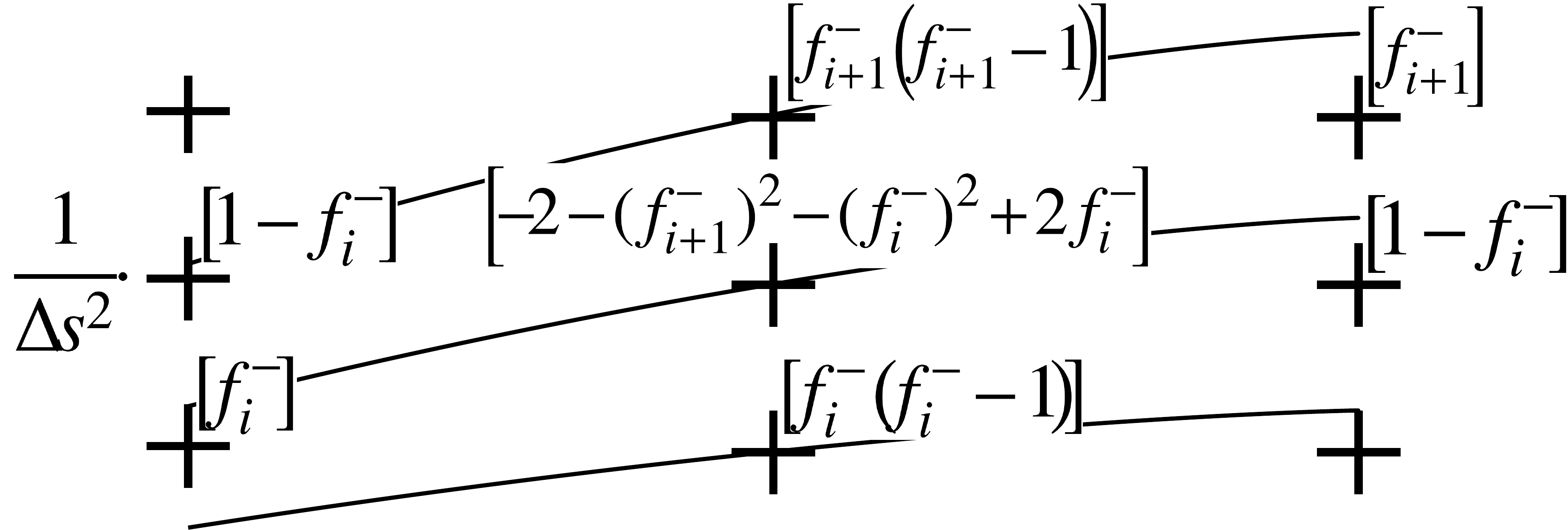}
\caption{Discrete parallel diffusion operator for non-homogeneous magnetic field. a) Naive scheme $\mathbf{D}_\parallel^{naive}$, b) support scheme $'+'$ discretisation $\mathbf{D}_\parallel^{supp,+}$, c) $'-'$ discretisation $\mathbf{D}_\parallel^{supp,-}$.} 
\label{fig_model_inhom}
\end{figure}

In section \ref{sec_model2d} the case for constant displacement has been discussed, i.e.~$f_{i}^\pm=f=const.$. We consider here the more general case where the displacement factors $f^\pm_i$ vary. A linear interpolation is applied for the computation of the discrete parallel gradient. 
\begin{align}
\left(\mathbf{Q}^\pm\mathbf{u}\right)_{i,k}=\pm\frac{1}{\Delta s}\left[\left(1-f_i^\pm\right)u_{n^\pm_i,k\pm1}+f^\pm_i u_{n^\pm_i\pm1,k\pm1}-u_{i,k}\right]
\end{align}
The expression for the discrete parallel diffusion according to the naive scheme is given in fig.~\ref{fig_model_inhom}a. The $'+'$ and the $'-'$ discretisation  for the support scheme yield in general different expressions. In fig.~\ref{fig_model_inhom}b the expression for $\mathbf{D}_\parallel^{supp,+}$ is given and in fig.~\ref{fig_model_inhom}c for $\mathbf{D}_\parallel^{supp,-}$. The given expressions are valid only for $n_i^\pm=const.$ The final operator $\mathbf{D}_\parallel^{supp}$ can be easily obtained as the average between both.

\begin{figure}[!htb]
\begin{center}
\includegraphics[width=1.0\linewidth]{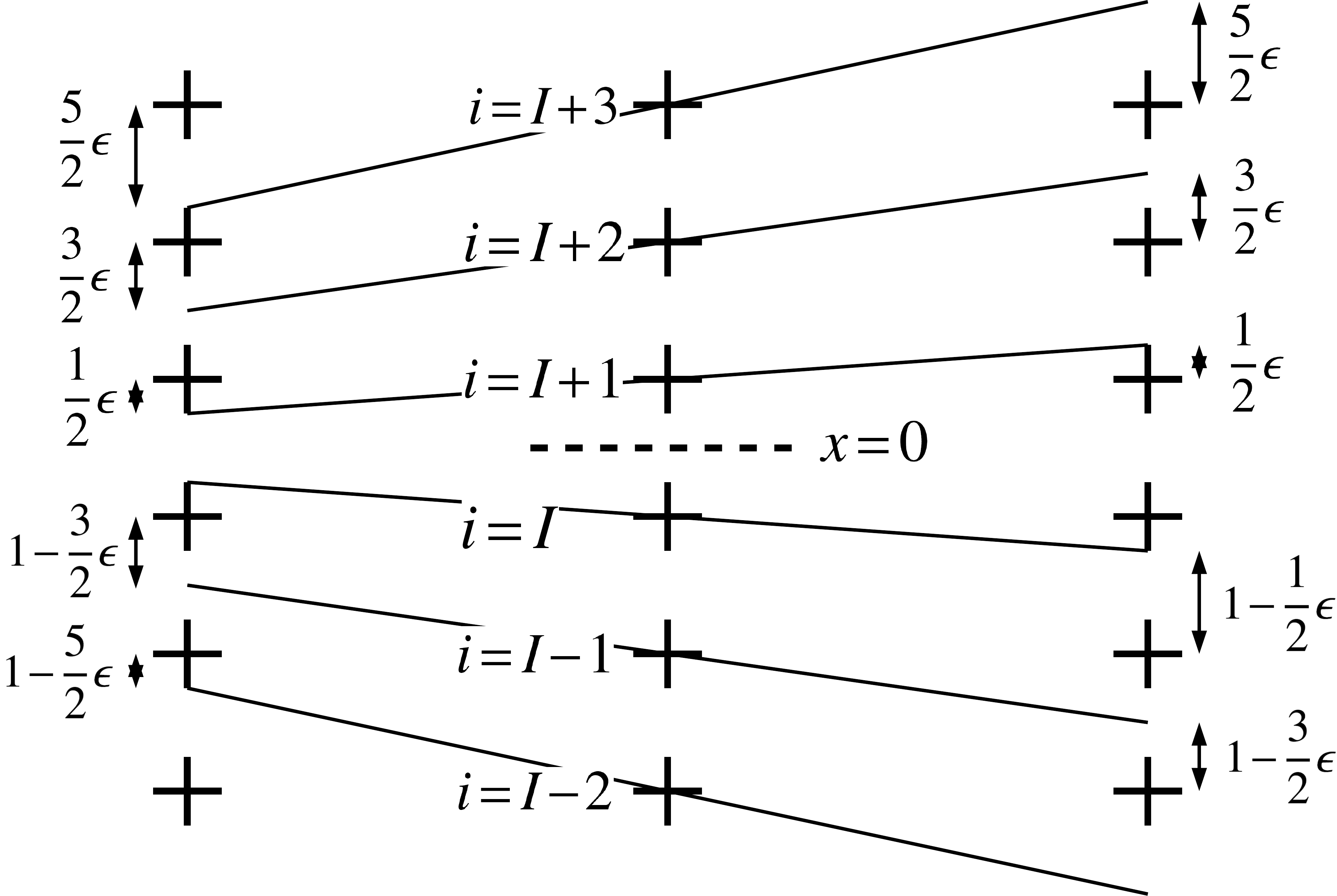}
\end{center}
\caption{Displacement factors $f_{i}^\pm$ for model problem with diverging field lines.} 
\label{fig_model_interpolchange}
\end{figure}

An interesting case to consider is $B^x=\alpha x$ (This mimics the situation at the X-point, where a convergent magnetic field in a third dimension $B^y=-\alpha y$ would ensure $\nabla\cdot\mathbf{B}=0$). In the limit $\epsilon:=\alpha\Delta z/B_0\ll 1$ the penetration points are $x_i^\pm=x_i\left(1\pm\epsilon\right)$, and the displacement factors can be obtained as (see fig.~\ref{fig_model_interpolchange}):
\begin{align}
n_i^\pm=&\begin{cases}
-1, &\text{for: }i\leq I, \\
0, &\text{for: }i> I,
\end{cases} \\
f_i^\pm=&\begin{cases}
1-\epsilon\left[\frac{1}{2}+\left(I-i\right)\right], &\text{for: }i\leq I, \\
\epsilon\left[\frac{1}{2}+\left(i-I\right)\right], &\text{for: }i> I
\end{cases},
\end{align}
where $x_I=-h/2$. Note that there is a jump present in $n_i^\pm$ at $i=I$. One may easily work out the expression for the discrete parallel diffusion operator. However, we give here only the result for a structure, which varies slowly in the $x$-direction, i.e.~$u_{i,k}=\tilde{u}_k$. The result for the naive scheme is:
\begin{align}
&\left(\mathbf{D}_\parallel^{naive}\tilde{\mathbf{u}}\right)_{i,k}=\frac{-2\tilde{u}_k+\tilde{u}_{k-1}+\tilde{u}_{k+1}}{\Delta s^2},
\end{align}
and for the support scheme:
\begin{align}
&\left(\mathbf{D}_\parallel^{supp}\tilde{\mathbf{u}}\right)_{i,k}=\notag\\
&\begin{cases}
\dfrac{-2\tilde{u}_k+\tilde{u}_{k-1}+\tilde{u}_{k+1}}{\Delta s^2}+\dfrac{\epsilon}{\Delta s}\dfrac{\tilde{u}_{k+1}-\tilde{u}_{k-1}}{2\Delta s} &\text{for }i\neq I,I+1, \vspace{0.2cm}\\
\left(1+\dfrac{\epsilon}{4}\right)\dfrac{-2\tilde{u}_k+\tilde{u}_{k-1}+\tilde{u}_{k+1}}{\Delta s^2}+\dfrac{\epsilon}{\Delta s}\dfrac{\tilde{u}_{k+1}-\tilde{u}_{k-1}}{2\Delta s} &\text{for }i=I,I+1.
\end{cases}
\label{model_dist_s1}
\end{align}
The first term represents the discrete analogue of $\nabla_\parallel^2u$ and the second term $\left(\nabla\cdot\mathbf{b}\right)\nabla_\parallel u$, since $\epsilon/\Delta s\approx\alpha/B_0\approx\nabla\cdot\mathbf{b}$. Note that for $i=I,I+1$ a small erroneous oscillation of size $\epsilon/4$ arises due to the jump in $n_i^\pm$. With the integration method for the discrete parallel gradient according to equation \refeq{par_grad_coordfree} this oscillation does not arise (see section \ref{app_integration}, equation \refeq{model_dist_sc}). 

Such an oscillation is also observed with GRILLIX (\textbf{S-1} and \textbf{S-3}). In practice, as this oscillation remains on the grid scale as long as $\left|n_{i+1}^\pm-n^\pm_i\right|\leq 1$, it can also be cured by applying additionally a small perpendicular high order diffusion. 

\subsection{Example for integration method}\label{app_integration}
\begin{figure}[!htb]
\begin{center}
\includegraphics[width=1.0\linewidth]{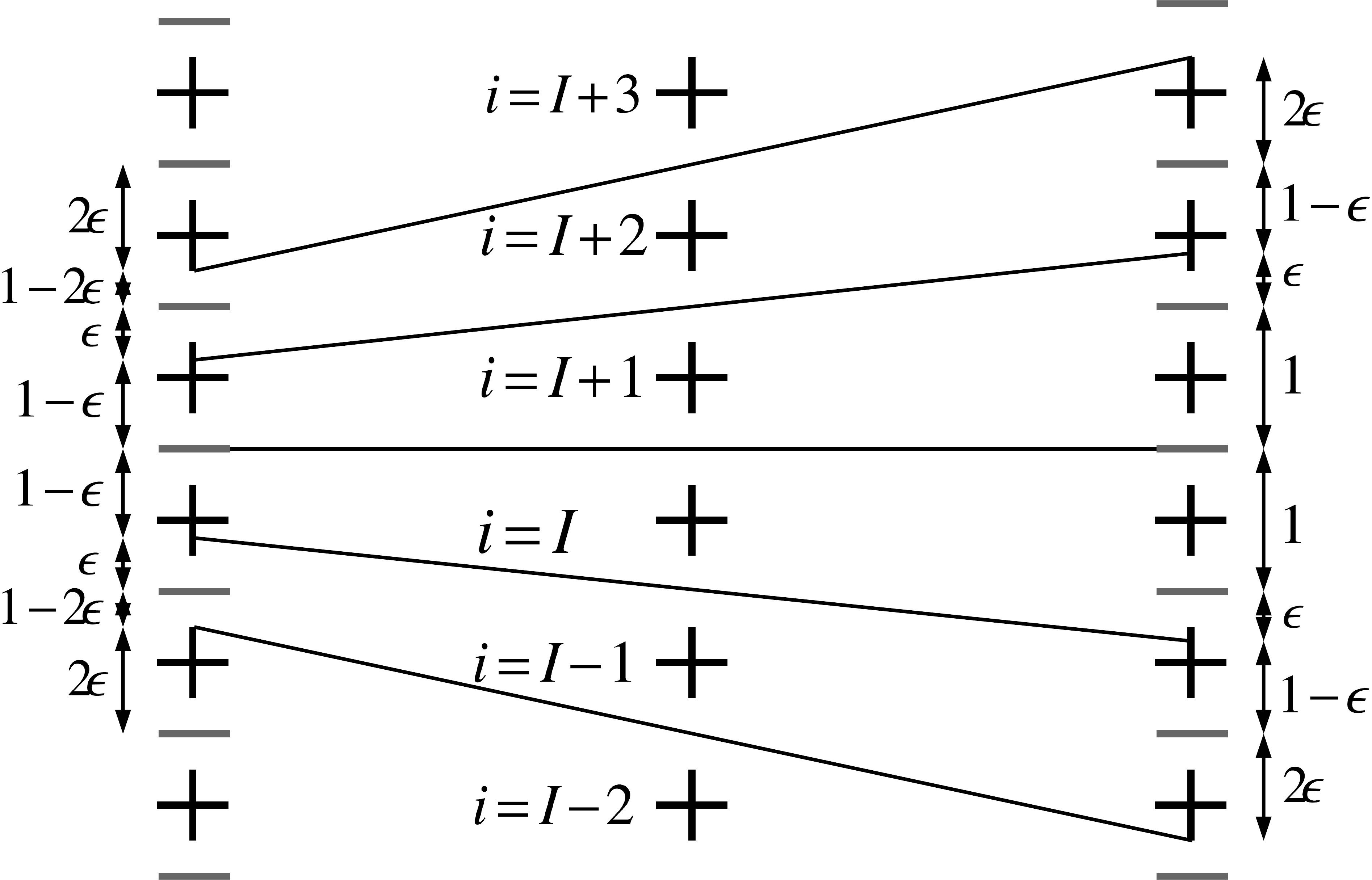}
\end{center}
\caption{Surface overlaps $\Delta A^\pm_{i,n}$ for model problem with diverging field lines.} 
\label{fig_model_coordfree}
\end{figure}

We use the integration method (equation \refeq{par_grad_coordfree}) for the discretisation of the parallel gradient: 
\begin{align}
\left(\mathbf{Q}^\pm\mathbf{u}\right)_{i,k}=\pm\frac{1}{h\Delta z}\left(\sum\limits_{n}\Delta A^\pm_{i,n}u_{n,k\pm1}-h u_{i,k}\right)
\end{align}
We consider the same example as described in \ref{app_inhom}. The surface overlaps are (see fig.~\ref{fig_model_coordfree}):
\begin{align}
\Delta A^+_{i,i}=&h\begin{cases}
1-\epsilon\left(I-i\right) & \text{for } i\leq I, \\
1-\epsilon\left(i-I-1\right) &\text{for } i>I,
\end{cases} \\
\Delta A^+_{i,i-1}=&h\begin{cases}
\epsilon\left(I-i+1\right) & \text{for } i\leq I, \\
0 &\text{for } i>I,
\end{cases} \\
\Delta A^+_{i,i+1}=&h\begin{cases}
0 & \text{for } i\leq I, \\
\epsilon\left(i-I\right) &\text{for } i>I,
\end{cases} \\
\Delta A^-_{i,i}=&h\begin{cases}
1-\epsilon\left(I-i+1\right) & \text{for } i\leq I, \\
1-\epsilon\left(i-I\right) &\text{for } i>I,
\end{cases} \\
\Delta A^-_{i,i-1}=&h\begin{cases}
0 & \text{for } i\leq I, \\
\epsilon\left(i-I-1\right) &\text{for } i>I,
\end{cases} \\
\Delta A^-_{i,i+1}=&h\begin{cases}
\epsilon\left(i-I-1\right) & \text{for } i\leq I, \\
0 &\text{for } i>I,
\end{cases}
\end{align}


One may again easily work out the expression for the discrete parallel diffusion operator, and we give again only the result for a structure which varies slowly in the $x$-direction i.e.~$u_{i,k}=\tilde{u}_k$.
\begin{align}
&\left(\mathbf{D}_\parallel^{supp}\tilde{\mathbf{u}}\right)_{i,k}=
\frac{-2\tilde{u}_k+\tilde{u}_{k-1}+\tilde{u}_{k+1}}{\Delta s^2}+\frac{\epsilon}{\Delta s}\dfrac{\tilde{u}_{k+1}-\tilde{u}_{k-1}}{2\Delta s}
\label{model_dist_sc}
\end{align}
In contrast to expression \refeq{model_dist_s1} no oscillation around $i=I$ is present.

\subsection{Third order polynomial interpolation}\label{app_polint}
\begin{figure}[!htb]
a)\newline
\includegraphics[width=1.0\linewidth]{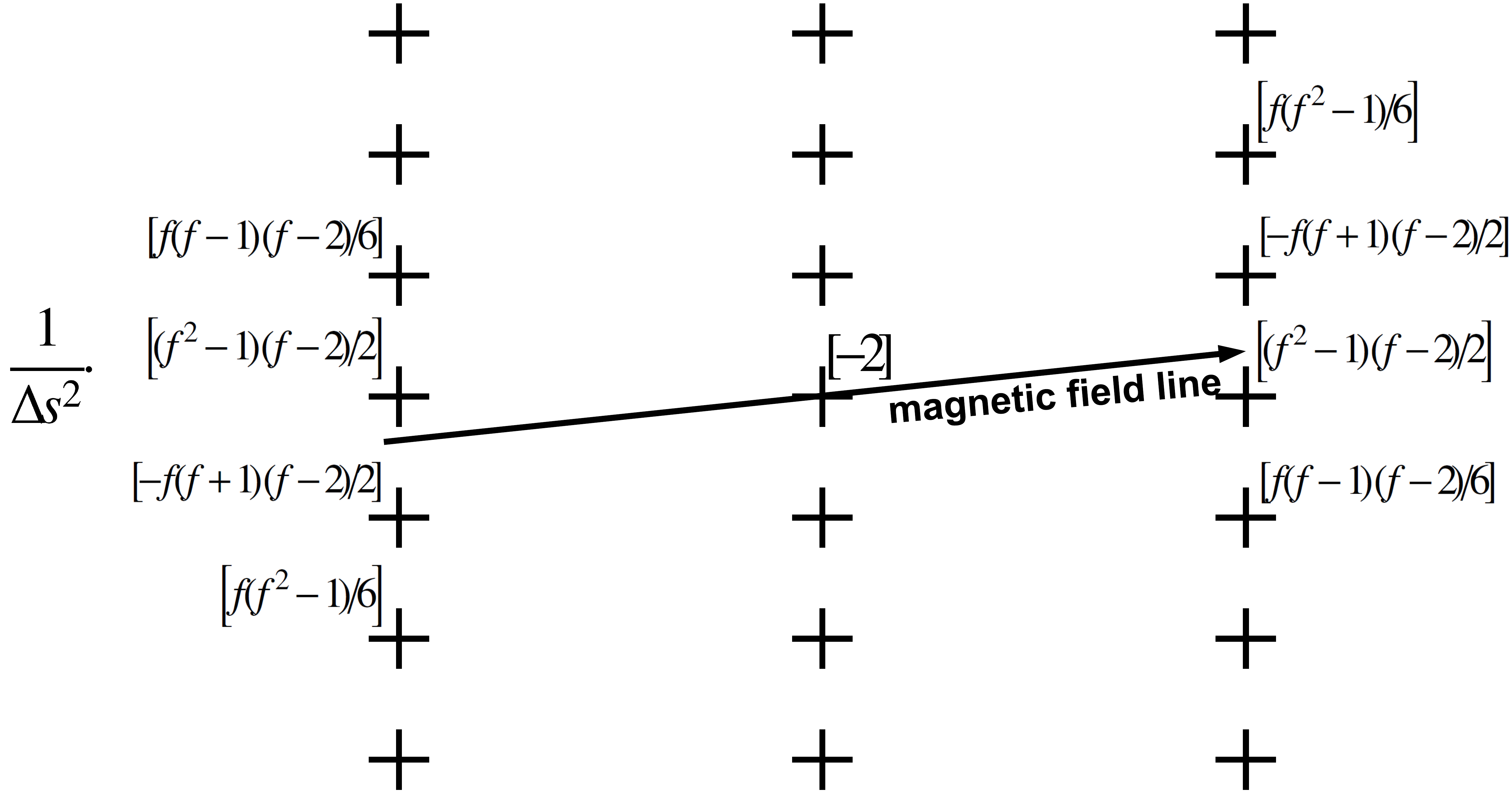}\newline
b)\newline
\includegraphics[width=1.0\linewidth]{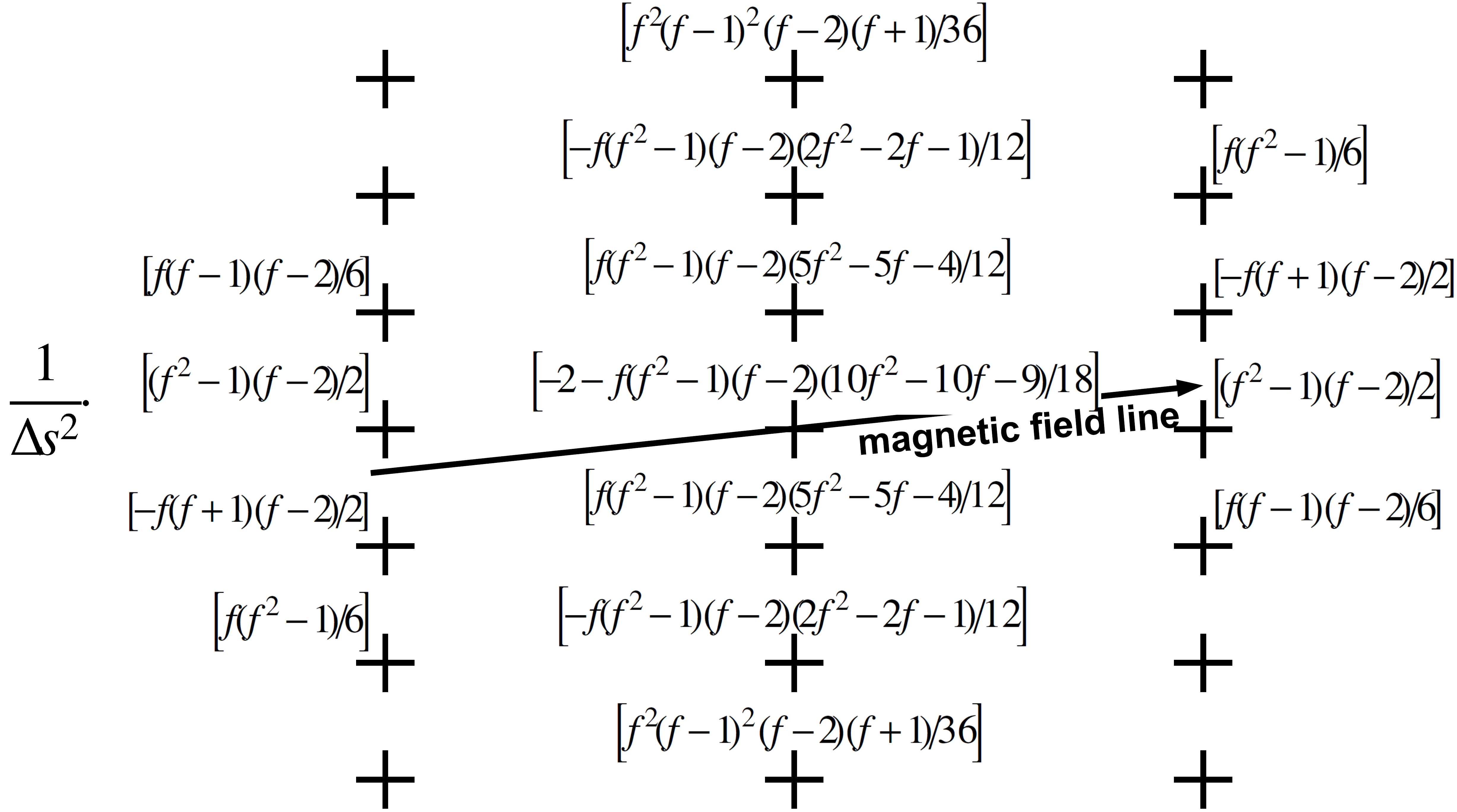}
\caption{Discrete parallel diffusion operator for two-dimensional model problem with third order polynomial interpolation. a) Naive scheme, b) Support scheme.} 
\label{fig_modelpol}
\end{figure}

We consider again a homogeneous magnetic field, i.e.~$n_i^{\pm}=0,\,f_i^\pm=f=const.$ but apply a third order polynomial interpolation for the computation of the quantity at the penetration points. The problem, which contains at least one inner grid point is a $7\times3$ grid. The parallel gradient matrices are:
\begin{align}
&\mathbf{Q}^+=-\left(\mathbf{Q}^-\right)^T=\notag\\
&\frac{1}{\Delta s}\left(\begin{smallmatrix}
 -1  &    &    &    &    &    &    & c  & b  & a  &    &    &    &    &    &    &    &    &    &    &   \\
     & -1 &    &    &    &    &    & d  & c  & b  & a  &    &    &    &    &    &    &    &    &    &   \\
     &    & -1 &    &    &    &    &    & d  & c  & b  & a  &    &    &    &    &    &    &    &    &   \\
     &    &    & -1 &    &    &    &    &    & d  & c  & b  & a  &    &    &    &    &    &    &    &   \\     
     &    &    &    & -1 &    &    &    &    &    & d  & c  & b  & a  &    &    &    &    &    &    &   \\         
     &    &    &    &    & -1 &    &    &    &    &    & d  & c  & b  &    &    &    &    &    &    &   \\  
     &    &    &    &    &    & -1 &    &    &    &    &    & d  & c  &    &    &    &    &    &    &   \\
     &    &    &    &    &    &    & -1 &    &    &    &    &    &    & c  &  b & a  &    &    &    &   \\     
     &    &    &    &    &    &    &    & -1 &    &    &    &    &    & d  & c  &  b & a  &    &    &   \\              
     &    &    &    &    &    &    &    &    & -1 &    &    &    &    &    & d  & c  &  b & a  &    &   \\                   
     &    &    &    &    &    &    &    &    &    & -1 &    &    &    &    &    & d  & c  &  b & a  &   \\
     &    &    &    &    &    &    &    &    &    &    & -1 &    &    &    &    &    & d  & c  &  b & a \\            
     &    &    &    &    &    &    &    &    &    &    &    & -1 &    &    &    &    &    & d  & c  & b \\  
     &    &    &    &    &    &    &    &    &    &    &    &    & -1 &    &    &    &    &    & d  & c \\   
 c   & b  & a  &    &    &    &    &    &    &    &    &    &    &    & -1 &    &    &    &    &    &   \\   
 d   & c  & b  & a  &    &    &    &    &    &    &    &    &    &    &    & -1 &    &    &    &    &   \\  
     & d  & c  & b  &  a &    &    &    &    &    &    &    &    &    &    &    & -1 &    &    &    &   \\      
     &    &  d & c  &  b &  a &    &    &    &    &    &    &    &    &    &    &    & -1 &    &    &   \\     
     &    &    & d  &  c &  b & a  &    &    &    &    &    &    &    &    &    &    &    & -1 &    &   \\   
     &    &    &    &  d & c  &  b &    &    &    &    &    &    &    &    &    &    &    &    & -1 &   \\  
     &    &    &    &    & d  & c  &    &    &    &    &    &    &    &    &    &    &    &    &    & -1     
\end{smallmatrix}\right)\notag
\end{align}
with:
\begin{align}
&a=\frac{f^2(f-1)}{6},& b=-\frac{f(f+1)(f-2)}{2},\notag\\
&c=\frac{(f^2-1)(f-2)}{2},& d=\frac{f(f-1)(f-2)}{6},
\end{align}
The only inner point is $i=4$, $k=2$, which corresponds to the 11th row of the matrices $\mathbf{Q}^\pm$. The discrete parallel diffusion operator for the naive scheme is shown in fig.~\ref{fig_modelpol}a and for the support scheme in fig.~\ref{fig_modelpol}b. Again, the stencil for the support scheme is bigger and if the penetration points coincide with grid points $(f=0,1)$ both schemes yield again the standard second order finite difference expression. 

The action of the parallel gradient on a mode $u=\exp\left(ik_xx+ik_zz\right)$ yields:
\begin{align}
\mathbf{D}_\parallel^{naive}\mathbf{u}\approx& \bigg[
-k_{\parallel}^2
+\frac{1}{12}k_{\parallel}^4\Delta s^2
-\frac{(k_xh)^4}{\Delta s^2}\frac{f(f-1)(f+1)(f-2)}{12} \notag \\
&+\mathcal{O}\left(\frac{(k_xh,k_z\Delta z)^6}{\Delta s^2}\right)\bigg]\mathbf{u}, 
\label{taylor_naive_pol}
\end{align}
\begin{align}
\mathbf{D}_\parallel^{supp}\mathbf{u}\approx& \bigg[
-k_{\parallel}^2
+\frac{1}{12}k_{\parallel}^4\Delta s^2
+k_{\parallel}\mathcal{O}\left(\frac{(k_xh,k_z\Delta z)^5}{\Delta s}\right) \notag \\
&+\mathcal{O}\left(\frac{(k_xh,k_z\Delta z)^8}{\Delta s^2}\right)\bigg]\mathbf{u}.
\label{taylor_supp_pol}
\end{align}
The error term $k_{\parallel}^4\Delta s^2/12$ arises in both schemes and represents the discretisation error in the parallel direction. This error could be eliminated by using a fourth order finite difference expression along magnetic field lines, where the stencil would cover also planes $k\pm2$. Although expression \refeq{taylor_supp_pol} might suggest that the error for the support scheme scales with respect to poloidal resolution like $\left(k_xh\right)^5$, it actually scales as the naive scheme also only like $\left(k_xh\right)^4$, since the parallel discretisation error is also dependent on $\left(k_xh\right)^4$, i.e.:
\begin{align}
k_{\parallel}^4\Delta s^2=\frac{\left(k_x\sin\theta+k_z\cos\theta\right)^4\left(\left(fh\right)^2+\Delta z^2\right)^2}{\Delta s^2},
\end{align}
which contains also terms $\propto\left(k_xh\right)^4/\Delta s^2$. Therefore, for $k_{\parallel}\neq 0$ modes the schemes \textbf{S-1}, \textbf{N-3} and \textbf{S-3} exhibit the same scaling for the error with respect to poloidal resolution. However, only those terms  where $k_{\parallel}$ can not be factored out, represent a directional error and are responsible for numerical perpendicular 'diffusion'. Therefore, for the naive scheme the numerical perpendicular diffusion scales like $\left(k_xh\right)^4/\Delta s^2$, whereas for the support scheme like $\left(k_xh\right)^8/\Delta s^2$.

  \bibliographystyle{elsarticle-num} 
  \bibliography{bibliography.bib}





\end{document}